\documentclass{article}

\usepackage{graphicx,amsmath,amsfonts,amscd,amssymb,bm,bbm,url,color,latexsym,fge,mathdots,mathtools,mathabx}
\usepackage[T1]{fontenc}
\usepackage{tgpagella}
\usepackage{fullpage}
\usepackage[small,bf]{caption}
\usepackage[inline]{enumitem}
\usepackage{mwe}
\usepackage{subcaption}
\usepackage{microtype}
\usepackage{algorithm}
\usepackage{algorithmicx}
\usepackage{algpseudocode}
\usepackage{hyperref}
\usepackage[nameinlink]{cleveref}
\usepackage{framed}
\usepackage{comment}
\usepackage{wrapfig}
\usepackage{tikz}
\usepackage{pgfplots}
\usepackage[margin=0.95in]{geometry}
\usepackage{tikz}
\usepackage{cleveref}

\usepackage{authblk}
\providecommand{\keywords}[1]{\textbf{\textit{Index terms---}} #1}

\allowdisplaybreaks

\hypersetup{
    colorlinks=true,%
    citecolor=blue,%
    filecolor=blue,%
    linkcolor=blue,%
    urlcolor=blue
}
\usepackage[toc, page]{appendix}
\renewcommand{\mathbf}{\boldsymbol}

\newcommand{\conv}{\circledast}
\newcommand{\cconv}{  \boxasterisk }
\newcommand{\mb}{\mathbf}
\newcommand{\mc}{\mathcal}

\newcommand{\bb}{\mathbb}

\newcommand{\set}[1]{\left\{ #1 \right\}}

\newcommand{\reals}{\bb R}

\newcommand{\eps}{\varepsilon}
\newcommand{\vphi}{\varphi}
\newcommand{\R}{\reals}

\newcommand{ \res }[2]{\mb \iota_{ #1 \rightarrow #2  }}

\newcommand{ \Brac }[1]{\left\lbrace #1 \right\rbrace}
\newcommand{ \brac }[1]{\left[ #1 \right]}
\newcommand{ \paren }[1]{ \left( #1 \right) }

\DeclareMathOperator{\supp}{supp}

\DeclareMathOperator{\sign}{sign}
\DeclareMathOperator{\grad}{grad}

\newcommand{\di}{{n_0}}
\newcommand{\Di}{{n}}
\newcommand{\sample}{{m}}

\newcommand{\wh}{\widehat}
\newcommand{\wc}{\widecheck}
\newcommand{\wt}{\widetilde}
\newcommand{\ol}{\overline}

\newcommand{\norm}[2]{\left\| #1 \right\|_{#2}}
\newcommand{\nrm}[1]{\left\Vert#1\right\Vert}
\newcommand{\abs}[1]{\left| #1 \right|}
\newcommand{\innerprod}[2]{\left\langle #1,  #2 \right\rangle}

\newcommand{\iter}[1]{^{(#1)}}

\newcommand*\samethanks[1][\value{footnote}]{\footnotemark[#1]}

\numberwithin{equation}{section}

\newcommand{\mr}{\mathrm}

\newcommand{\shift}[2]{{\mathrm{s}_{#2}}\left[#1\right]}
\newcommand{\Shift}[2]{{\wh{\mathrm{s}}_{#2}}\left[#1\right]}

\newcommand{\Psit}[1]{\Psi_\text{#1}}
\newcommand{\phit}[1]{\vphi_\text{#1}}
\newcommand{\bhat}[1]{\wh{\mb{#1}}}

\newcommand{\edited}[1]{{\color{black}{#1}}}
\pgfplotsset{compat=1.14}

\begin{document}

\title{Short-and-Sparse Deconvolution -- A Geometric Approach}
\author[$\dagger$]{Yenson Lau\thanks{\normalsize These authors contributed equally to this work.}$^,$}
\author[$\sharp$]{Qing Qu\samethanks$^,$}
\author[$\dagger$]{Han-Wen Kuo}
\author[$\Diamond$]{Pengcheng Zhou}
\author[$\flat$]{Yuqian Zhang}
\author[$\dagger$,$\ddagger$]{John Wright}

\affil[$\dagger$]{Department of Electrical Engineering and Data Science Institute, Columbia University}
\affil[$\sharp$]{Center for Data Science, New York University}
\affil[$\Diamond$]{Department of Statistics and Center for Theoretical Neurosicence, Columbia University}
\affil[$\flat$]{Department of Computer Science, Cornell University}
\affil[$\ddagger$]{Department of Applied Physics and Applied Mathematics, Columbia University }

 \maketitle

%\author[1]{Author C\thanks{C.C@university.edu}}
%\author[2]{Author D\thanks{D.D@university.edu}}
%\author[2]{Author E\thanks{E.E@university.edu}}

%\author{Yenson Lau$^*$, Qing Qu$^*$, Han-Wen Kuo, Pengcheng Zhou, Yuqian Zhang, John Wright}
%\author{Qing Qu\thanks{ Department of Electrical Engineering, Columbia University, New York, 10027. Email: \{qq2105,yz2409,jw2966\}@columbia.edu.}, Yuqian Zhang$^\ast$, and John Wright$^\ast$ %\and
%\IEEEauthorblockN{Qing Qu}
%\IEEEauthorblockA{Department of Electrical Engineering\\
%Columbia University\\
%%New York, NY 10027\\
%qq2105@columbia.edu }
%\and
%\IEEEauthorblockN{Yuqian Zhang}
%\IEEEauthorblockA{Department of Electrical Engineering\\
%Columbia University\\
%%New York, NY 10027\\
%yz2409@columbia.edu }
%\and
%\IEEEauthorblockN{John Wright}
%\IEEEauthorblockA{Department of Electrical Engineering\\
%Columbia University\\
%%New York, NY 10027\\
%jw2966@columbia.edu }
%}

\begin{abstract}

%The problem of locating repetitive unknown motifs in signals appears ubiquitously in neuroscience, computational imaging, microscopy/biomedical data analytics, and more. These problems can be cast as a (multiple-kernel) blind deconvolution problem, whose task is to simultaneously recover a motif $\mb a_0$ and an activation signal $\mb x_0$ from their convolution $\mb y = \mb a_0 \conv \mb x_0$. Although the task is ill-posed in the most general form, problems in practice often exhibits extra low-dimensional structures -- \emph{short} motif $\mb a_0$ and \emph{sparse} activation signal $\mb x_0$. In this work, we investigate the \emph{short-and-sparse} (SaS) deconvolution problems. We consider the most natural nonconvex formulation, and constraint the problem over the sphere. By studying the geometric structure of the nonconvex landscape, we show how to develop practical optimization methods, that solves the problem to the target solution $(\mb a_0,\mb x_0)$ up to a shift ambiguity. The new geometric insights lead us to new optimization strategies that even works in very challenging scenarios. Moreover, We show that the new geometric insights can also be extended to deal with the problem of recovering multiple unknown motifs (a.k.a. convolutional dictionary learning), where a superposition of multiple convolution signals is observed. We demonstrate the effectiveness of the proposed methods on problems such as calcium imaging, spike sorting, super-resolution microscopy imaging, and microscopy data analysis.

Short-and-sparse deconvolution (SaSD) is the problem of extracting localized, recurring motifs in signals with spatial or temporal structure. Variants of this problem arise in applications such as image deblurring, microscopy, neural spike sorting, and more. The problem is challenging in both theory and practice, as natural optimization formulations are nonconvex. Moreover, practical deconvolution problems involve smooth motifs (kernels) whose spectra decay rapidly, resulting in poor conditioning and numerical challenges. This paper is motivated by recent theoretical advances \cite{zhang2017global,kuo2019geometry}, which characterize the optimization landscape of a particular nonconvex formulation of SaSD. This is used to derive a \emph{provable} algorithm which exactly solves certain non-practical instances of the SaSD problem. We leverage the key ideas from this theory (sphere constraints, data-driven initialization) to develop a \emph{practical} algorithm, which performs well on data arising from a range of application areas. We highlight key additional challenges posed by the ill-conditioning of real SaSD problems, and suggest heuristics (acceleration, continuation, reweighting) to mitigate them. Experiments demonstrate both the performance and generality of the proposed method.

% \noindent\yl{What is SaS and what is its relationship to the discourse community? What makes this problem interesting / challenging? What do we contribute? What should readers take away from this?}

% Short-and-sparse (SaS) deconvolution problems are frequently encountered in signal processing applications such as image deblurring, microscopy, neural spike sorting, and more.
\end{abstract}

\keywords{sparse blind deconvolution, convolutional dictionary learning, computational imaging, nonconvex optimization, alternating descent methods.}

\newpage

\section{Introduction}
% !TEX root = ../main.tex

%\jw{posting on the arxiv preserves / publicizes commented text. please sanitize all the files before posting}

%\jw{this section is final aside from henry's polish on the figure. please do not make any other changes (even small) without discussing them first.}

\begin{figure*}[htbp]
  \centering
  \input{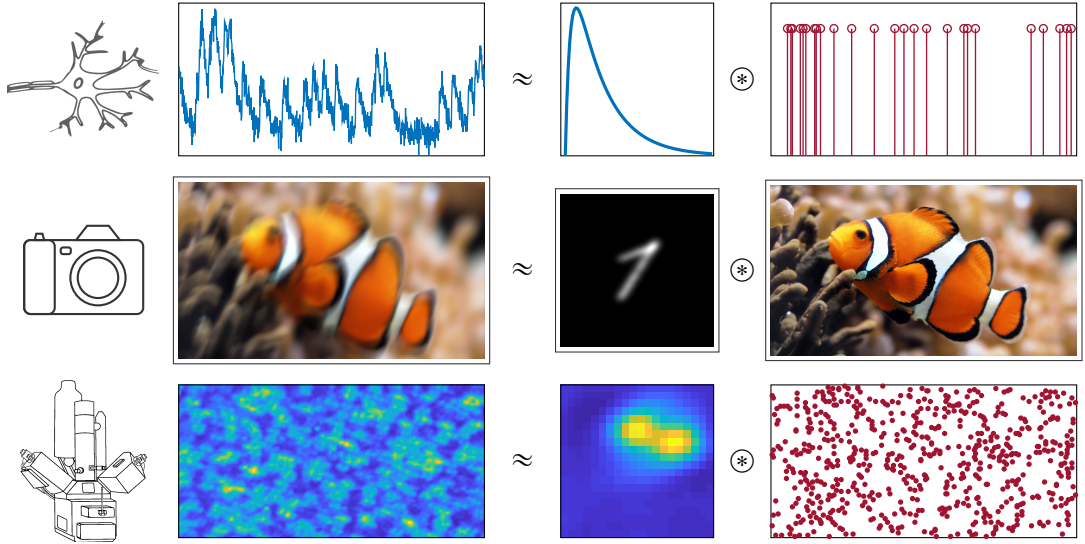}
  \caption{\textbf{Natural signals with short-and-sparse structure.}
  In calcium imaging (top), each neuronal spike induces a fluoresence pattern measuring a transient increase in calcium concentration.
  In photography (middle), photos with sharp edges (sparse in the gradient domain) are often obfuscated by blurring due to shaking the camera.
  In scanning tunneling microscopy (bottom), dopants embedded in some base material produce individual electronic signatures.
  For each of these cases, the observed signal can be modeled as a convolution between a {\em short} kernel and a {\em sparse} activation map.}
  \label{fig:SaSD-examples}
\end{figure*}

Signals in medical/scientific/natural imaging can often be modeled as superpositions of basic, recurring motifs \edited{(see \Cref{fig:SaSD-examples} for an illustration)}. For example, in calcium imaging \cite{stosiek2003vivo,grienberger2012imaging}, the excitation of neurons produces short pulses of fluorescence, repeating at distinct firing times.
In the material and biological sciences, repeated motifs often encode crucial information about the subject of interest; e.g., in nanomaterials these motifs correspond to defects in the crystal lattice due to doping \cite{Cheung2018DictionaryLI}. In all of these applications, the motifs of interest are {\em short}, and they are {\em sparsely} distributed within the sample of interest. Signals with this short-and-sparse structure also arise in natural image processing: when a blurry image is taken due to the resolution limit or malfunction of imaging procedure, it can be modeled as a short blur pattern applied to a visually plausible sharp image \cite{chan1998total,rust2006stochastic,levin2011understanding}.

Mathematically, an observed signal $\mb y$ with this \emph{short-and-sparse} (SaS) structure can be modeled as a {\em convolution}\footnote{For simplicity we follow the convention of \cite{kuo2019geometry} and use cyclic convolution throughout this paper, unless otherwise specified. The choice is superficial; any algorithms and results discussed here should also apply to linear convolution with minor modifications.} of a \emph{short} signal $\mb a_0\in\R^{\di}$ and a much longer \emph{sparse} signal $\mb x_0\in \R^\sample $  $( \sample \gg \di)$:
\begin{equation}  \label{eqn:sas-mdl}
  \mb y \quad =\quad \mb a_0 \;\;\conv \;\; \mb x_0.
\end{equation}
In all of the above applications, the signals $\mb a_0$ and $\mb x_0$ are not known ahead of time. The {\em short-and-sparse deconvolution} (SaSD) problem asks us to recover these two signals from the observation $\mb y$. This is a challenging \emph{inverse} problem: natural optimization formulations are {\em nonconvex} and have many equivalent solutions. The kernel $\mb a_0$ is often smooth, and hence attenuates high frequencies.
Although study of the SaSD problem stretches back several decades and across several disciplines \cite{haykin1994blind,loke1995least,kundur1996blind}, the need for efficient, reliable, and general purpose optimization methods remains.

One major challenge associated with developing methods for SaSD arises from our relatively limited understanding of the global geometric structure of nonconvex optimization problems. Our goal is to recover this ground truth $(\mb a_0,\mb x_0)$ (perhaps up to some trivial ambiguities), which typically requires us to obtain a globally optimal solution to a nonconvex optimization problem. This is impossible in general. Fortunately, recent theoretical evidence \cite{zhang2018structured,kuo2019geometry} guarantees that the SaSD problem can solved efficiently under certain idealized assumptions. Using an appropriate selection of optimization domain and a specific initialization scheme,
these results yield provable methods that solve certain instances of SaSD in polynomial time.

Unfortunately, practical SaSD problems raise additional challenges beyond the assumptions in theory, causing the provable methods \cite{zhang2018structured,kuo2019geometry} to fail on real problem instances. While the emphasis of \cite{zhang2017global,kuo2019geometry} is on theoretical guarantees, here we focus on practical computation. We show how to combine ideas from this theory with heuristics that better address the properties of practical deconvolution problems, to build a novel method that performs well on data arising in a range of application areas. Many of our design choices are natural and have a strong precedent in the literature. We will show how these natural choices help to cope with the (complicated!) geometry of practically occurring deconvolution problems. A critical issue in moving from theory to practice is the poor conditioning of naturally-occurring deconvolution problems: we show how to address this with a combination of ideas from sparse optimization, including momentum, continuation, and reweighting. The end result is a general purpose method, which we demonstrate on data for spike recovery \cite{friedrich2017fast} and neuronal localization \cite{pnevmatikakis2016simultaneous} from calcium imaging data, as well as fluorescence microscopy \cite{rust2006stochastic}.

\paragraph{Organization of the paper.} The remainder of the paper is organized as follows. \Cref{sec:problem} introduces key aspects of SaSD, and \Cref{sec:geometry} shows how they play out in a theoretical analysis of SaSD, culminating in a provable algorithm grounded in geometric intuition. In \Cref{sec:algorithm}, we discuss how to combine this intuition with additional heuristics to create
%efficient \jw{should be practical, not efficient}
 practical methods. \Cref{sec:exp_synthetic} revisits and demonstrates these ideas in a simulated setting.
\Cref{sec:exp} illustrates the performance of our method on data drawn from a number of applications. Finally, \Cref{sec:conclusion} reviews the literature, and poses interesting future directions.
%\jw{ i just cut the last sentence about the appendices}
%Notation is introduced in Appendix \ref{app:basic}, and we refer interested readers to Appendix \ref{app:algorithm} and Appendix \ref{app:details} for more technicalities. \yl{last sentence needs work.} \jw{suggest we just cut it.}

\paragraph{Reproducible research.} The code for implementations of our algorithms can be found online:
\begin{center}
	\url{https://github.com/qingqu06/sparse_deconvolution}.
\end{center}
For more details of our work on SaSD, we refer interested readers to our project website
\begin{center}
	\url{https://deconvlab.github.io/}.
\end{center}

\section{Two Key Intuitions for SaS Deconvolution}
\label{sec:problem}
% !TEX root = ../main.tex

%\yl{This section is now closed; please let John or I know \emph{before} making any changes.}

We begin by describing two basic intuitions for SaS deconvolution, which play an important role in the geometry of optimization and the design of efficient methods.

\begin{figure*}[h]
\centering
\input{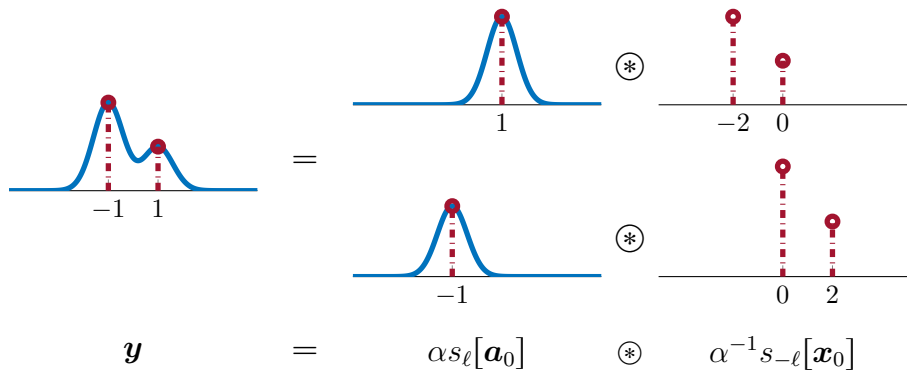}
\caption{ \textbf{Scaling-shift symmetry.} The SaS convolution model exhibits a scaled shift symmetry: $ \alpha s_\ell[\mb a_0]$ and $ \alpha^{-1} s_{-\ell}[\mb x_0]$ have the same convolution as $\mb a_0$ and $\mb x_0$. Therefore, the ground truth $(\mb a_0,\mb x_0)$ can only by identified up to some scale and shift ambiguity.}
\label{fig:symmetry-demo}
\end{figure*}

\paragraph{Symmetry structure.} The SaS model exhibits a basic {\em scaled shift symmetry}: for any nonzero scalar $\alpha$ and cyclic shift  $s_\ell\brac{ \cdot }$
\begin{align*}
 \mb y \;=\;  \mb a_0 \conv \mb x_0 \;=\; \paren{ \pm \alpha s_\ell \brac{ \mb a_0  }  }  \; \conv \; \paren{ \pm \alpha^{-1} s_{-\ell} \brac{ \mb x_0  }  }.
\end{align*}
In other words, shifting $\mb a_0$ to the right by $\ell$ samples and shifting $\mb x_0$ to the left by the same amount leaves the convolution $\mb a_0 \conv \mb x_0$ unchanged (see Figure \ref{fig:symmetry-demo}). We can therefore only expect to recover the ground truth $(\mb a_0,\mb x_0)$ up to some scaling and some shift. As a result, natural optimization formulations for SaSD exhibit multiple global minimizers, corresponding to these scaled shifts of the ground truth. Due to the existence of multiple discrete global minimizers, natural formulations are {\em nonconvex}. Fortunately, this symmetry structure often creates leads to benign objective landscapes for optimization; two such examples for SaSD are \cite{zhang2018structured,kuo2019geometry}.

\begin{figure}[t!]
	\centering
	\input{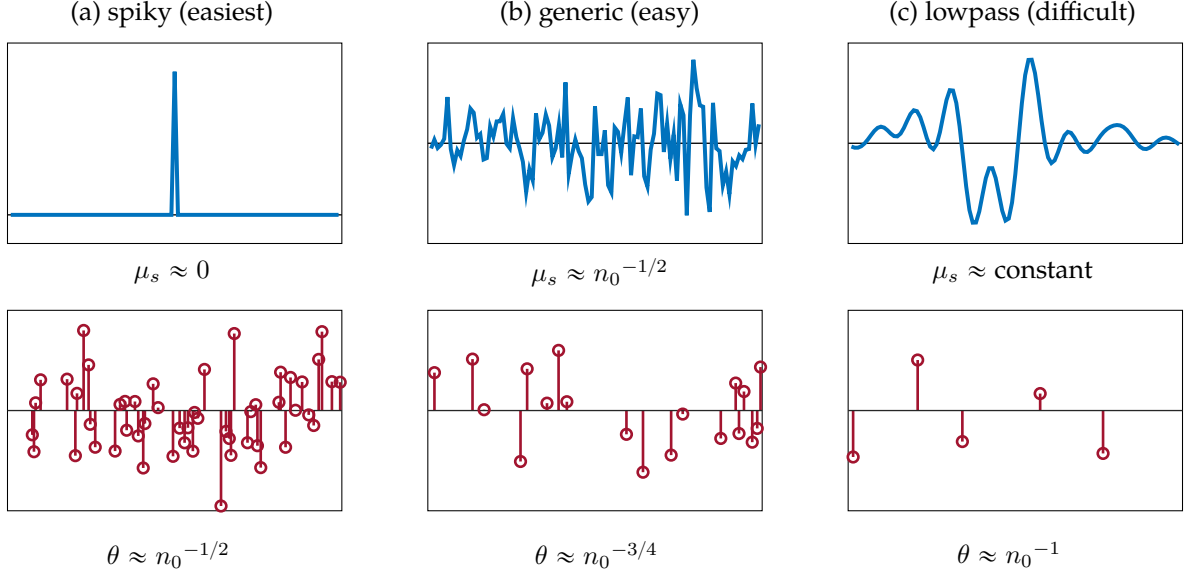}
	\par\vspace{.1in}
	\caption{ \textbf{Sparsity-coherence tradeoff} \cite{kuo2019geometry}: examples with varying coherence parameter $\mu_s(\mb a_0)$ and sparsity rate $\theta$ (i.e., probability a given entry is nonzero). Smaller shift-coherence $\mu_s(\mb a_0)$ allows SaSD to be solved with higher $\theta$, and vice versa. In order of increasing difficulty:
  (a) when $\mb a_0$ is a Dirac delta function, $\mu_s(\mb a_0) = 0$; (b) when $\mb a_0$ is sampled from a uniform distribution on the sphere $\bb S^{\di-1}$, its shift-coherence is roughly $\mu_s(\mb a_0) \approx \di^{-1/2}$
  %, where the allowable sparsity rate $\theta$ decreases as $\mu_s(\mb a_0)$ increases
  ; (c) when $\mb a_0$ is low-pass, $\mu_s(a_0)\rightarrow\textrm{const.}$ as $\di$ grows.}
\label{fig:sparsity-coherence}
\end{figure}

\paragraph{Sparsity-coherence tradeoff.}

Clearly, not all SaSD problems are equally easy to solve. Problems with denser $\mb x_0$ are more challenging. Moreover, there is a basic tradeoff between the sparsity of the spike train $\mb x_0$ and the properties of the kernel $\mb a_0$. If $\mb a_0$ is smooth (e.g., Gaussian), then each occurrence of $\mb a_0$ would, on average, need to be relatively far apart to be distinguishable; i.e.\ $\mb x_0$ would have to be sparser.
%\jw{here -- yenson, we should discuss}
Conversely, denser instances of $\mb x_0$ should be allowable if $\mb a_0$ is ``spikier''.\footnote{Similar tradeoffs occur in non-blind deconvolution where $\mb a_0$ is known (e.g.\ \cite{candes2014towards}) and in other inverse problems.}
One way of formalizing this tradeoff is through the \emph{shift-coherence} of the kernel $\mb a_0$, which measures the ``similarity'' between $\mb a_0$ and its cyclic-shifts:
\begin{align}\label{eqn:self-coherence}
	\mu_s(\mb a_0) \doteq \max_{\ell \not =0  }   \abs{ \innerprod{ \frac{ \mb a_0}{ \norm{\mb a_0}{2} } }{  \frac{ \mathrm{s}_\ell \brac{\mb a_0} }{ \norm{\mb a_0}{2} }  }  } \in \brac{0,1}.
\end{align}
As $\mu_s(\mb a_0)$ increases, the shifts of $\mb a_0$ become more correlated and hence closer together on the sphere. \cite{kuo2019geometry} uses this quantity to study the behavior of a particular nonconvex formulation of SaSD. For generic choices of $\mb x_0$, such as $\mb x_0\sim\mc{BG}(\theta)$ drawn from a Bernoulli-Gaussian distribution, the {\em sparsity-coherence tradeoff} of \cite{kuo2019geometry} guarantees recoverability when the sparsity rate $\theta$ is sufficiently small relative to $\mu_s(\mb a_0)$. Intuitively speaking, this implies that SaSD problems with smaller $\mu_s(\mb a_0)$ tend to be ``easier'' to solve (\Cref{fig:sparsity-coherence}).

In the next section, we will use the idealized formulation of \cite{kuo2019geometry} to illustrate how these basic properties of the SaSD problem shape the landscape of optimization. In later sections, we will borrow these ideas to develop practical, general purpose methods. The major challenge in moving from theory to practice is in coping with highly coherent $\mb a_0$: in most practical applications, $\mb a_0$ is smooth and hence $\mu_s(\mb a_0)$ is large.

\section{Problem Formulation and Nonconvex Geometry}
\label{sec:geometry}
% !TEX root = ../main.tex

In this section, we summarize some recent algorithmic theory characterizing the optimization landscape of an idealized nonconvex formulation for SaSD \cite{zhang2017global,kuo2019geometry}, with the goal of applying the geometric intuition from this theory towards designing practical optimization methods.

\subsection{The Bilinear Lasso and its marginalization}
A natural idea for solving SaSD is to minimize a reconstruction loss $\psi(\mb a,\mb x)$ between $\mb a\conv\mb x$ and $\mb y$, plus a sparsity-promoting regularizer $g(\mb x)$ on $\mb x$. This can be achieved, for instance, by minimizing the squared reconstruction error in combination with an $\ell_1$-penalty on $\mb x$,
\begin{align}\label{eqn:bilinear-lasso}
  \boxed{\min_{\mb a,\mb x}\;\;  \Psit{BL}(\mb a, \mb x)\;\;\doteq\;\;
  \tfrac{1}{2}\norm{\mb y - \mb a\conv\mb x}{2}^2  \;+\; \lambda \norm{\mb x}{1}
  ,\qquad \mathrm{s.t.}\quad \mb a \in \bb S^{\Di-1}.}
\end{align}
This \emph{Bilinear Lasso} problem (BL) resembles the Lasso estimator in statistics \cite{tibshirani1996regression}, and is a nonconvex optimization problem.
The sparsity of the solution for $\mb x$ is controlled by the regularization penalty $\lambda$: a larger $\lambda$ leads to sparser $\mb x$, and vice versa\footnote{\cite{kuo2019geometry} suggests a good choice $\lambda \in \mc O(1/\sqrt{\theta n})$, where $\theta\in (0,1)$ denotes the sparsity level.}. We constrain $\mb a$ onto the sphere $\bb S^{\Di -1 }$, which reduces the scaling ambiguity into a sign ambiguity.
We also increase the dimension of $\mb a$ to $\Di = 3 \di -2$; this creates an objective landscape that allows various descent methods to recover a full shift of $\mb a_0$ and avoid any shift-truncation effects, upon the application of a simple data initialization scheme.

In this paper, we will also frequently refer to the \emph{marginalized} Bilinear Lasso cost
\begin{align}\label{eqn:lasso}
  \phit{BL} (\mb a) \; \doteq\; \min_{\mb x} \; \Psit{BL}(\mb a, \mb x)
  \; = \; \min_{\mb x}\; \tfrac{1}{2}\norm{\mb y - \mb a\conv\mb x}{2}^2  \;+\; \lambda \norm{\mb x}{1}.
\end{align}
It is clear that minimizing $\Psit{BL}(\mb a, \mb x)$ is equivalent to minimizing $\phit{BL}(\mb a)$ over $\mb a\in \bb S^{\Di-1}$.

\begin{figure*}[!t]
\centering
\captionsetup[sub]{font=normalsize,labelfont={bf,sf}}
\begin{minipage}[t][6cm]{0.33\textwidth}
\subcaption{a single shift $ \shift{ \mb a_0 }{\ell_1}$ }\label{subfig:one-shift}
\centering
	\includegraphics[width = 0.8\linewidth]{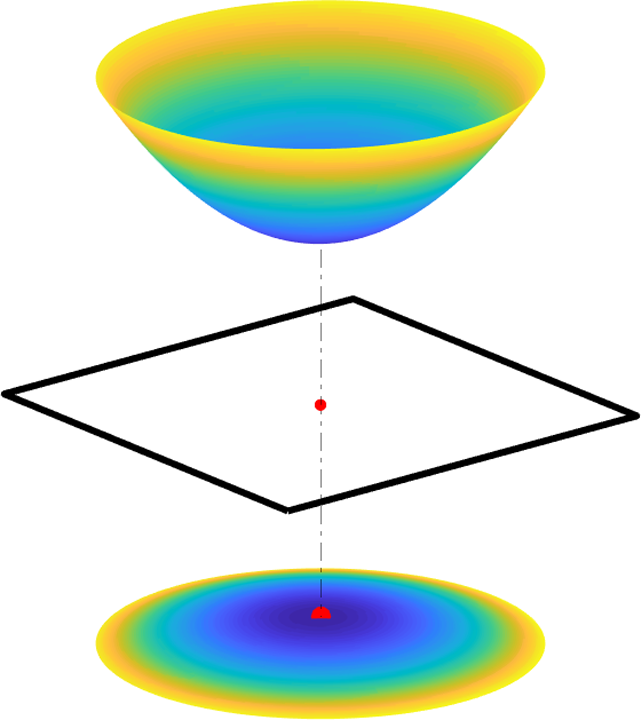}
\end{minipage}
\begin{minipage}[t][6cm][t]{0.33\textwidth}
\subcaption{two shifts $\shift{ \mb a_0 }{\ell_1}$, $\shift{ \mb a_0 }{\ell_2}$}\label{subfig:two-shifts}
\centering
	\includegraphics[width = 0.8\linewidth]{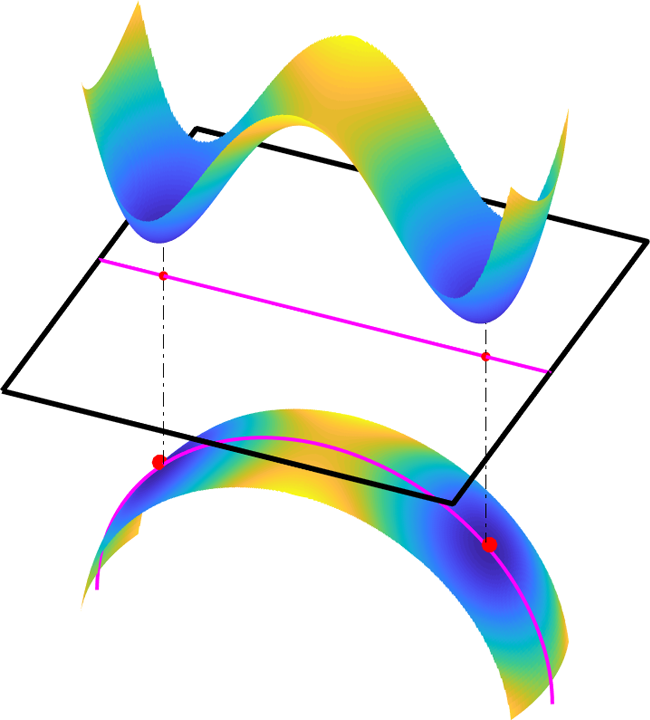}
\end{minipage}
\begin{minipage}[t][6cm][t]{0.33\textwidth}
\subcaption{multiple shifts}\label{subfig:multi-shifts}
%$\Brac{\Shift{ \mb a_0 }{\ell_k}}_{k=1}^4$
\centering
	\includegraphics[width = 0.8\linewidth]{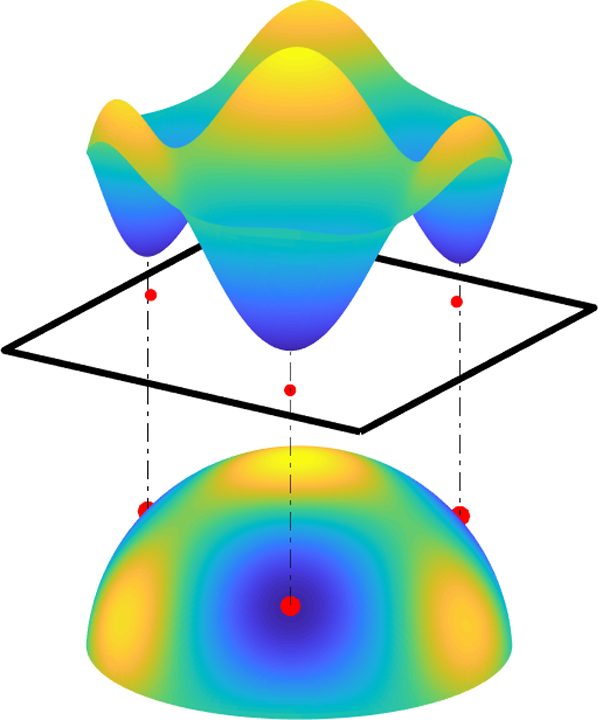}
\end{minipage}
\caption{ \textbf{Geometry of Approximate Bilinear Lasso loss $\phit{ABL}(\mb a)$} near superpositions of shifts of $\mb a_0$ \cite{kuo2019geometry}.
{\bf Top:} function values of $\phit{ABL}(\mb a)$ visualized as height. {\bf Bottom:} heat maps of $\phit{ABL}(\mb a)$ on the sphere $\bb S^{\Di-1}$.
{\bf (a)} the region near a single shift is strongly convex; {\bf (b)} the region between two shifts contains a saddle-point, with negative curvature pointing towards each shift and positive curvature pointing away; {\bf (c)} region near the span of several shifts of $\mb a_0$.}
\label{fig:shift-curvature}
\end{figure*}

\subsection{Structured nonconvexity and geometric properties}\label{subsec:geometry}
%\yl{There is some contention about shifting. When we use $\shift{\mb a_0}\ell$ in Section 2 it is technically a cyclic shift of $\iota\mb a_0$. Here there is a new notation being introduced: $\Shift{\mb a_0}\ell\doteq\iota_n^*s_{\iota_{n_0}}$. I am not sure if this new notation should be introduced at all, and if this issue is addressed perhaps it should be in a footnote instead.}

To understand the nonconvex optimization landscape of the Bilinear Lasso, it is natural to study the marginalized objective in \Cref{eqn:lasso}. The benefit of this approach is twofold: (i) for a fixed $\mb a$, the Lasso problem in \Cref{eqn:lasso} is convex w.r.t. $\mb x$, and (ii) the \emph{short} kernel $\mb a$ lives on a low dimensional manifold --- the space $\mb a\in \bb S^{\Di -1}$ is where measure concentrates when $\mb x_0$ is generic random and has high dimension ($\sample \gg \di$). Unfortunately, $\phit{BL}(\mb a)$ remains challenging for analysis; a major culprit is that the Lasso estimator in \Cref{eqn:lasso} does not usually admit closed-form solutions.

\paragraph{Approximate Bilinear Lasso.}
When $\mb a$ is incoherent ($\mu_s(\mb a)\approx0$), however, we approximately have $\norm{\mb a \conv \mb x}2^2 \approx \norm{\mb x}2^2$.
%This allows the term $\norm{\mb a \conv \mb x}2^2$ to be replaced by a simpler approximation $\norm{\mb x}2^2$},
Carrying this approximation through to \Cref{eqn:bilinear-lasso} yields an {\em Approximate Bilinear Lasso} (ABL) objective\footnote{As our focus here is on solving the Bilinear Lasso, we intentionally omit the concrete form of $\Psit{ABL}(\mb a)$ and $\phit{ABL}(\mb a)$. Readers may refer to Section 2 of \cite{kuo2019geometry} for more details.}
$\phit{ABL}(\mb a,\mb x) = \min_{\mb x} \Psit{ABL} (\mb a,\mb x)$, which satisfies $\phit{ABL}(\mb a) \approx \phit{BL}(\mb a)$
% \begin{align}\label{eqn:drop-quadratic}
% 	\phit{ABL}(\mb a)\;\; \approx\;\; \phit{BL}(\mb a)
%   %,\qquad\qquad \text{when}\qquad \mu_s(\mb a)\;\approx\;0;
% \end{align}
whenever $\mu_s(\mb a)\approx0$ \cite{kuo2019geometry}. For the purposes of our discussion, this objective serves as a valid simplification of the Bilinear Lasso when the true kernel is itself incoherent ($\mu_s(\mb a_0)\approx0$). Although such incoherence assumptions are stringent and impractical, $\phit{ABL}(\mb a)$ admits a simple analytical form and is more amenable to analysis as a result.

\paragraph{Geometry of $\phit{ABL}$ in the span of a few shifts.}
\newcommand{\saz}[1]{s_{#1}[\mb a_0]}

Under the assumptions that $\mb a_0$ is incoherent and $\mb x_0$ is generic, $\phit{ABL}(\mb a)$ enjoys a number of nice properties on the sphere $\bb S^{n-1}$. In particular, Kuo et al. \cite{kuo2019geometry} provides a geometrical characterization of the optimization landscape $\phit{ABL}(\mb a)$ near the span of several shifts\footnote{When optimizing over $\bb S^{\Di-1}$, $\Di=3\di-2$, we denote $\ell$-th (full) shift with the abuse of notation $\saz{\ell} = [\mb 0_\ell; \mb a_0; \mb 0_{\Di-\ell-\di}] \in \bb S^{\Di-1}$
%to denote a length-$\Di$ vector whose $\ell$-th to $(\ell+\di-1)$-th entries contain $\mb a_0$ with zeroes elsewhere
, for $\ell\in\set{0,\dots,\Di-\di}$. Each shift is a length-$m$ cyclic shift of $\mb a_0$, truncated to a length-$\Di$ window without removing any entries from $\mb a_0$.}
of $\mb a_0$:

\begin{enumerate}[leftmargin=*]
  \item {\em Near a single shift of $\mb a_0$.\;} Within a local neighborhood of each shift $\saz{\ell}$, the optimization landscape of $\phit{ABL}(\mb a)$ exhibits \emph{strong convexity} (\Cref{subfig:one-shift}), with a {\em unique} minimizer corresponding to a shift
  %\footnote{Here, as our optimization space is $\Di =3\di -2$, we use $\Shift{\mb a_0}{\ell}$ instead for denoting a zero-pad shift, where $\Shift{\mb a_0}{\ell} = \mathrm{s}_\ell[\begin{bmatrix} \mb 0_{\di-1} \\ \mb a_0 \\  \mb 0_{\di-1} \end{bmatrix}]$ for $-\di+1 \leq \ell \leq \di-1$. We refer readers to Appendix \ref{app:basic} for more details.}
  $\saz{\ell}$.

  \item {\em In the vicinity of two shifts.\;} Near the span of two shifts,
  \begin{align*}
     \mc S_{ \Brac{ \ell_1,\ell_2} } \;=\;\big\{\; \alpha_1 \saz{\ell_1} + \alpha_2 \saz{\ell_2} \;:\; \alpha_1,\; \alpha_2 \in \bb R \;\big\} \;\bigcap\; \bb S^{\Di-1},
  \end{align*}
  the only local minimizers are approximately $\saz{\ell_1}$ and $\saz{\ell_2}$.
  A saddle point $\mb a_s$ exists at the symmetric superposition of the shifts (i.e.\ $\mb a_s = \alpha_1 \saz{\ell_1} + \alpha_2 \saz{\ell_2}$ with $\alpha_1 \approx \alpha_2$), but can be escaped
  by taking advantage of the large negative curvature present\footnote{Here, negative curvature means that the Hessian exhibits negative eigenvalues, such that the function can be decreased by following the negative eigenvector direction.} (\Cref{subfig:two-shifts}).

  \item {\em In the vicinity of multiple shifts.\;} The geometric properties for two shifts carry over to those of multiple shifts of $\mb a_0$. Any local minimizers over
  \begin{align}\label{eqn:geometry-subspace}
  	\mc S_{\mc I} \;\doteq\; \big\{\;\textstyle\sum_{\ell \in \mc I} \;\alpha_{\ell} s_{\ell}\brac{\mb a_0}  \;:\; \alpha_{\ell} \in  \bb R\; \big\} \;\bigcap\; \bb S^{\Di-1}
  \end{align}
  are again close to signed shifts (\Cref{subfig:multi-shifts}). Any saddle-points present sit at symmetric superpositions of two or more shifts, and exhibit strong negative curvature in directions towards the participating shifts. Additionally, the function value of $\phit{ABL}(\mb a)$ increases when moving away from $\mc S_{\mc I}$.
\end{enumerate}
\cite{kuo2019geometry} proves that these geometric properties of $\phit{ABL}$ hold for sufficiently small\footnote{It is sufficient for $\abs{\mc I} = \mc O(\theta \di)$, where $\theta$ is the probability that any entry of $\mb x_0$ is nonzero \cite{kuo2019geometry}.} $\abs{\mc I}$ whenever the sparsity-coherence tradeoff $\di\theta \lessapprox \mu_s^{-1/2}(\mb a_0)$ is satisfied. This bound is stringent, however, and shows that the ABL formulation is unsuited for practical applications where $\mu_s(\mb a_0)$ often approaches one as $\di$ grows.
%\edited{Moreover, the regularizing effect created by problem symmetries is a fairly general phenomenon \cite{sun2015nonconvex} and, as \cite{zhang2017global,kuo2019geometry} have shown, its influence extends far beyond local minimizers.} \yl{<=== Suggest removing last sentence to avoid over-generalization.}

{\paragraph{Algorithmic implications and data-driven initialization.}}
The benign optimization landscape of $\phit{ABL}(\mb a)$ provides strong implications for optimization. Indeed, if we could initialize $\mb a$ near $\mc S_{\mc I}$, iterates of many local descent methods such as \cite{goldfarb1980curvilinear,conn2000trust,boumal2016global,nesterov2006cubic} can exploit gradient and negative curvature to remain near $\mc S_{\mc I}$, and eventually converge to the target solution -- a signed-shift of $\mb a_0$.
Finding a good initialization is also deceptively simple: since $\mb x_0$ is sparse, {\em any length-$\di$ truncation of the observation $\mb y$ is itself approximately a superposition of a few shifts of $\mb a_0$},
\begin{align}\label{eqn:linear-combine-y}
   \mb y \;=\; \sum_{ \ell \in \supp(\mb x_0) } 	 (x_0)_\ell \cdot \shift{\mb a_0}{\ell}.
\end{align}
%where $\wh{\mb y}_i = \begin{bmatrix} y_i& y_{i+1} & \cdots & y_{i+\di -1} \end{bmatrix}^\top$ is a length-$\di$ truncation of $\mb y$ starting from the entry $y_i$.
Therefore, if we simply chose $\di$ consecutive entries of $\mb y$, (e.g.\ $[y_i,y_{i+1},\dots,y_{i+\di-1}],\; i\in [m-\di]$) randomly from the observation $\mb y$ and initialize $\mb a_0$ by setting
\begin{align}\label{eqn:initialization}
	\mb a\iter0 \;=\;  \mc P_{\bb S^{\Di-1}} \paren{ \begin{bmatrix}
 \mb 0_{\di-1} \;;\; y_i,y_{i+1},\dots,y_{i+\di-1} \;;\; \mb 0_{\di-1}
 \end{bmatrix}
 },
\end{align}
then $\mb a\iter0 \in \bb R^{\Di}$ is close to a subsphere $\mc S_{\mc I}$ spanned by roughly $\mc O(\di \theta)$ shifts of $\mb a_0$. Moreover, any truncation effects are absorbed by the zero-padding in \Cref{eqn:initialization}. In \cite{kuo2019geometry}, this initialization scheme is improved and made rigorous, and interested readers may refer to Appendix \ref{subsec:misc} for details.

\begin{figure*}[!htbp]
\centering
\captionsetup[sub]{font=normalsize,labelfont={bf,sf}}
\begin{minipage}[c]{.4\textwidth}
\subcaption{Approximate Bilinear Lasso $\phit{ABL}(\mb a)$}\label{subfig:geometry-DQ}
\centering
	\includegraphics[width = .6\linewidth]{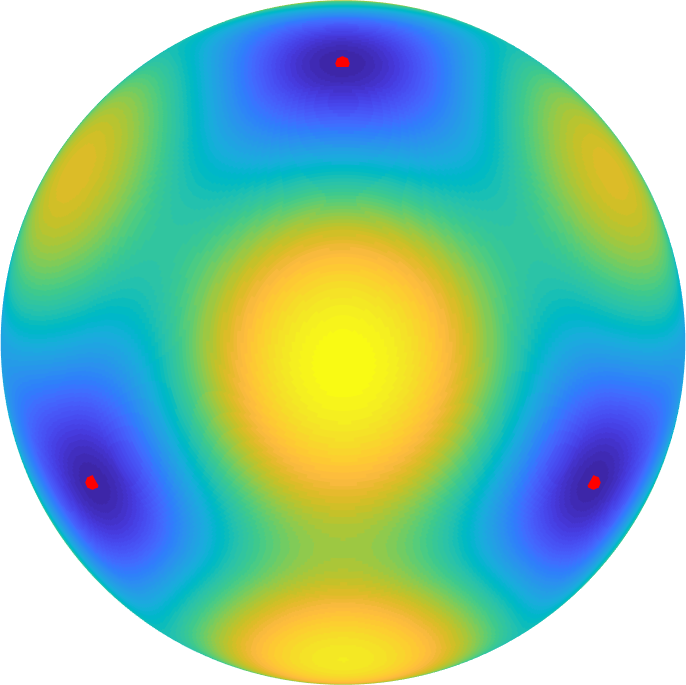}
\end{minipage}
\hspace{0.1\textwidth}
\begin{minipage}[c]{.4\textwidth}
\subcaption{Bilinear Lasso $\phit{BL}(\mb a)$}\label{subfig:geometry-lasso}
\centering
	\includegraphics[width = .6\linewidth]{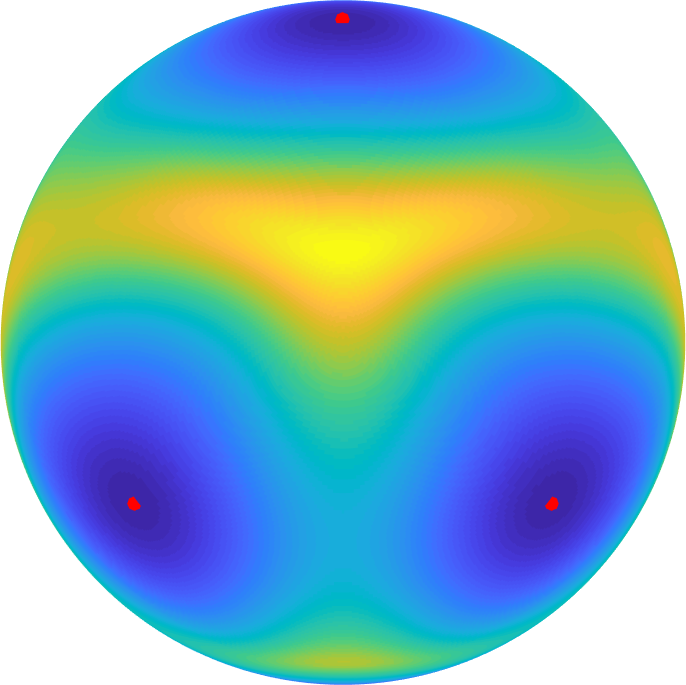}
\end{minipage}
\caption{\textbf{Approximate Bilinear Lasso vs. Bilinear Lasso:} Given an incoherent truth kernel $\mb a_0 \sim \mc U\paren{\bb S^{\di-1}}$, we plot the heat maps of objective landscapes of {\bf (a)} the Approximate Bilinear Lasso and {\bf (b)} Bilinear Lasso losses, restricted to the subsphere spanned by $\mb a_0,\; \shift{\mb a_0}{1}$, and $\shift{\mb a_0}{2}$, shown as red dots on the heat map. The curvature properties of both objective landscapes are empirically similar at key locations, e.g.,\ near and between shifts.}
\label{fig:geometry-loss}
\end{figure*}

\paragraph{Optimization over the sphere.} For both the Bilinear Lasso and ABL, a unit-norm constraint on $\mb a$ is enforced to break the scaling symmetry between $\mb a_0$ and $\mb x_0$.
Choosing the $\ell_2$-norm, however, has surprisingly strong implications for optimization. The ABL objective, for example, is piecewise concave whenever $\mb a$ is sufficiently far away from any shift of $\mb a_0$, but the sphere induces positive curvature near individual shifts to create strong convexity. These two properties combine to ensure recoverability of $\mb a_0$.
In contrast, enforcing $\ell_1$-norm constraints often leads to spurious minimizers for deconvolution problems \cite{levin2009understanding,benichoux2013fundamental,zhang2017global}.

\paragraph{Implications for the Bilinear Lasso.} The ABL is an example of a formulation for SaSD posessing a (regionally) benign optimization landscape, which guarantees that efficient recovery is possible when $\mb a_0$ is incoherent.
Applications of sparse deconvolution, however, are often motivated by sharpening or resolution tasks \cite{huang2009super,candes2014towards,campisi2016blind} where the motif $\mb a_0$ is smooth and coherent ($\mu_s(\mb a_0)$ is large). The ABL objective is a poor approximation of the Bilinear Lasso in such cases and therefore fails to yield practical algorithms.

In such cases, the Bilinear Lasso should be optimized directly, and \Cref{fig:geometry-loss} shows that its loss surface does indeed share similar symmetry breaking properties with the ABL objective. In the next section, we apply the geometric intuition gained from the ABL formulation, in combination with a number of computational heuristics, to create an optimization method for SaSD that performs well in general problem instances.

%Furthermore, we address some nonidealities that appear in high-coherence situations by using several computational heuristics, and this allows us to create optimization methods for SaSD that performs well in general problem instances.
 %\input{sec/geometry_old}

\section{Designing Practical Nonconvex Optimization Algorithms}
\label{sec:algorithm}
% !TEX root = ../main.tex

Several algorithms for SaSD type problems have been developed for specific applications, such as image deblurring \cite{levin2009understanding,briers2013laser,campisi2016blind}, neuroscience \cite{rey2015past,friedrich2017fast,song2018spike},
and image super-resolution \cite{baker2002limits,shtengel2009interferometric,yang2010image}.
In this section, however, we will instead leverage the intuition from \Cref{sec:geometry} and build optimization methods for the Bilinear Lasso
\begin{align}\label{eqn:bilinear-lasso-1}
    \min_{\mb a,\; \mb x}\;\;  \Psit{BL}(\mb a, \mb x)\;\;\doteq\;\; \underbrace{\tfrac{1}{2}\norm{\mb y - \mb a\conv\mb x}{2}^2 }_{
    \text{smooth } \psi(\mb a,\mb x)
    } \;+\; \underbrace{ \lambda \cdot \norm{\mb x}{1} }_{
    \text{nonsmooth } g(\mb x)
    },\quad \text{s.t.}\quad \mb a \in \bb S^{\Di-1},
\end{align}
that perform well in general settings for SaSD, as the Bilinear Lasso more accurately accounts for interactions between $\mb a \conv\mb x$ when $\mb a_0$ is shift-coherent. In such situations, optmization of $\Psit{BL}$ will also suffer from slow convergence and poor resolution of $\mb x_0$, which we will address in this section with a number of heuristics. This leads to an efficient and practical algorithms for solving sparse deconvolution problems.

% Several algorithms for SaSD type problems have been developed for specific applications, such as image deblurring \cite{levin2009understanding,briers2013laser,campisi2016blind}, neuroscience \cite{rey2015past,friedrich2017fast,song2018spike},
% and image super-resolution \cite{baker2002limits,shtengel2009interferometric,yang2010image},
% or are augmented with additional structure \cite{wipf2014revisiting,ling2017blind,walk2017blind}. Here, we will instead attempt to leverage recent developments in algorithmic theory in SaSD (\Cref{sec:geometry}) to build algorithms that perform well in general practical settings.

\subsection{Solving the Bilinear Lasso via alternating descent method}\label{subsec:adm-algorithm}
Efficient global optimization of the nonconvex objective in \Cref{eqn:bilinear-lasso-1} is a nontrivial task, largely due to the existence of {spurious local minima} and {saddle points}. In the following, we introduce a simple first-order method dealing with these issues. As suggested by our discussion of the geometry of the Dropped Quadratic in \Cref{sec:geometry}, we avoid such spurious minimizers using a {data-driven initialization} scheme introduced in \Cref{subsec:geometry}. On the other hand, our study in \Cref{sec:geometry} implies that all saddle points exhibit large negative curvature and can hence be effectively escaped by first-order methods\footnote{ In \cite{zhang2017global} and \cite{kuo2019geometry}, they employed second-order trust-region \cite{conn2000trust,boumal2016global} and curvilinear search \cite{goldfarb1980curvilinear,goldfarb2017using} methods for solving SaSD. Although second-order methods can also escape strict saddle points by directly exploiting the Hessian, they are much more expensive computationally and hence not practical for large datasets.} \cite{lee2017first,jin2017escape,gilboa2018efficient}.

Starting from the data-driven initialization, we optimize the Bilinear Lasso using a first-order \emph{alternating descent method} (ADM). The basic idea of our ADM algorithm is to alternate between taking first-order descent steps on $\Psi(\mb a,\mb x)$ w.r.t. one variable while the other is fixed:

\paragraph{Fix $\mb a$ and take a descent step on $\mb x$.} At each iteration $k$, with fixed $\mb a^{(k)}$, ADM first descends the objective $\Psit{BL}(\mb a,\mb x)$ by taking a {\em proximal gradient} step w.r.t.\ $\mb x$ with an appropriate stepsize $t_k$
\begin{align} \label{eqn:main-xiter}
  \mb x\iter{k+1} \;\leftarrow\;
    \text{prox}_{g}^{\lambda t_k} \paren{ \mb x\iter k -
    t_k\cdot\nabla_{\mb x\,} \psi\paren{\mb a\iter k, \mb x\iter k}},
\end{align}
where $\text{prox}_{g}(\cdot)$ denotes the proximal operator of $g(\cdot)$ \cite{nesterov2007gradient}. Since the subproblem of minimizing $\Psit{BL}(\mb a, \mb x)$ only w.r.t.\ $\mb x$ is the Lasso problem, the proximal step taken in \Cref{eqn:main-xiter} here is classical\footnote{The \Cref{eqn:main-xiter} can also be rewritten and interpreted as $\mb x\iter{k+1} = \mb x\iter k - t_k \mc G_{t_k} \paren{ \mb x\iter k }$ with the \emph{composite gradient mapping} $\mc G_{t_k}$ \cite{nesterov2007gradient}. $\mc G_{t_k}$ behaves like the ``gradient'' on the smooth Moreau envelope of $\Psit{BL}(\mb a, \mb x)$, as a function of $\mb x$.} \cite{beck2009fast,parikh2014proximal}.
% and can be interpreted as a descent step\footnote{The \Cref{eqn:main-xiter} can also be rewritten and interpreted as $\mb x\iter{k+1} = \mb x\iter k - t_k \mc G_{t_k} \paren{ \mb x\iter k }$ with the \emph{composite gradient mapping} $\mc G_{t_k}$ \cite{nesterov2007gradient}. $\mc G_{t_k}$ behaves like the ``gradient'' on the smooth Moreau envelope of $\Psit{BL}(\mb a, \mb x)$, as a function of $\mb x$.} of the smooth envelope of $\Psit{BL}(\mb a, \mb x)$ with fixed $\mb a$.

\paragraph{Fix $\mb x$ and take a descent step on $\mb a$.} Next, we fix the iterate $\mb x\iter{k+1}$ and we take a \emph{Riemannian gradient} step \cite{absil2009} w.r.t.\ $\mb a$ over the sphere $\bb S^{\Di-1}$, with stepsize $\tau_k>0$,
\begin{align} \label{main-aiter}
  \mb a\iter{k+1} \;\leftarrow \; \mc P_{\bb S^{\Di-1}} \paren{ \mb a\iter k
    - \tau_k \cdot \grad_{\mb a} \psi\paren{\mb a\iter k, \mb x\iter{k+1}}},
\end{align}
where $\grad_{\mb a}\psi(\mb a, \mb x)$ denotes the Riemannian gradient of $\psi(\mb a, \mb x)$ w.r.t. $\mb a$, and $\mc P_{\bb S^{\Di-1}}\paren{ \cdot }$ is a projection operator onto the sphere $\bb S^{\Di-1}$.
The Riemannian gradient $\grad_{\mb a}\psi(\mb a, \mb x)$ can be interpreted as the standard gradient projected to the (Euclidean) tangent space\footnote{The tangent space is a $n-1$ dimensional Euclidean linear space, containing all the tangent vectors at $\mb a \in \bb S^{n-1}$. We refer the readers to \cite[Section 3]{absil2009} for more concrete definitions.} of $\bb S^{\Di-1}$ at point $\mb a$, and the projection operator $\mc P_{\bb S^{\Di-1}}\paren{ \cdot }$ ensures that our iterate stays on the sphere\footnote{The Riemannian gradient step is a specific \emph{manifold retraction} operator on the sphere, which takes a point from the tangent space at some point $\mb a$ and pushes it to a new point on the manifold. We refer interested readers to Section 3 of \cite{absil2009} for more details.}.

ADM simply alternates between steps of \Cref{eqn:main-xiter} and \Cref{main-aiter} until convergence, and can seamlessly incorporate other acceleration techniques that we will discuss in the later part of this section. We refer readers to Appendix \ref{subsec:adm} for more implementation details.

The geometric intuition gained in \Cref{sec:geometry} is based on the {marginalized} objective $\phit{BL}(\mb a)$ over the sphere $\bb S^{\Di-1}$, whereas here we simply a descent step on $\Psit{BL}(\mb a, \mb x)$ w.r.t.\ $\mb x$ rather than minimize $\mb x$ explicitly to reduce computational complexity per iteration. Nonetheless, the sequence of gradients $\nabla_{\mb a}\Psit{BL}(\mb a\iter k, \mb x\iter k)$ on $\mb a$ approximates $\nabla\phit{BL}(\mb a \iter k)$ as $k\rightarrow\infty$, since ADM is guaranteed to converge to some stationary point \cite{bolte2014proximal,pock2016inertial}. Therefore, ADM on $\Psit{BL}(\mb a,\mb x)$ eventually becomes equivalent to Riemannian gradient descent on $\phit{BL}(\mb a)$.

% \subsection{Key geometric ingredients for solving SaSD}
\subsection{Heuristics for improving the geometry of Bilinear Lasso}
\label{subsec:acceleration}

Although the Bilinear Lasso is able to account for the interactions between $\mb a_0$ and $\mb x_0$ under high coherence, the smooth term $\norm{\mb a\conv\mb x - \mb y}2^2$ nonetheless becomes ill-conditioned as $\mu(\mb a_0)$ increases, leading to slow convergence for practical problem instances. Here we will discuss a number of heuristics which will help to obtain faster algorithmic convergence and produce better solutions in such settings.

\begin{figure*}[!htbp]
\centering
\captionsetup[sub]{font=normalsize,labelfont={bf,sf}}
\begin{minipage}[c]{0.49\textwidth}
\subcaption{Standard gradient descent}\label{subfig:without_momentum}
\centering
  \includegraphics[width = \linewidth]{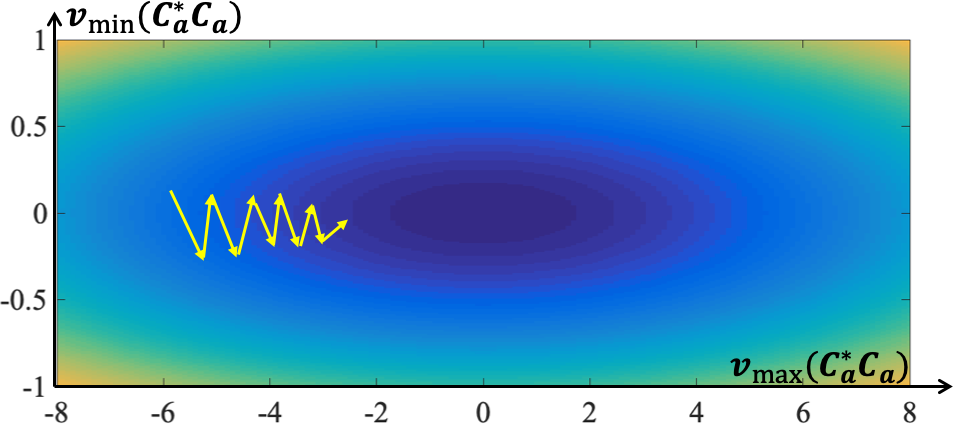}
\end{minipage}
\begin{minipage}[c]{0.49\textwidth}
\subcaption{With momentum acceleration}\label{subfig:with_momentum}
\centering
  \includegraphics[width = \linewidth]{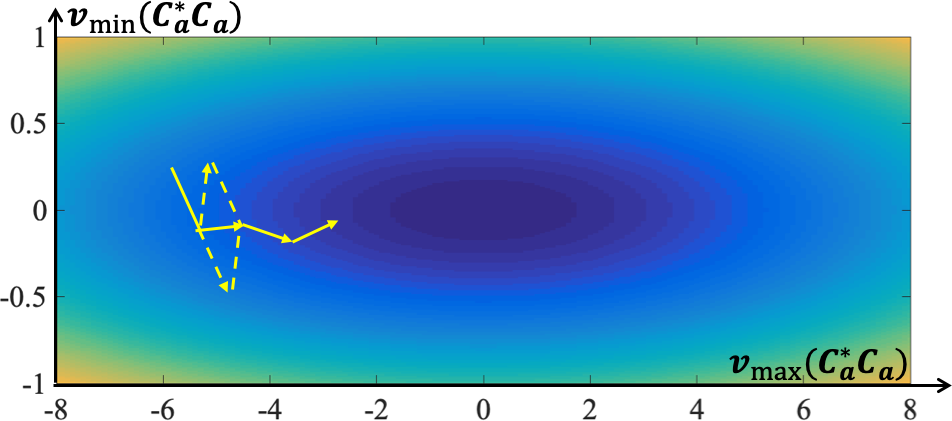}
\end{minipage}
\caption{\textbf{Momentum acceleration.} The left figure shows the behavior of standard gradient descent which oscillates on functions of ill-conditioned Hessian; the right figure shows that by incorporating the previous steps the momentum acceleration alleviates the oscillation effects and achieves faster convergence.}
\label{fig:momentum}
\end{figure*}

\subsubsection{Accelerating first-order descent under high coherence}
\label{subsubsec:main-momentum}

When $\mu_s(\mb a_0)$ is large, the Hessian of $\Psit{BL}$ becomes ill-conditioned as $\mb a$ converges to single shifts. the objective landscape contains ``narrow valleys'' in which first-order methods tend to exhibit severe oscillations (\Cref{subfig:without_momentum}) \cite{nesterov2013introductory}. For a nonconvex problem such as the Bilinear Lasso, iterates of first-order methods could encounter many narrow and flat valleys along the descent trajectory, resulting in slow convergence.

One remedy here is to add \emph{momentum} \cite{polyak1964some,beck2009fast} to standard first-order iterations. For example, when updating $\mb x$, we could modify the iterate in \Cref{eqn:main-xiter} by
\begin{align} \label{eqn:main-xiter-momentum}
    \mb w\iter k \;&=\; \mb x\iter k + \beta \cdot \underbrace{\paren{ \mb x\iter k - \mb x\iter{k-1} }}_{ \text{inertial term} }, \\
  \mb x\iter{k+1} \;&=\; \mathrm{prox\,}_{t_k g} \paren{
    \mb w\iter k - t_k \nabla_{\mb x} \psi\paren{\mb a\iter k, \mb w\iter k} }.
\end{align}
Here, the \emph{inertial term} incorporates the momentum from previous iterations, and $\beta\in (0,1)$ controls the inertia\footnote{Setting $\beta=0$ here removes momentum and reverts to standard proximal gradient descent.}. In a similar fashion, we can modify the iterate \cite{absil2009} for updating\footnote{It modifies iPALM \cite{pock2016inertial} to perform updates on $\mb a$ via retraction on the sphere.} $\mb a$ in \Cref{main-aiter}. We term the new algorithm \emph{inertial alternating descent method} (iADM), and we refer readers to Appendix \ref{subsec:momentum} for more details.

As illustrated in \Cref{subfig:with_momentum}, the additional inertial term improves convergence by substantially reducing oscillation effects for ill-conditioned problems. The acceleration of momentum methods for convex problems are well-known in practice\footnote{In the setting of strongly convex and smooth function $f(\mb z)$, the momentum method improves the iteration complexity from $\mc O\paren{\kappa \log(1/\eps)}$ to $\mc O\paren{\sqrt{\kappa} \log(1/\eps)}$ with $\kappa$ being the condition number, while leaving the computational complexity approximately unchanged \cite{bubeck2015convex}.}. Recently, momentum methods has also been proven to improve convergence for nonconvex and nonsmooth problems \cite{pock2016inertial,jin2017accelerated}.

\begin{figure*}[!htbp]
\centering
\captionsetup[sub]{font=normalsize,labelfont={bf,sf}}
\begin{minipage}[c]{0.25\textwidth}
\subcaption{$\lambda = 5 \times 10^{-1}$}\label{subfig:geometry-lambda-0.5}
\centering
	\includegraphics[width = \linewidth]{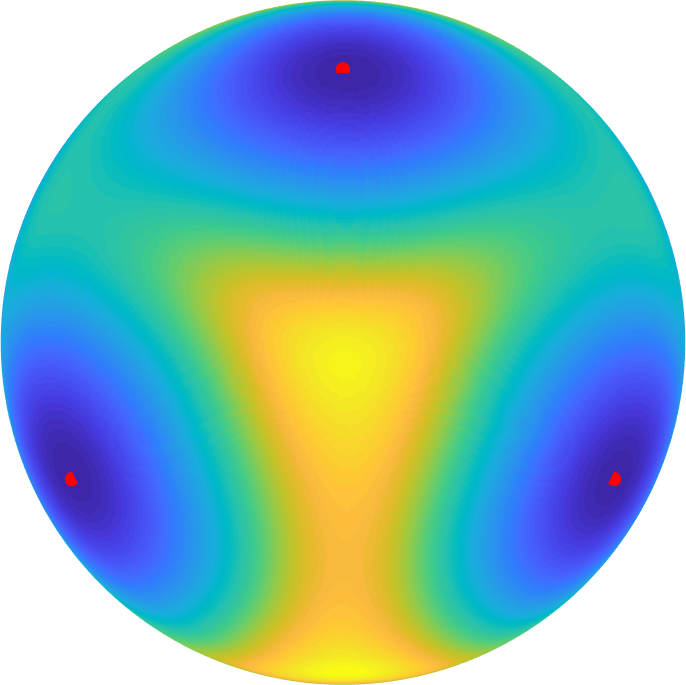}
\end{minipage}
\hspace{0.05\textwidth}
\begin{minipage}[c]{0.25\textwidth}
\subcaption{$\lambda = 5 \times 10^{-2}$}\label{subfig:geometry-lambda-0.05}
\centering
	\includegraphics[width = \linewidth]{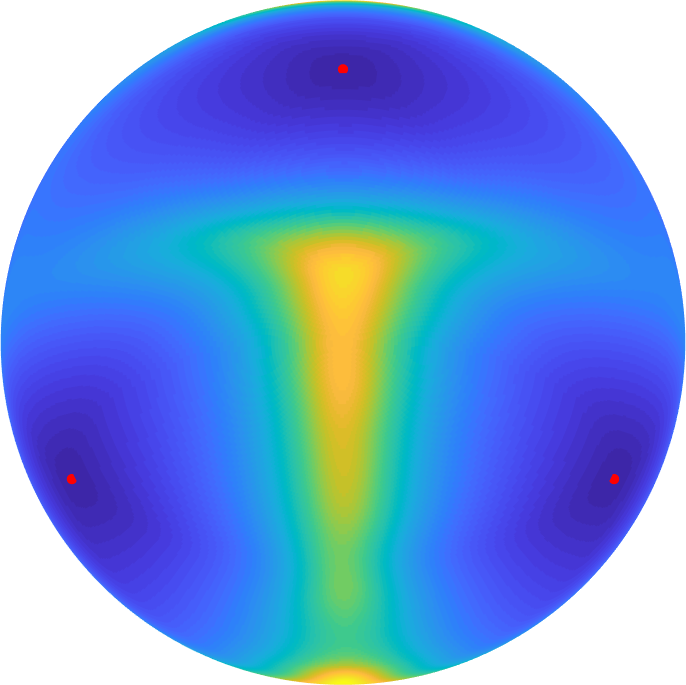}
\end{minipage}
\hspace{0.05\textwidth}
\begin{minipage}[c]{0.25\textwidth}
\subcaption{$\lambda = 5 \times 10^{-3}$}\label{subfig:geometry-lambda-0.005}
\centering
	\includegraphics[width = \linewidth]{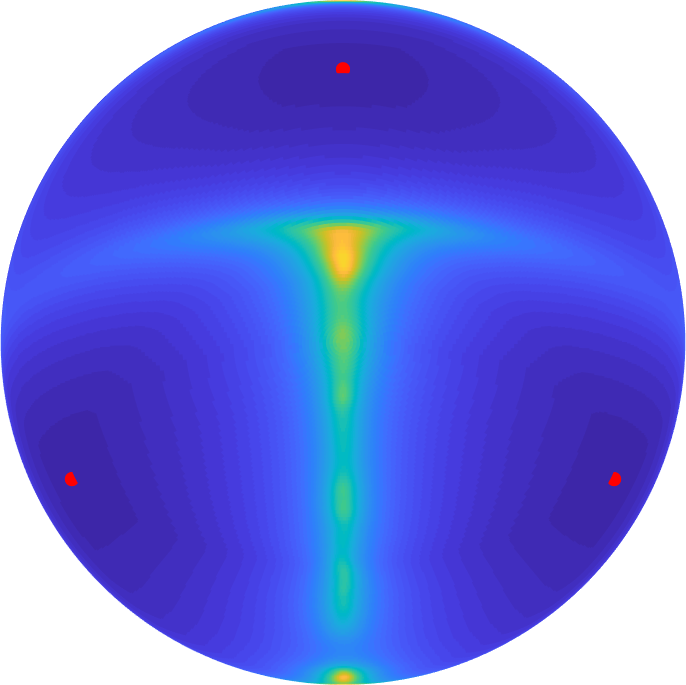}
\end{minipage}
\caption{\textbf{Low-dimensional functional landscape of Bilinear Lasso with varying $\lambda$.} Each subfigure shows the objective $\phit{BL}(\mb a)$, with $\mb a$ restricted to the subsphere $\mc S_{\{0,1,2\}}$ defined in \Cref{eqn:geometry-subspace}, with varying choices of $\lambda>0$. The kernel $\mb a_0$ is incoherent and drawn uniformly from the sphere. The red dots denote the location of the kernel and its shifts. }
\label{fig:homotopy-lambda}
\end{figure*}

\subsubsection{A practical method for SaSD based on homotopy continuation}\label{subsec:main-homotopy}

It is also possible to improve optimization by modifying the objective $\Psit{BL}$ directly through the sparsity penalty $\lambda$. Variations of this idea appear in both \cite{zhang2017global} and \cite{kuo2019geometry}, and can also help to mitigate the effects of large shift-coherence in practical problems.

When solving \eqref{eqn:bilinear-lasso} in the noise-free case, it is clear that larger choices of $\lambda$ encourage sparser solutions for $\mb x$. Conversely, smaller choices of $\lambda$ place local minimizers of the marginal objective $\phit{BL}(\mb a) \doteq \min_{\mb x} \Psit{BL}(\mb a, \mb x)$ closer to signed-shifts of $\mb a_0$ by emphasizing reconstruction quality.
When $\mu(\mb a_0)$ is large, however, $\phit{BL}$ becomes ill-conditioned as $\lambda\rightarrow0$ due to the poor spectral conditioning of $\mb a_0$, leading to severe flatness near local minimizers (\Cref{fig:homotopy-lambda}) and the creation spurious local minimizers when noise is present. At the expense of precision, larger values of $\lambda$ limit $\mb x$ to a small set of support patterns and simplify the landscape of$\phit{BL}$. It is therefore important both for fast convergence and accurate recovery for $\lambda$ to be chosen appropriately.

When problem parameters -- such as the severity of noise, or $p_0$ and $\theta$ -- are not known a priori, a \emph{homotopy continuation method} \cite{hale2008fixed,wright2009sparse,xiao2013proximal} can be used to obtain a {\em range} of solutions for SaSD. Using the initialization \eqref{eqn:initialization}, a rough estimate $(\bhat a\iter 1, \bhat x\iter 1)$ is first obtained by solving \eqref{eqn:bilinear-lasso} with iADM using a large choice for $\lambda\iter 1$; this estimate is refined by gradually decreasing $\lambda\iter n$ to produce the {\em solution path} $\big\{(\bhat a\iter n,\bhat x\iter n; \lambda\iter n)\big\}$. By ensuring that $\mb x$ remains sparse along the solution path, homotopy provides the objective $\Psit{BL}$ with (restricted) strong convexity w.r.t.\ both $\mb a$ and $\mb x$ throughout optimization \cite{agarwal2010fast}.
As a result, homotopy achieves linear convergence for SaSD where sublinear convergence is expected otherwise (\Cref{fig:convergence_incoherent,fig:convergence_coherent}).

\begin{algorithm}
\begin{algorithmic}
\caption{Solving SaSD with homotopy continuation} \label{alg:sasbd-homotopy}
\renewcommand{\algorithmicrequire}{\textbf{Input:}}
\renewcommand{\algorithmicensure}{\textbf{Output:}}
\Require~~\
Measurement $\mb y\in\bb R^m$; momentum parameter $\beta \in[0,1)$; initial and final sparse penalties $\lambda_0$, $\lambda_\star$ ($\lambda_0>\lambda_\star$); decay penalty parameter $\eta \in (0,1)$; precision factor $\delta \in (0,1)$ and tolerance $\eps_\star$.
\Ensure~~\
final solution $(\mb a_\star,\mb x_\star)$.

\State {\em Set} iteration number $K  \leftarrow  \big\lfloor {\log( \lambda_\star/\lambda_0 ) }\;/\;{ \log\eta }  \big\rfloor$.

\State {\em Initialize} $\bhat a\iter0\in\bb R^{\Di}$ using \Cref{eqn:initialization}, $\bhat x\iter0 =\mb 0_m$, and $\lambda^{(0)} = \lambda_0$, $\eps^{(0)} = \delta  \lambda^{(0)}$;

\For { $k \;=\; 1,\dots,K  $ }
\State {\em Solve} $$\min_{\mb a \in \bb S^{\Di-1} ,\mb x}\;\Psi_{\lambda\iter{k-1} }(\mb a,\mb x) \doteq \frac{1}{2} \norm{\mb y - \mb a \conv \mb x}{2}^2 + \lambda^{(k-1)} \norm{\mb x}{1} $$
to precision $\eps\iter{k-1} = \delta \lambda\iter{k-1}$ via iADM, using $\paren{\bhat a\iter{k-1}, \bhat x\iter{k-1}}$ as warm start
\begin{align*}
   \paren{\bhat a\iter k, \bhat x\iter k} \;\leftarrow\; \text{iADM} \paren{\;\bhat a\iter{k-1}, \bhat x\iter{k-1}; \mb y, \lambda\iter {k-1}, \beta\;}.
\end{align*}

\State {\em Update} $\lambda\iter{k} \leftarrow \eta \lambda\iter{k-1}$.
\EndFor
\State {\em Final round:} starting from $\paren{\bhat a\iter K, \bhat x\iter K}$, optimize $\Psi_{\lambda_\star }(\mb a,\mb x)$ with penalty $\lambda_\star$ to precision $\eps_\star$ via
\begin{align*}
   \paren{\bhat a_\star, \bhat x_\star} \;\leftarrow\; \text{iADM} \paren{\;\bhat a\iter K, \bhat x\iter K; \mb y, \lambda_\star, \beta\;}.
\end{align*}
\end{algorithmic}
\end{algorithm}

\paragraph{Algorithm for SaSD.} We summarize our discussion by presenting a practical algorithm for solving SaSD (\Cref{alg:sasbd-homotopy}), which initializes $\mb a$ using \Cref{eqn:initialization} and subsequently find a local minimizer of the Bilinear Lasso using homotopy continuation, combined with the accelerated first-order iADM method, with an appropriate choice of $\lambda$. However, we note should be possible to substitute iADM with any first or second-order descent method (e.g. the Riemannian trust-region method \cite{absil2007trust,Cheung2018DictionaryLI}). We compare some of these different choices in \Cref{sec:exp_compare_alg}.

For \Cref{alg:sasbd-homotopy}, we usually set  $\beta = 0.9$ to incorporate sufficient momentum for iADM; setting $\beta$ too large, however, can cause iADM to diverge. The stepsizes $t_k$ and $\tau_k$ in iADM are obtained by backtracking (linesearch) \cite{nocedal2006numerical,pock2016inertial}. We often set the initial penalty $\lambda_0 = \norm{ \mb C_{ \res{n}{m} \bhat a\iter0}^* \mb y }{\infty}$ large enough to ensure sparse $\mb x$, and choose $\lambda_\star$ based on problem dimension and noise level (often $\lambda_\star = 0.1/\sqrt{n}$ is good choice). Typically, a good choice is to set the decaying parameter $\eta = 0.9$ and the precision factor $\delta = 0.1$. We refer readers to Appendices for more implementation details.

%For ease of exposition, \Cref{alg:sasbd-homotopy} is based a fairly idealistic SaS model for the observed signal $\mb y$
%$$\mb y = \mb a_0 \conv \mb x_0 + \mb n,$$ where the convolution operator $\conv$ is cyclical and the noise model is absent up to \Cref{sec:geometry}, and the Bilinear Lasso is also practical under additive Gaussian noise settings $\mb n\sim\mc N(0,\sigma\mb I)$ \cite{wainwright2009sharp}. In practical applications, standard convolution is often preferred and additional nuisances or structure may be present in the data. We address some of these in the remainder of this section. \yl{Explain that momentum and homotopy aren't strictly required, but will greatly improve performance.}

\subsection{Extension for convolutional dictionary learning}\label{subsec:cdl}

\begin{figure*}[!htbp]
\centering
\input{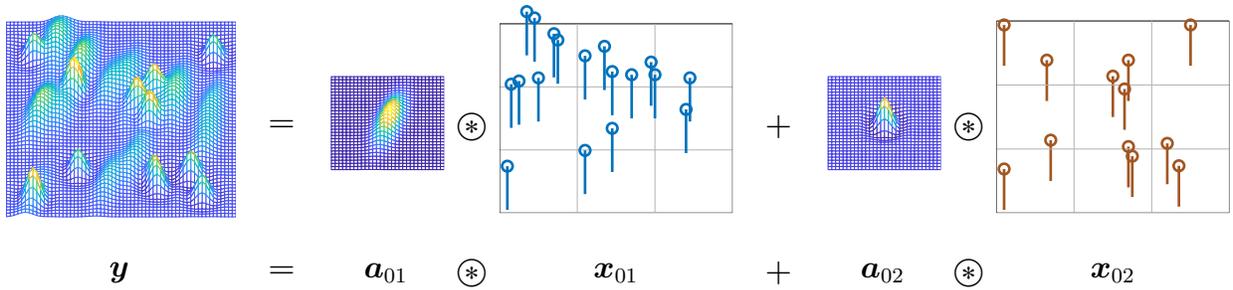}
\caption{\textbf{Convolutional dictionary learning.} Simultaneous recovery for multiple unknown kernels $\Brac{ \mb a_{0k} }_{k=1}^N$ and sparse activation maps $\Brac{\mb x_{0k}}_{k=1}^N$ from $\mb y \;=\; \sum_{k=1}^N \mb a_{0k} \conv \mb x_{0k}$.}
\label{fig:demo_CDL}
\end{figure*}
The optimization methods we introduced for SaSD here can be naturally extended to tackle sparse blind deconvolution problems with multiple unknown kernels/motifs (a.k.a. convolutional dictionary learning \cite{chun2017convolutional,garcia2018convolutional}), which have broad applications in microscopy data analysis \cite{yellin2017blood,zhou2014classification,Cheung2018DictionaryLI} and neural spike sorting \cite{ekanadham2011blind,rey2015past,song2018spike}. As illustrated in \Cref{fig:demo_CDL}, the new observation $\mb y$ in this task is the sum of $N$ convolutions between short kernels $\Brac{ \mb a_{0k} }_{k=1}^N$ and sparse maps $\Brac{\mb x_{0k}}_{k=1}^N$,
\begin{align*}
   \mb y \;\;=\;\; \sum_{k=1}^N \;\mb a_{0k}\; \conv\; \mb x_{0k},\qquad \mb a_{0k} \;\in \;\bb R^{\di},\quad \mb x_{0k} \;\in \;\bb R^{\sample},\quad (1\leq k\leq N).
\end{align*}
The natural extension of SaSD, then, is to recover $\Brac{ \mb a_{0k} }_{k=1}^N$ and $\Brac{\mb x_{0k}}_{k=1}^N$ up to signed, shift, and permutation ambiguities, leading to the SaS convolutional dictionary learning (SaS-CDL) problem. The SaSD problem can be seen as a special case of SaS-CDL with $N = 1$. Based on the Bilinear Lasso formulation in \Cref{eqn:bilinear-lasso-1} for solving SaSD, we constrain all kernels $\mb a_{0k}$ over the sphere, and consider the following nonconvex objective:
\begin{align}\label{eqn:bilinear-lasso-CDL}
	\min_{\Brac{\mb a_k}_{k=1}^N,\; \Brac{\mb x_k}_{k=1}^N}\;  \frac{1}{2} \norm{ \mb y - \sum_{k=1}^N \mb a_k \conv \mb x_k }2^2  + \lambda \sum_{k=1}^N \norm{\mb x_k}{1},\quad \text{s.t.}\quad \mb a_k \in \bb S^{\Di-1} \quad(1\leq k \leq N).
\end{align}
When the kernels $\Brac{ \mb a_{0k} }_{k=1}^N$ are incoherent enough to each other, we anticipate that all local minima are near signed shifts of the ground truth. Similar to the idea of solving the Bilinear Lasso in \Cref{eqn:bilinear-lasso-1}, we optimize \Cref{eqn:bilinear-lasso-CDL} via ADM and its variants, by taking alternating descent steps on $\Brac{\mb a_k}_{k=1}^N$ and $\Brac{\mb x_k}_{k=1}^N$ with one fixed. We refer readers to Appendix \ref{app:algorithm} and Appendix \ref{app:details} for more technical details.

\subsection{Additional modifications for practical settings}\label{subsec:structure}
Here briefly summarize some prevalent issues that appear in with real datasets and how the our SaS model and associated deconvolution method can be adjusted to deal with these additional challenges.

\begin{itemize}[leftmargin=*]
\item \textbf{Linear vs. cyclic convolution.} In this work, we follow the convention of \cite{kuo2019geometry} and mainly discuss SaSD in the context of cyclic convolution. The linear convolution, however, is a better model for many practical SaSD tasks (e.g.\ involving natural images or time series). Despite this, there is no loss of generality as any statements about cyclic convolution can easily be carried over to linear convolution; by zero-padding $\mb x$ appropriately, one can always rewrite a linear convolution as a cyclic convolution. This is also convenient practically as convolution operations should be implemented using Fast Fourier transform techniques (which map directly to cyclic convolution) to reduce computational complexity for each iteration. 

\item \textbf{Resolution of $\mb x_0$ under noise.} We introduce a reweighting technique \cite{candes2008enhancing} to deal with noisy datasets. The method adaptively sets large penalty on small entries of $\mb x$ to suppress noisy small entries, and set small penalty on large entries to promote sparse solutions of $\mb x$. We refer readers to Appendix \ref{subsec:misc} for more algorithmic details.
\item \textbf{Dealing with extra data structure.} In many problems such as calcium imaging \cite{pnevmatikakis2016simultaneous} and spike sorting \cite{song2018spike}, the sparse spike train $\mb x_0$ is usually nonnegative. As we shall see in \Cref{sec:exp_synthetic}, by enforcing nonnegative constraint on $\mb x$ for ADM, it often enables recovery of denser $\mb x_0$. Additionally, measurement in practice often contains unknown low frequency DC component $\mb b$, such that $\mb y = \mb a_0 \conv \mb x_0 +\mb b$. We add an extra minimization in ADM to deal with $b$. We refer readers to Appendix \ref{subsec:misc} for more technical details.

%\item \textbf{Shift correction.} The shift symmetry implies that we can only solve sparse deconvolution problems up to a shift ambiguity. However, as predicting the precise activation locations $\mb x_0$ could be very important in certain scenarios, post-processing is needed to correct the shift ambiguity. We show how to deal with this practical issue in Appendix \ref{subsec:misc}.

%\begin{figure*}[!htbp]
%\centering
%\includegraphics[width = 0.8\linewidth]{figs/shift_correction.png}
%\caption{\textbf{Shift correction.} An illustration of shift correction by using the Frobenius norm.}
%\label{fig:shift_correction}
%\end{figure*}
%
% YL: since we have managed to avoid shift-truncations in our exposition thus far, we will ignore shift correction.
%\item \textbf{Shift correction.} The shift symmetry implies that we can only solve sparse deconvolution problems up to a shift ambiguity. However, as predicting the correct activation locations $\mb x_0$ could be very important in certain scenarios, post-processing is needed to correct the shift ambiguity. Again, we show how to deal with issue in Appendix \ref{subsec:misc}.
\end{itemize}

\section{Synthetic Experiments}

\label{sec:exp_synthetic}
% !TEX root = ../main.tex
In this section, we experimentally demonstrate several core ideas presented in this work on both incoherent and coherent kernels. Incoherent kernels are randomly drawn by $\mb a_0 \sim \mr{Uniform}(\bb S^{\di-1})$, which leads to
$\mu_s(\mb a_0) \in \mc O\paren{\sqrt{ \frac{ \log \di }{ \di } } } $
diminishing w.r.t. dimension $\di$. Coherent kernels are descretized from the Gaussian window function $\mb a_0 = \mb g_{n_0,0.5}$, where
$\mb g_{\di,\sigma} \doteq \mc P_{\bb S^{\di-1}} \big(\big[\exp\big(-{\tfrac{(2i-\di-1)^2}{\sigma^2(\di-1)^2}}\big)\big]_{i=1}^\di\big)$; in this case $\mu_s(\mb a_0) \rightarrow 1$ as $\di$ grows.
This allows us to illustrate some of the difficulties of optimization encountered by the Bilinear Lasso under high coherence, as well as the effectiveness of heuristics proposed in \Cref{sec:algorithm} for alleviating these difficulties.

\subsection{Recovery of true kernel under coherence}
\label{subsec:phase-transition}

\begin{figure*}[!htbp]
\centering
\captionsetup[sub]{font=normalsize,labelfont={bf,sf}}
\begin{minipage}[c]{0.35\textwidth}
\subcaption{incoherent kernel}
\centering
  \includegraphics[width = 0.7\linewidth]{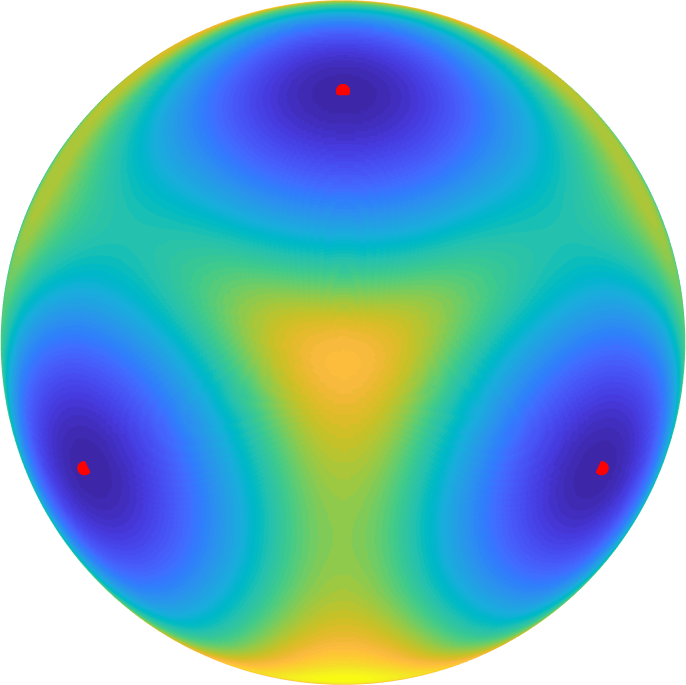}
\end{minipage}
\hspace{0.1\textwidth}
\begin{minipage}[c]{0.35\textwidth}
\subcaption{coherent kernel}
\centering
  \includegraphics[width = 0.7\linewidth]{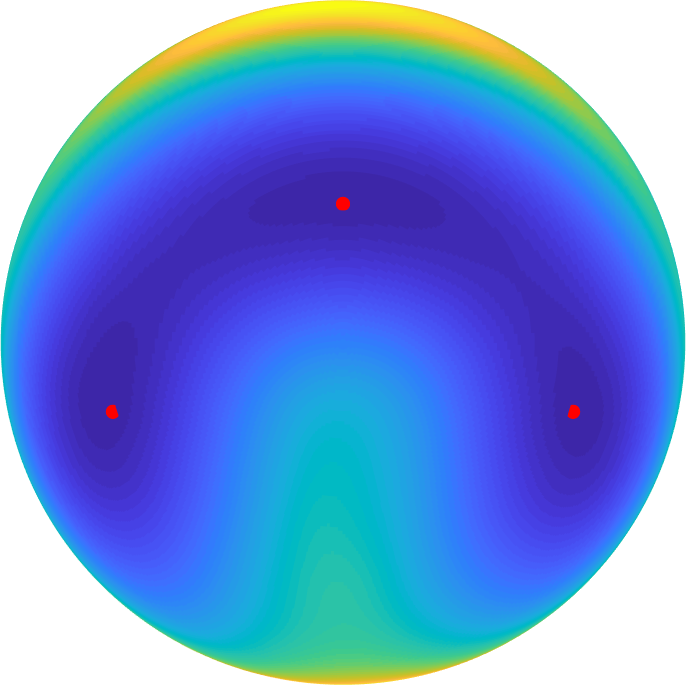}
\end{minipage}
\caption{\textbf{Incoherent vs. coherent kernels.} The subfigures from left to right present the optimization landscape of $\phit{BL} (\mb a)$ w.r.t. $\mb a \in \bb S^{\Di-1}$ defined in \eqref{eqn:lasso}, restricted to a subspace spanned by three shifts of $\mb a_0$. The left figure shows the landscape of incoherent kernel, and the right one presents that of the coherent kernel. The red dots denote the location of the shifts of ground truth $\mb a_0$.}
\label{fig:geometry-coherent-vs-incoherent}
\end{figure*}

\paragraph{Low-dimensional plots of function landscapes.}
As $\mu_s(\mb a_0)$ increases, the shifts of $\mb a_0$ lie closer together on the sphere. We show how this affects the optimization landscape of the Bilinear Lasso $\phit{BL} (\mb a)$ over $\mb a \in \bb S^{\Di-1}$ by plotting the objective restricted in the subsphere spanned by three shifts\footnote{For incoherent kernel, we generate the kernel $\mb a_0$ with the last two entries zero, and consider the subspace spanned by $\mb a_0, \mathrm{s}_1 \brac{\mb a_0}, \mathrm{s}_2 \brac{\mb a_0}$. For the coherent kernel, we consider the subspace spanned by $\mb a_0, \mathrm{s}_{\lceil \di/3 \rceil } \brac{\mb a_0}, \mathrm{s}_{\lceil 2\di/3 \rceil} \brac{\mb a_0}$.} of $\mb a_0\in\bb S^{\di-1}$ with $\di = 20$, $\sample = 2 \times 10^3 $, $\theta = \di^{-3/4}$, and $\lambda = 0.5$.
From \Cref{fig:geometry-coherent-vs-incoherent}, we see that $\phit{BL}$ exhibits clear symmetry breaking structure between the shifts of $\mb a_0$ in the incoherent case. As $\mu_s(\mb a_0)$ increases, however, adjacent shifts of $\mb a_0$ lie close together and symmetry breaking becomes more difficult.
Practically speaking, recovering a precise shift of $\mb a_0$ becomes less important when recovering smooth, highly coherent kernels. Nonetheless \Cref{fig:geometry-coherent-vs-incoherent} suggests that the target minimizers of $\phit{BL}$ become non-discretized in these cases.

\paragraph{Recovery performance.} Next, we corroborate our observation of sparsity-coherence tradeoff by comparing recovery performance for incoherent vs.\ coherent kernels. We fix $\sample = 100\di$, and plot the probability for successful recovery, which occurs if
\begin{align*}
  \min_{\ell\in [2\di]} \; \Brac{1 - \abs{\innerprod{ \mb a_0 }{ \res{\di}{n}^* s_\ell \brac{ \mb a_\star}   }  } } \leq 10^{-2},
\end{align*}
w.r.t. dimension $\di$ and sparsity level $\theta$. For each $(\di,\theta)$, we randomly generate ten independent instances of the data $\mb y= \mb a_0 \conv \mb x_0$. Here $\mb a_\star$ denotes the optimal solution produced by minimizing $\Psit{BL}$ with $\lambda = 10^{-2}/ \sqrt{ \theta \di }$.

\begin{figure*}[!htbp]
\centering
\captionsetup[sub]{font=normalsize,labelfont={bf,sf}}
\begin{minipage}[c]{0.49\textwidth}
\subcaption{incoherent $\mb a_0$ and $\mb x_0 \sim_{i.i.d.} \mc {BR}(\theta)$}
\label{fig:pt_incoherent}
\centering
  \includegraphics[width = \linewidth]{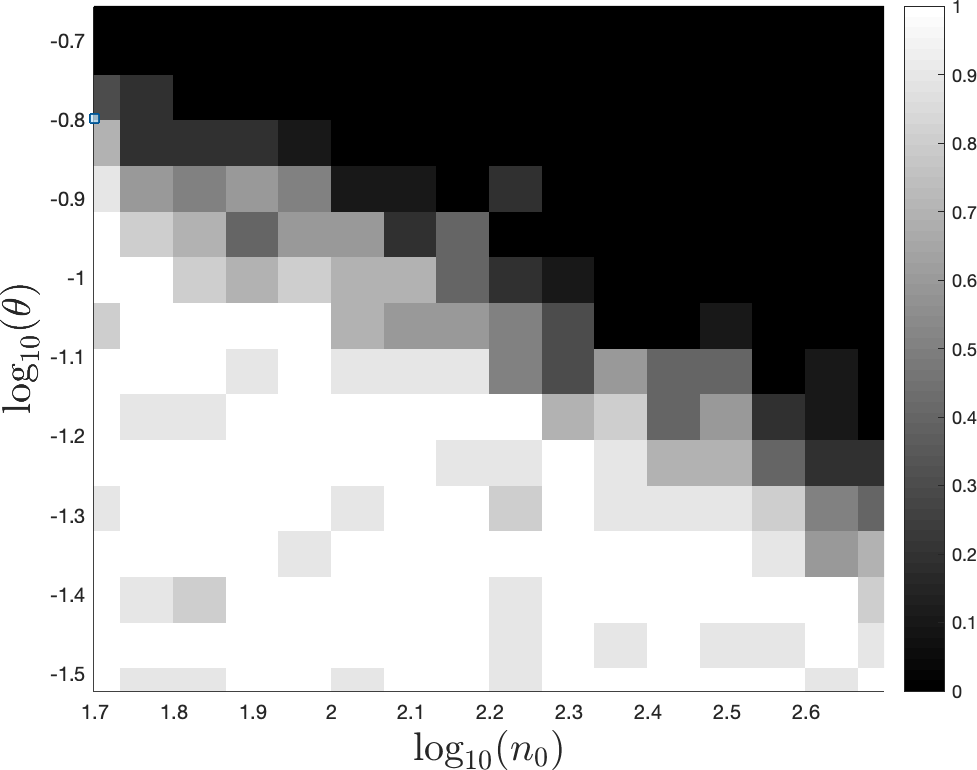}
\end{minipage}
\begin{minipage}[c]{0.49\textwidth}
\subcaption{coherent $\mb a_0$ and $\mb x_0 \sim_{i.i.d.} \mc {BR}(\theta)$}
\label{fig:pt_coherent}
\centering
  \includegraphics[width = \linewidth]{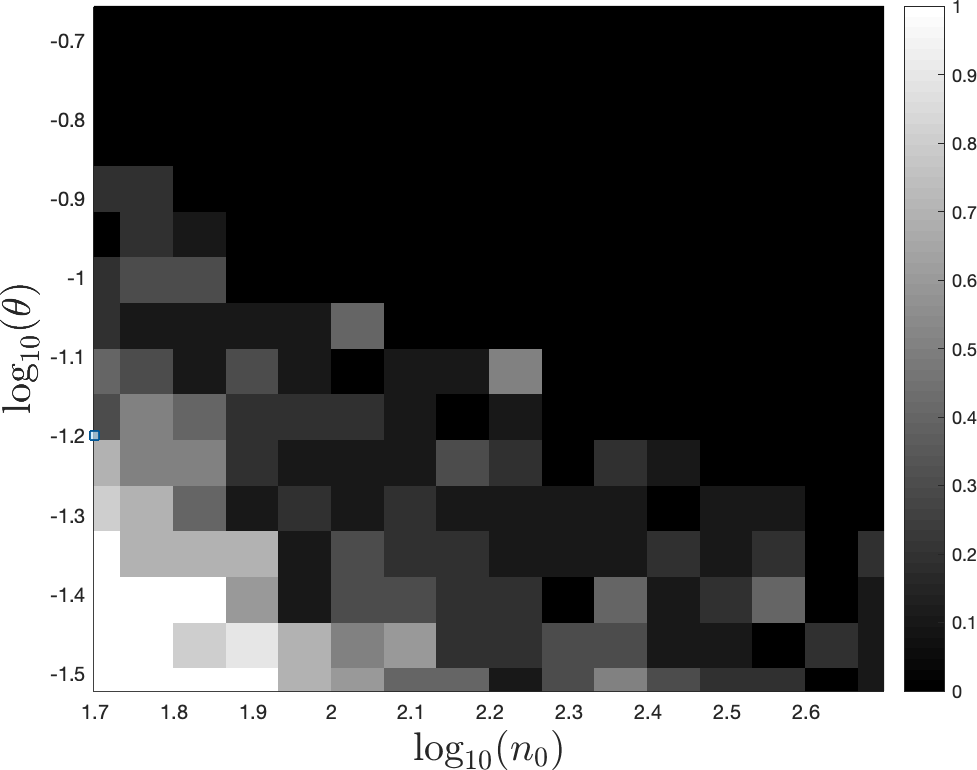}
\end{minipage}
\begin{minipage}[c]{0.49\textwidth}
\subcaption{incoherent $\mb a_0$ and $\mb x_0 \sim_{i.i.d.} \mc {B}(\theta)$}
\centering
  \includegraphics[width = \linewidth]{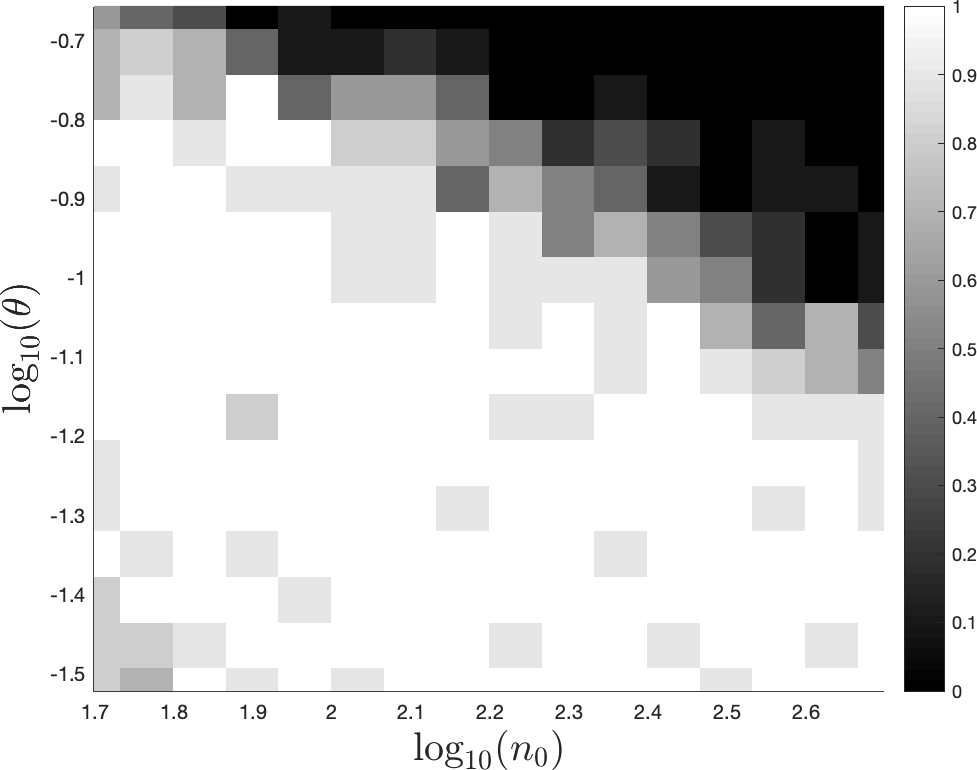}
  \label{fig:pt_incoherent_n}
\end{minipage}
\begin{minipage}[c]{0.49\textwidth}
\subcaption{coherent $\mb a_0$ and $\mb x_0 \sim_{i.i.d.} \mc {B}(\theta)$}
\label{fig:pt_coherent_n}
\centering
  \includegraphics[width = \linewidth]{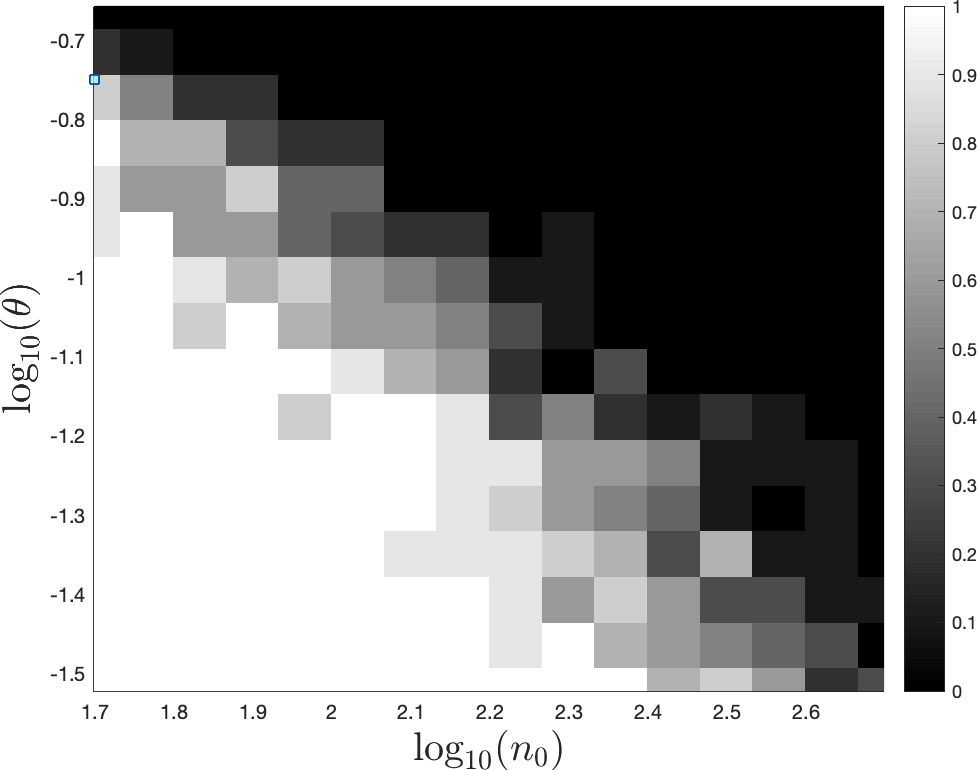}
\end{minipage}
\caption{ \textbf{phase transitions for solving SaS-BD:} (a) shows the case when $\mb a_0$ is incoherent, and $\mb x_0 \sim_{i.i.d.} \mc {BR}(\theta)$; (b) shows the case when $\mb a_0$ is coherent, and $\mb x_0 \sim_{i.i.d.} \mc {BR}(\theta)$; (c) shows the case when $\mb a_0$ is incoherent, and $\mb x_0 \sim_{i.i.d.} \mc {B}(\theta)$; (d) shows the case when $\mb a_0$ is coherent, and $\mb x_0 \sim_{i.i.d.} \mc {B}(\theta)$. For signal $\mb x_0\sim_{i.i.d.} \mc {B}(\theta)$, positivity constraint is enforced. For each subfigure, brighter means higher probability of successful recovery, while darker means higher probability of failure. } \label{fig:phase-transition}
\end{figure*}

From \Cref{fig:phase-transition}, we see that successful recovery is likely when sparsity $\theta$ is sufficiently small compared to $n_0$ in general. Furthermore, recovering $\mb a_0$ in the coherent setting is noticably more difficult than the incoherent setting, and typically requires lower sparsity rates $\theta$. Finally, enforcing extra structure such as nonnegativity in appropriate settings enables recovery with denser of $\mb x_0$ (\Cref{fig:pt_incoherent,fig:pt_incoherent_n}).

\subsection{Demonstration of data-driven initialization and homotopy acceleration}\label{sec:exp_compare_alg}
Our next experiments study the effectiveness of the data-driven (DD) initialization from \Cref{eqn:initialization} and the heuristics introduced in \Cref{sec:algorithm}, namely momentum acceleration and homotopy. Throughout this subsection, we set the kernel length $\di = 100$ and the number of samples $\sample = 10^4$. We generate the data $\mb y = \mb a_0 \conv \mb x_0+b \mb 1_m$ with both coherent and incoherent $\mb a_0$, $\mb x_0 \sim \mc {BR}(\theta)$ with sparsity level $\theta = \di^{-3/4}$, and $b$ is a constant unkown bias. No noise is added. We stop each algorithm either when the preset maximum iteration is reached, or when differences between two consecutive iterates (in $\ell_2$ norm) is smaller than threshold $10^{-6}$.

\begin{figure*}[!htbp]
\centering
\captionsetup[sub]{font=normalsize,labelfont={bf,sf}}
\begin{minipage}[c]{0.45\textwidth}
\subcaption{function value convergence}
\centering
  \includegraphics[width = \linewidth]{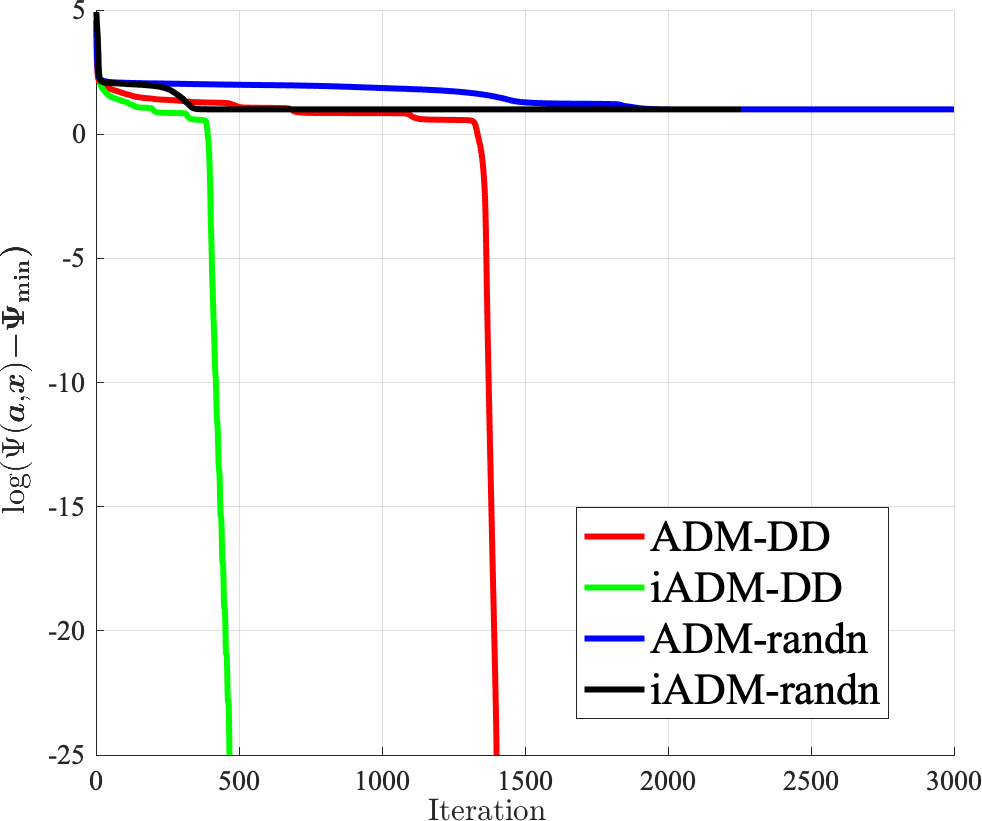}
\end{minipage}
\hspace{0.2in}
\begin{minipage}[c]{0.45\textwidth}
\subcaption{iterate convergence}
\centering
  \includegraphics[width = \linewidth]{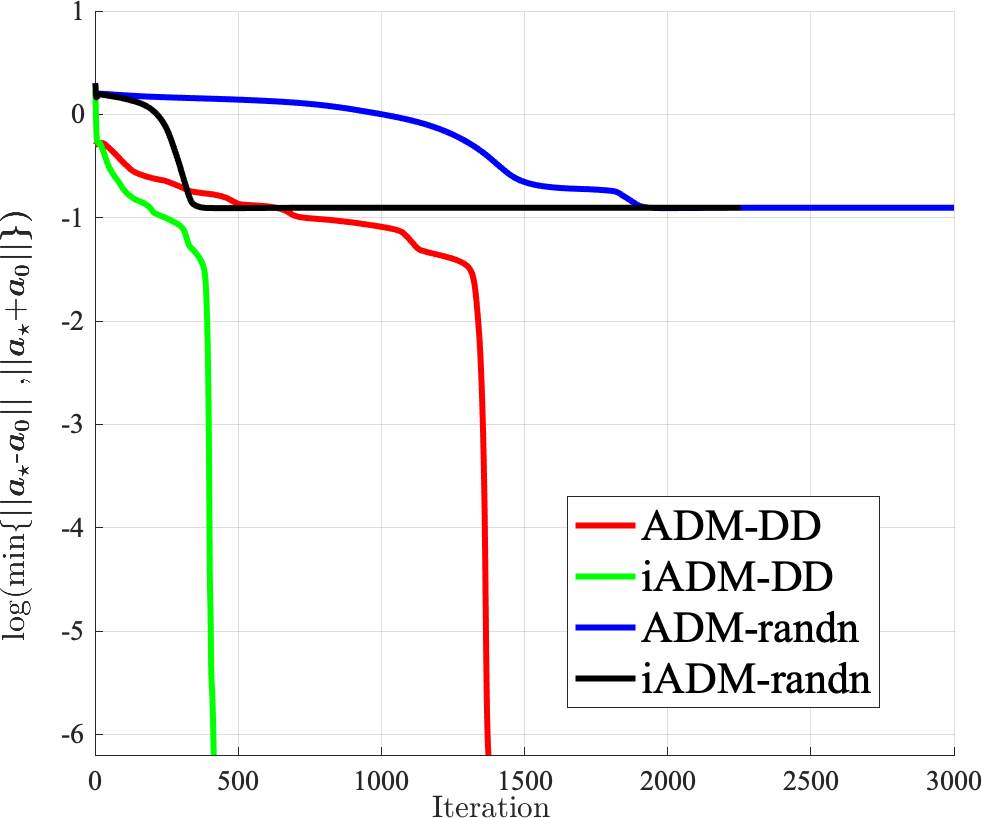}
\end{minipage}
\caption{\textbf{Comparison of initialization methods for solving SaS-BD on incoherent random kernel $\mb a_0$:} (a) shows the function value $\Psit{BL}(\mb a, \mb x)$ convergence; (b) shows the iterate convergence on $\mb a$, where $\mb a_\star$ denotes a shift correction of each iterate $\mb a$. Here, ADM-DD and iADM-DD denote the ADM and iADM methods using data-driven initialization, and ADM-randn and iADM-randn denote the ADM and iADM methods using initializations drawn uniformly random from the sphere $\bb S^{\di-1}$. }
\label{fig:initialization_incoherent}
\end{figure*}

\begin{figure*}[!htbp]
\centering
\captionsetup[sub]{font=normalsize,labelfont={bf,sf}}
\begin{minipage}[c]{0.45\textwidth}
\subcaption{function value convergence}
\centering
  \includegraphics[width = \linewidth]{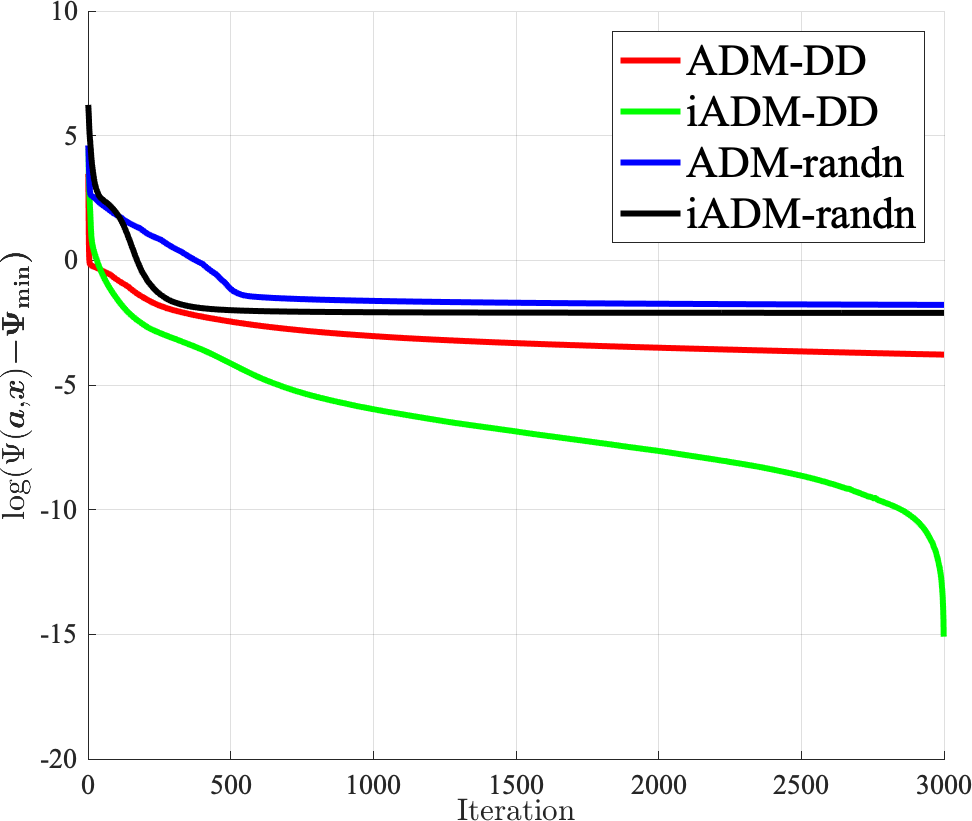}
\end{minipage}
\hspace{0.2in}
\begin{minipage}[c]{0.45\textwidth}
\subcaption{iterate convergence}
\centering
  \includegraphics[width = \linewidth]{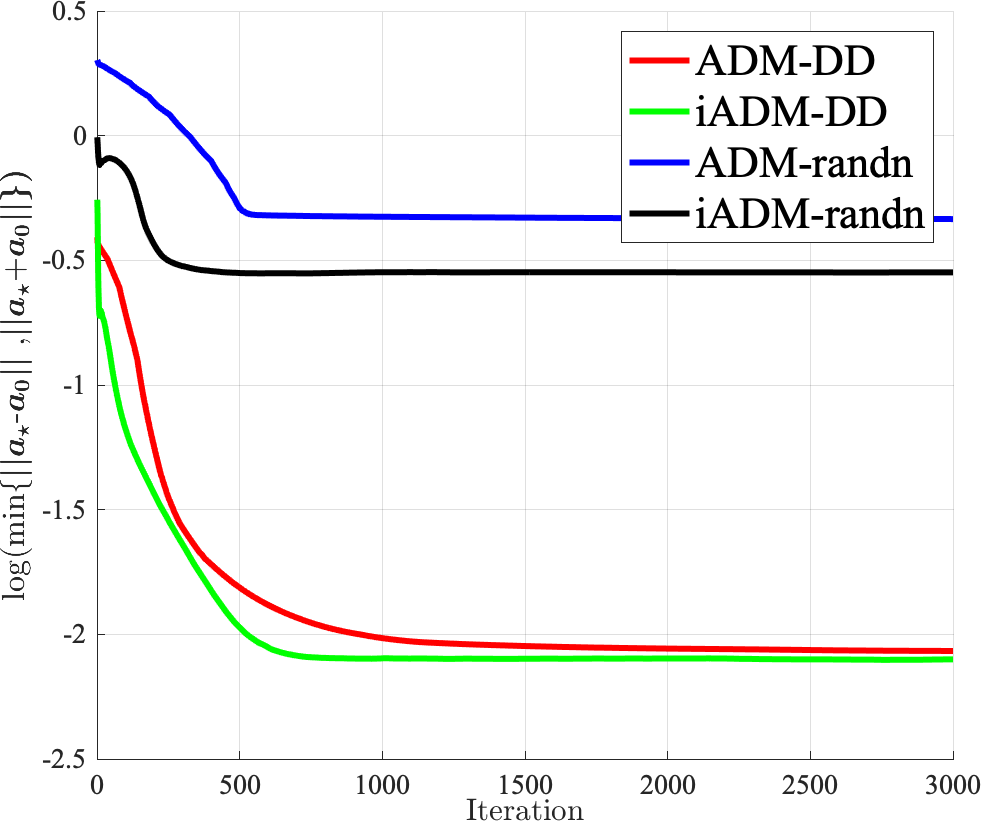}
\end{minipage}
\caption{\textbf{Comparison of initialization methods for solving SaS-BD on coherent smooth Gaussian kernel $\mb a_0$:} (a) shows the function value $\Psit{BL}(\mb a, \mb x)$ convergence; (b) shows the iterate convergence on $\mb a$, where $\mb a_\star$ denotes a shift correction of each iterate $\mb a$. Here, ADM-DD and iADM-DD denote the ADM and iADM methods using data-driven initialization, and ADM-randn and iADM-randn denote the ADM and iADM methods using initializations drawn uniformly random from the sphere $\bb S^{\di-1}$.  }
\label{fig:initialization_coherent}
\end{figure*}

\paragraph{Effectiveness of data-driven initialization.}
We compare the ADM and iADM methods using the data-driven initialization \Cref{eqn:initialization} vs. uniform random initializations for $\mb a$. From \Cref{fig:initialization_incoherent,fig:initialization_coherent}, we see that both methods converge faster to solutions of higher quality with data-driven initialization, as a result of $\mb a\iter 0$ being initialized near the superposition of a few shifts of $\mb a_0$.

\begin{figure*}[!htbp]
\centering
\captionsetup[sub]{font=normalsize,labelfont={bf,sf}}
\begin{minipage}[c]{0.45\textwidth}
\subcaption{function value convergence}
\centering
  \includegraphics[width = \linewidth]{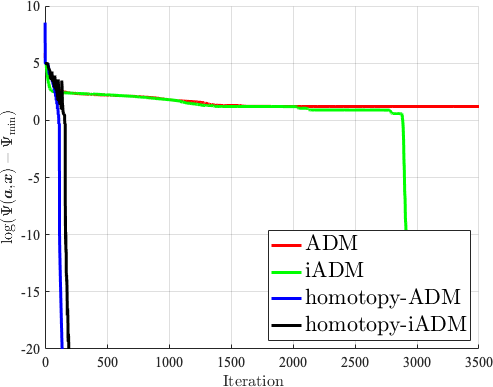}
\end{minipage}
\hspace{0.2in}
\begin{minipage}[c]{0.45\textwidth}
\subcaption{iterate convergence}
\centering
  \includegraphics[width = \linewidth]{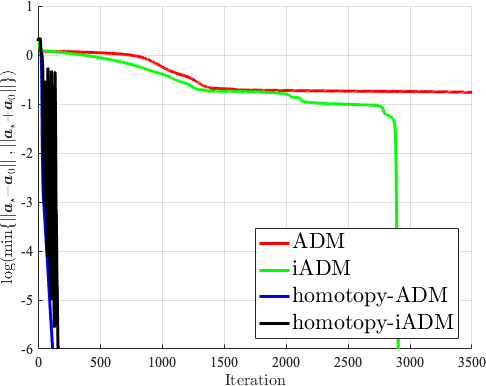}
\end{minipage}
\caption{\textbf{Comparison of algorithm convergence for solving SaS-BD on incoherent random kernel $\mb a_0$:} (a) shows the function value $\Psit{BL}(\mb a, \mb x)$ convergence; (b) shows the iterate convergence on $\mb a$, where $\mb a_\star$ denotes a shift correction of each iterate $\mb a$. The algorithms we compared here are ADM, iADM, and its homotopy accelerations. }
\label{fig:convergence_incoherent}
\end{figure*}

\begin{figure*}[!htbp]
\centering
\captionsetup[sub]{font=normalsize,labelfont={bf,sf}}
\begin{minipage}[c]{0.45\textwidth}
\subcaption{function value convergence}
\centering
  \includegraphics[width = \linewidth]{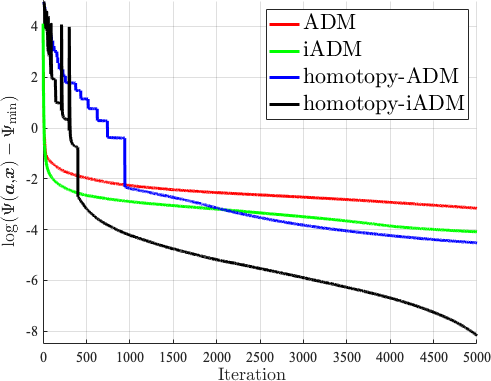}
\end{minipage}
\hspace{0.2in}
\begin{minipage}[c]{0.45\textwidth}
\subcaption{iterate convergence}
\centering
  \includegraphics[width = \linewidth]{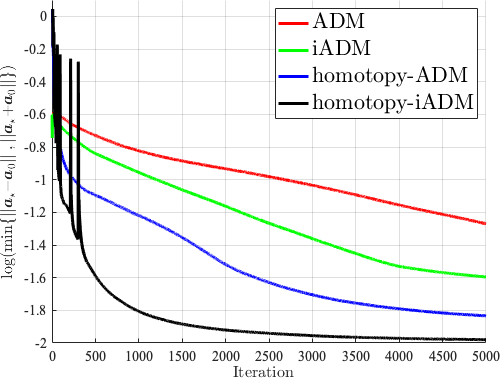}
\end{minipage}
\caption{\textbf{Comparison of algorithm convergence for solving SaS-BD on coherent smooth Gaussian kernel $\mb a_0$:} (a) shows the function value $\Psit{BL}(\mb a, \mb x)$ convergence; (b) shows the iterate convergence on $\mb a$, where $\mb a_\star$ denotes a shift correction of each iterate $\mb a$. The algorithms we compared here are ADM, iADM, and its homotopy accelerations. }
\label{fig:convergence_coherent}
\end{figure*}

\paragraph{Convergence with acceleration and homotopy.} Next we compare the convergence speeds of the ADM, with and without momentum (iADM) and homotopy continuation. We use \Cref{eqn:initialization} to initialize $\mb a$, and $\mb x$ is initialized as zero. From \Cref{fig:convergence_incoherent,fig:convergence_coherent}, we see applying acceleration and homotopy leads to in significant improvements over vanilla ADM in terms of convergence rate, especially when $\mb a_0$ is coherent.

\subsection{Comparison with existing methods}
Finally, we compare iADM, and iADM with homotopy, against a number of existing methods for minimizing $\phit{BL}$. The first is \emph{alternating minimization} \cite{kuo2019geometry}, which at each iteration $k$ minimizes $\mb a\iter k$ with $\mb x\iter k$ fixed using accelerated (Riemannian) gradient descent with backtracking, and vice versa. The next method is the popular \emph{alternating direction method of multipliers} (ADMM) \cite{boyd2011distributed}. Finally, we compare against iPALM \cite{pock2016inertial} with backtracking, using the unit ball constraint on $\mb a_0$ instead of the unit sphere.

\begin{figure}[ht]
\centering
\captionsetup[sub]{font=normalsize,labelfont={bf,sf}}
\begin{minipage}[c]{.48\textwidth}
\subcaption{Incoherent $\mb a_0$} 
\centering
  \includegraphics[width = \linewidth]{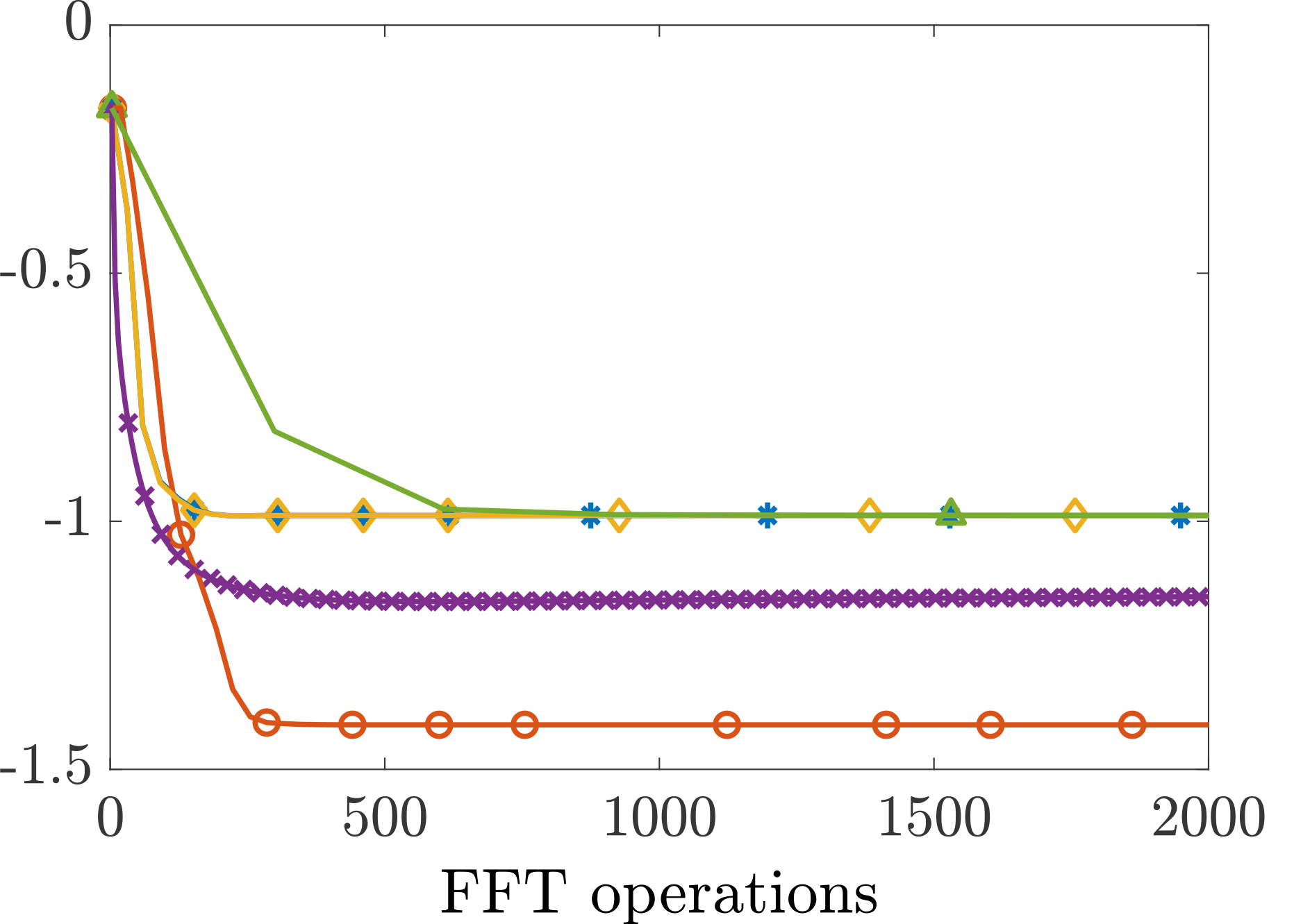}
\end{minipage}
\hfill
\begin{minipage}[c]{.48\textwidth}
\subcaption{Coherent $\mb a_0$} 
\centering
  \includegraphics[width = \linewidth]{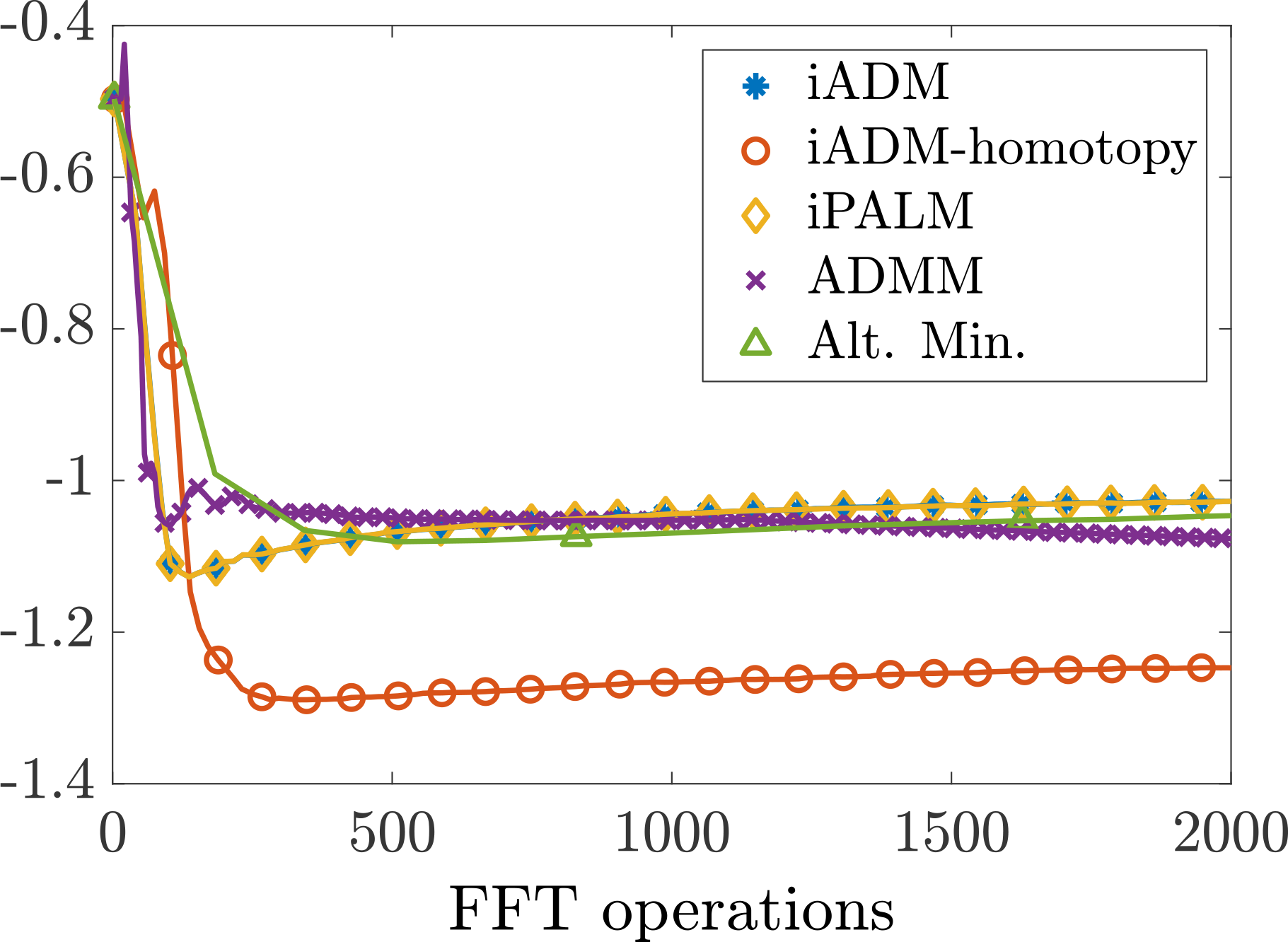}
\end{minipage}

\caption{\textbf{Algorithmic comparison.} (a) Convergence of various methods minimizing $\Psit{BL}$ with incoherent $\mb a_0$ over FFT operations used (for computing convolutions). The y-axis denotes the log of the angle between $\mb a\iter k$ and the nearest shift of $\mb a_0$, and each marker denotes five iterations. (b) Convergence for coherent $\mb a_0$.}
\label{fig:alg-compare}
\end{figure}

For each method, we deconvolve signals with $n_0=50, m=100$, and $\theta=n_0^{-3/4}$ for both coherent and incoherent $\mb a_0$. For both iADM, iADM with homotopy, and iPALM we set $\alpha=0.3$.
For homotopy, we set $\lambda\iter1=\max_\ell\lvert\langle s_\ell[\mb a\iter0], \mb y\rangle\rvert$, $\lambda^\star=\tfrac{0.3}{\sqrt{n_0\lambda}}$, and $\delta=0.5$. Furthermore we set $\eta=0.5$ or $\eta=0.8$ and
for ADMM, we set the slack parameter to $\rho=0.7$ or $\rho=0.5$ for incoherent and coherent $\mb a_0$ respectively. From \Cref{fig:alg-compare}, we can see that ADMM performs better than iADM in the incoherent case, but becomes less reliable in the coherent case. In both cases, iADM with homotopy is the best performer. Finally, we observe roughly equal performance between iPALM and iADM.

%!TEX root=../iclr.tex

\section{Experiments for Real Applications}

\label{sec:exp}
% !TEX root = ../main.tex
In this section, we demonstrate experimentally the effectiveness of the proposed methods for both SaSD and SaS-CDL on a wide variety of applications in computational imaging and neuroscience. Our goal here is not necessarily to outperform state of the art methods, which are often tailored to specific applications. Rather, we hope to provide evidence that the intuition and heuristics highlighted in \Cref{sec:geometry,sec:algorithm} are widely applicable, and can serve as a robust starting point for tackling SaS problems broadly in areas of imaging science.

\subsection{Sparse deconvolution of time sequences in neuroscience}
% Most neurons respond to external stimuli by firing action potentials which serve as a means of communication with other neurons. Therefore, the detection of neuronal spiking activity is a prerequisite for understanding the mechanism of brain function. In the following, we present two concrete examples showing the effectiveness of our sparse deconvolution methods for solving neuronal spiking detection and classification problems.
% !TEX root = ../main.tex

\subsubsection{Sparse deconvolution of calcium imaging}
It is well known that neurons process and transmit information via discrete spiking activity. Whenever a neuron fires, it produces a transient change in chemical concentrations in the immediate environment. Transients in calcium (Ca$^{2+}$) concentration, for example, can be measured using calcium fluoresence imaging. The resulting fluoresence signal can be modeled as the convolution between the short transient response $\mb a_0$ and the spike train in the form of nonnegative, sparse map $\mb x_0$,
\begin{align}\label{eqn:calcium-deconvolution-model}
   \underbrace{\mb y}_{\text{raw fluorescence trace}} = 	 \underbrace{ \mb a_0 }_{ \text{transient response } } \; \conv \; \underbrace{ \mb x_0 }_{ \text{action potentials } } \;+\; \underbrace{b \mb 1_m }_{ \text{ bias }} + \underbrace{\mb n}_{\text{noise}}, \qquad \mb x_0 \geq \mb 0.
\end{align}
The task of recovering the spike train $\mb x_0$ from such SaS signals are frequently of interest in the neuroscience, and can naturally be cast as a SaSD problem. An advantage of this approach is its ability to estimate transient response (which is rarely known a priori) simultaneously. This is important when neurons exhibit dense bursts of spiking activity, which is an especially challenging setting for deconvolution tasks.

\paragraph{Simulated data.}

\begin{figure*}[!htbp]
\centering
\captionsetup[sub]{font=normalsize,labelfont={bf,sf}}
\begin{minipage}[c]{0.4\textwidth}
\subcaption{AR(1) model}
\centering
	\includegraphics[width = \linewidth]{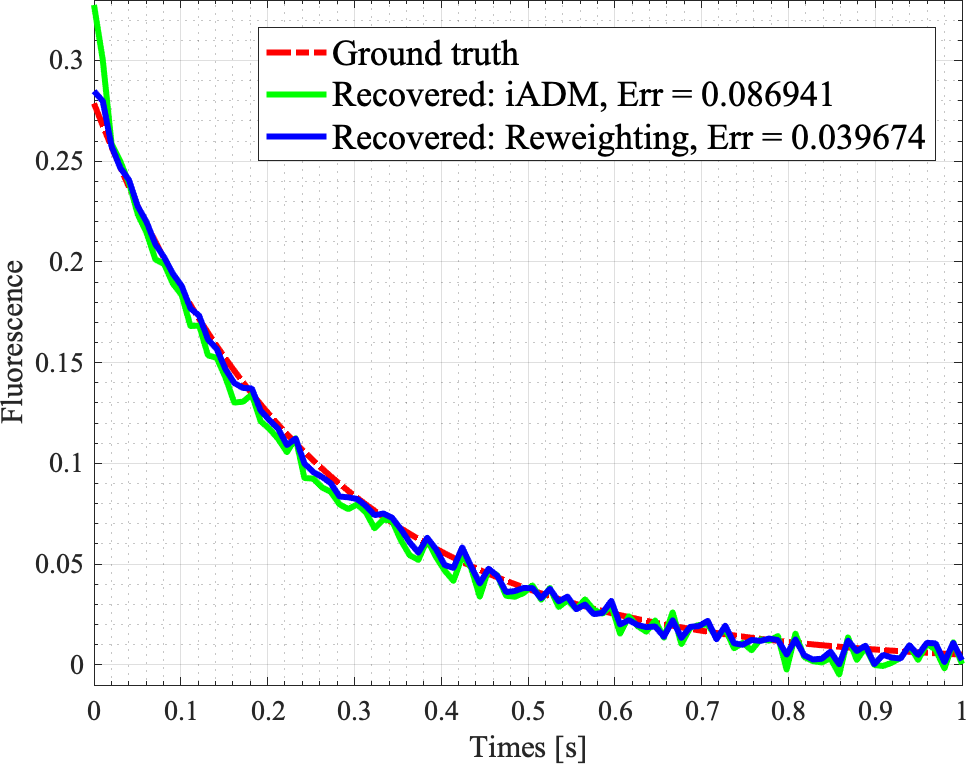}
\end{minipage}
\hspace{0.2in}
\begin{minipage}[c]{0.4\textwidth}
\subcaption{AR(2) model}
\centering
	\includegraphics[width = \linewidth]{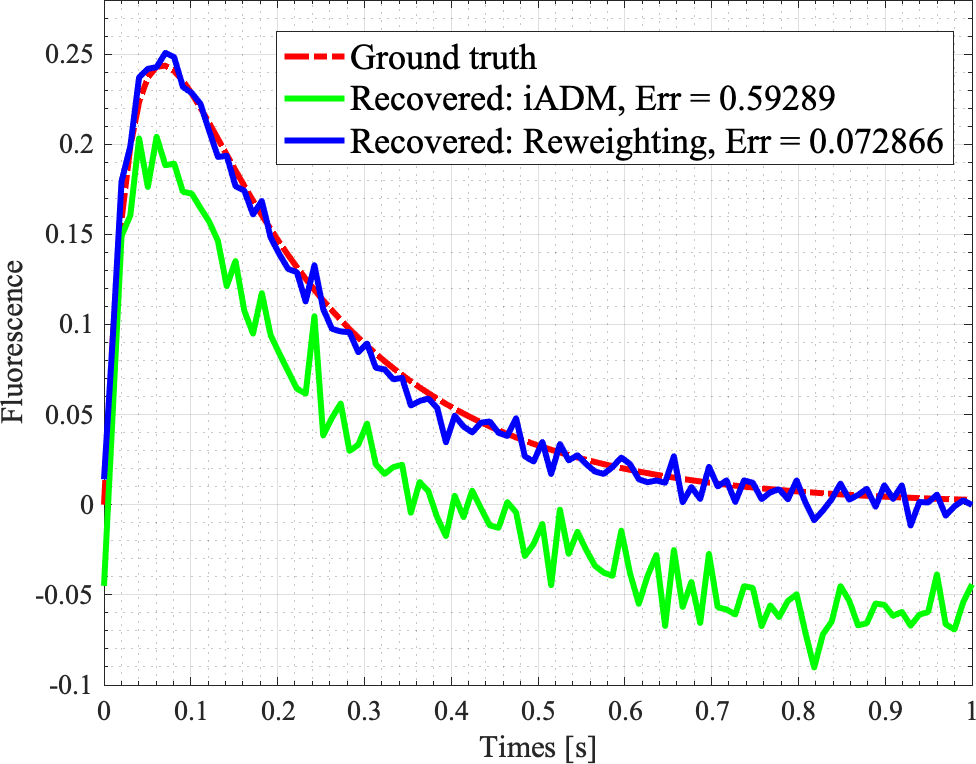}
\end{minipage}
\caption{\textbf{Recovery of transient response $\mb a_0$ for calcium imaging.} The left figure denotes kernel $\mb a_0$ for the AR(1) model, and the right figure shows the kernel $\mb a_0$ for AR(2) model.}
\label{fig:calcium-kernel-simulated}
\end{figure*}

Recent work \cite{vogelstein2010fast,pnevmatikakis2016simultaneous,friedrich2017fast} suggests that the calcium dynamics $\mb y$ can be well approximated by using a autoregressive (AR) process of order $r$,
\begin{align*}
	y(t) = \sum_{i=1}^r \gamma_i y(t-i) + x_0(t) + b + n_s(t),
\end{align*}
where $x_0(t)$ is the number of spikes that the neuron fired at $t$-th timestep, $n_s(t)$ is noise, and $\Brac{\gamma_i}_{i=1}^r$ are autoregressive parameters. \cite{pnevmatikakis2016simultaneous,friedrich2017fast} showed that the AR($r$) model is equivalent to \Cref{eqn:calcium-deconvolution-model} with a parameterized kernel $\mb a_0$. The order $r$ is chosen to be a small positive integer, usually $r = 1$ or $r=2$. When $r=1$, the AR(1) kernel is a one-sided exponential function
\begin{align}\label{eqn:kernel-ar1}
	a_0(t) = \exp\paren{ -t/\tau},\qquad t\geq 0,
\end{align}
for some $\tau>0$. The AR(1) model serves as a good approximation of the calcium dynamics when the temporal resolution of imaging sensors is low. In contrast, the AR(2) model serves as a more accurate model for high temporal resolution calcium dynamics, with
\begin{align}\label{eqn:kernel-ar2}
   a_0(t) =
 	\exp\paren{ -t/\tau_1 } - \exp\paren{ -t /\tau_2}, \qquad t \geq 0,
\end{align}
where $\tau_1$ and $\tau_2$ are some parameters with $\tau_1>\tau_2>0$. As illustrated in \Cref{fig:calcium-kernel-simulated}, for high temporal resolution calcium dynamics, the AR(2) model tends to be a better model which captures the short rise-time of calcium transients by the difference of two exponential functions.

Here we demonstrate the effectiveness of the proposed methods on synthetic data for both AR(1) and AR(2) models. We generate a sequence of simulated calcium dynamics $\mb y$ with length $T = 100(s)$ and sampling rate $f = 100Hz$ (i.e.\ $\sample = 10^4$ samples in total). We generate the kernel $\mb a_0 \in \bb R^{\di}$ with length $T=1(s)$ (i.e.\ $\di = 100$): for the AR(1) model, we set $\tau = 0.25$ in \Cref{eqn:kernel-ar1}; for AR(2) model, we set $\tau_1 = 0.2$ and $\tau_2=0.03$ in \Cref{eqn:kernel-ar2}. Each kernel is normalized so they lie on the sphere. The sparse spike train $\mb x_0$ is generated from Bernoulli distribution $\mb x_0 \sim_{i.i.d.} \mc B(\theta)$ with sparsity rate $\theta = \di^{-4/5}$. We set the bias $b = 1$ and noise $\mb n \sim \mc N\paren{ \mb 0, \sigma^2 \mb I },\ \sigma = 5\times 10^{-2}$ in \Cref{eqn:calcium-deconvolution-model}.

\begin{figure*}[!htbp]
\centering
\captionsetup[sub]{font=small,labelfont={bf,sf}}
\begin{minipage}[c]{\textwidth}
\subcaption{raw data vs. estimated calcium dynamics}
\centering
	\includegraphics[width = \linewidth]{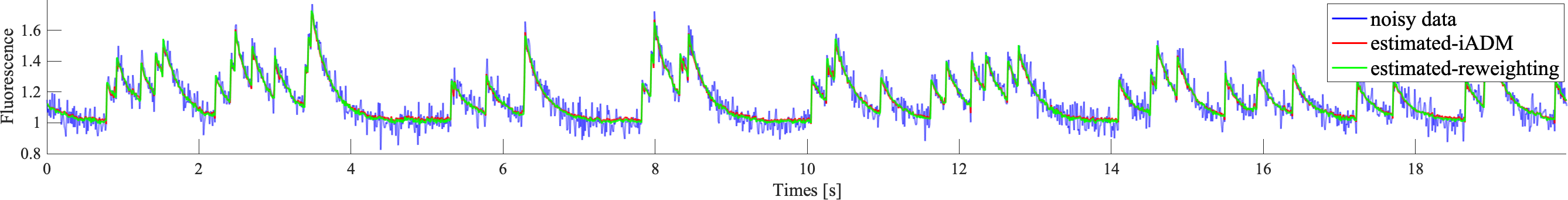}
\end{minipage}
\begin{minipage}[c]{\textwidth}
\subcaption{spike train, iADM algorithm, $ \min_{\ell} \norm{\mb x_0 - \mathrm{s}_\ell \brac{ \mb x_\star} }{2} = 6.2541 $}
\centering
	\includegraphics[width = \linewidth]{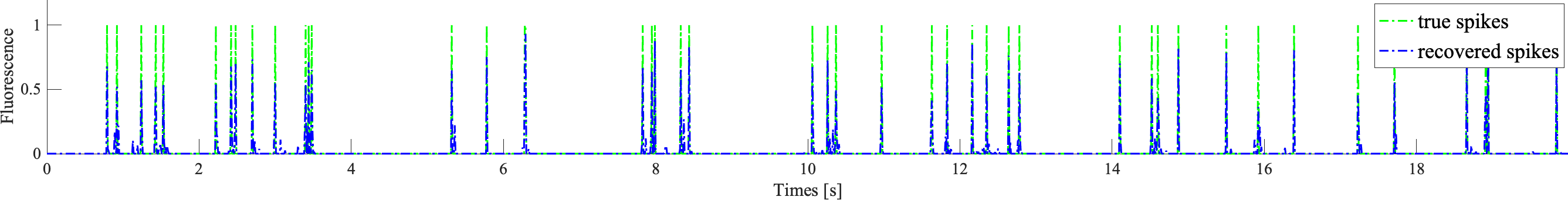}
\end{minipage}
\begin{minipage}[c]{\textwidth}
\subcaption{spike train, reweighted-iADM algorithm, $ \min_{\ell} \norm{\mb x_0 - \mathrm{s}_\ell \brac{ \mb x_\star} }{2} = 2.7989 $}
\centering
	\includegraphics[width = \linewidth]{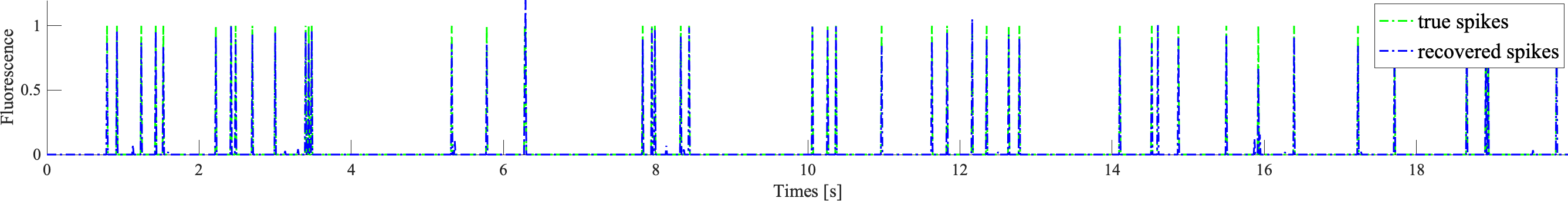}
\end{minipage}
\caption{\textbf{Estimation of spike train $\mb x_0$ for AR(1) model.} The first figure shows the estimation of calcium dynamics, the second figure shows the estimation of the spiking train $\mb x_0$ by iADM algorithm, and the third figure demonstrates the reweighting variant of iADM. $ \min_{\ell} \norm{\mb x_0 - \mathrm{s}_\ell \brac{ \mb x_\star} }{2}$ denotes the distance between the target $\mb x_0$ and estimated solution $\mb x_\star$. As we observe, the proposed methods can accurately predict the spiking locations even when spikes overlap.}
\label{fig:spiketrain_ar1}
\end{figure*}

\begin{figure*}[!htbp]
\centering
\captionsetup[sub]{font=small,labelfont={bf,sf}}
\begin{minipage}[c]{\textwidth}
\subcaption{raw data vs. estimated calcium dynamics}
\centering
	\includegraphics[width = \linewidth]{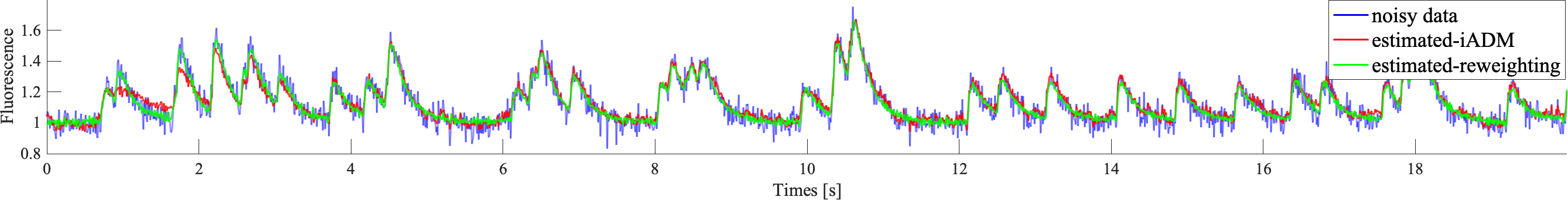}
\end{minipage}
\begin{minipage}[c]{\textwidth}
\subcaption{spike train, iADM algorithm, $ \min_{\ell} \norm{\mb x_0 - \mathrm{s}_\ell \brac{ \mb x_\star} }{2} = 14.2118 $}
\centering
	\includegraphics[width = \linewidth]{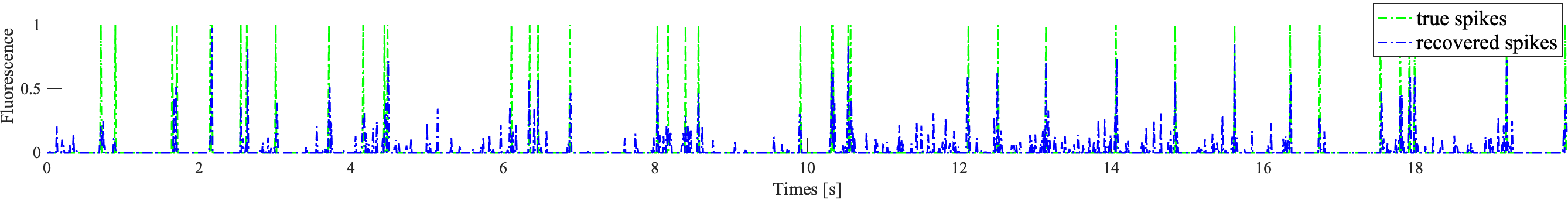}
\end{minipage}
\begin{minipage}[c]{\textwidth}
\subcaption{spike train, reweighted-iADM algorithm, $ \min_{\ell} \norm{\mb x_0 - \mathrm{s}_\ell \brac{ \mb x_\star} }{2} = 15.6756$}
\centering
	\includegraphics[width = \linewidth]{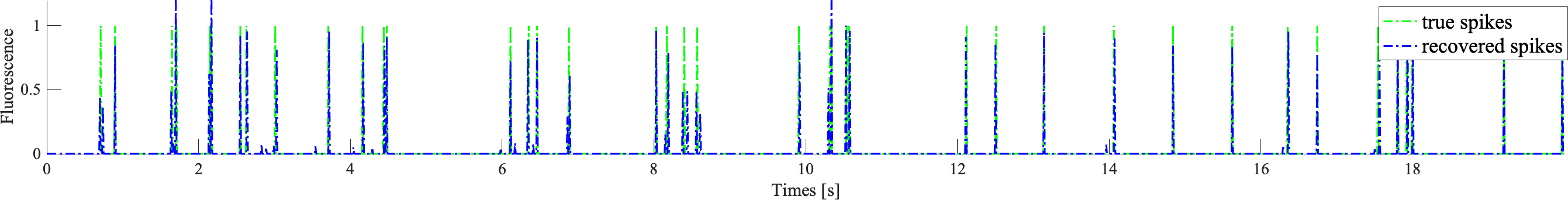}
\end{minipage}
\caption{\textbf{Estimation of spike train $\mb x_0$ for AR(2) model.} The first figure shows the estimation of calcium dynamics, the second figure shows the estimation of the spiking train $\mb x_0$ by iADM algorithm, and the third figure demonstrates the reweighting variant of iADM. We use $ \min_{\ell} \norm{\mb x_0 - \mathrm{s}_\ell \brac{ \mb x_\star} }{2}$ to denote the distance between the target $\mb x_0$ and estimated solution $\mb x_\star$. In comparison with the original iADM algorithm, the reweighting method is very effective in suppressing noise.}
\label{fig:spiketrain_ar2}
\end{figure*}

We test and compare the proposed iADM and its reweighted variant (see Appendix \ref{subsec:homotopy}) for deconvolving the data, with $\lambda = 10^{-1}$. Reweighting is especially effective under noise contamination (\Cref{subsec:structure}), as demonstrated by \Cref{fig:calcium-kernel-simulated} where it provides more accurate predictions of the unknown neuron kernels for both AR(1) and AR(2) models.
From \Cref{fig:spiketrain_ar1,fig:spiketrain_ar2}, we can clearly see that deconvolution is more difficult under the AR(2) model. In such cases reweighting can significantly improve resolution of spiking activity, allowing accurate estimation of firing times even in under dense bursts.

\begin{figure*}[!htbp]
\centering
\captionsetup[sub]{font=normalsize,labelfont={bf,sf}}
\centering
\begin{minipage}[c]{0.4\textwidth}
\subcaption{iADM}
\centering
	\includegraphics[width = \linewidth]{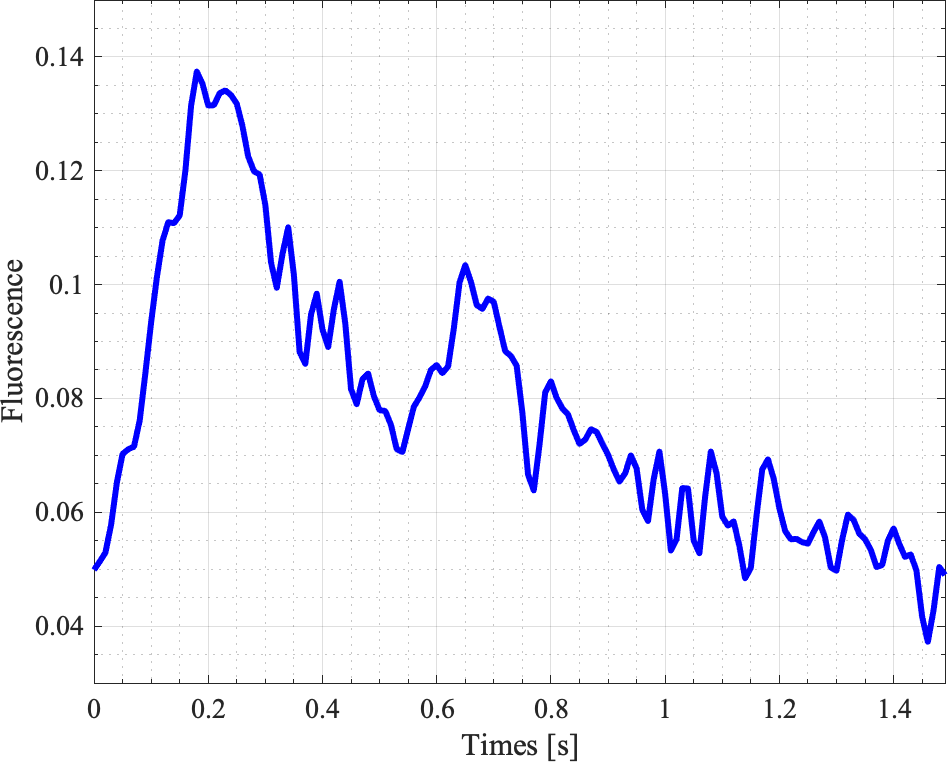}
\end{minipage}
\hspace{0.1in}
\begin{minipage}[c]{0.4\textwidth}
\subcaption{reweighted iADM}
\centering
	\includegraphics[width = \linewidth]{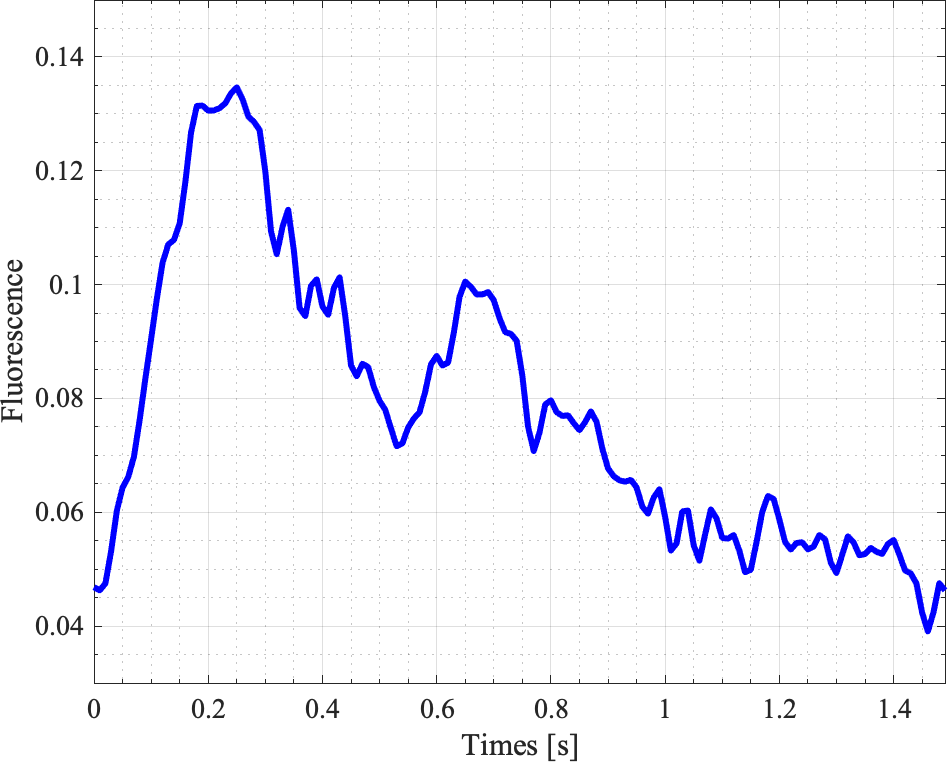}
\end{minipage}
\caption{\textbf{Recovery of transient response $\mb a_0$ for real dataset.} Left figure shows the recovered kernel by the iADM algorithm, right figure shows the recovered kernel by its reweighting variant.}
\label{fig:calcium_kernel_real}
\end{figure*}

\begin{figure*}[!htbp]
\centering
\captionsetup[sub]{font=small,labelfont={bf,sf}}
\begin{minipage}[c]{\textwidth}
\subcaption{raw data vs. estimated calcium dynamics}
\centering
	\includegraphics[width = \linewidth]{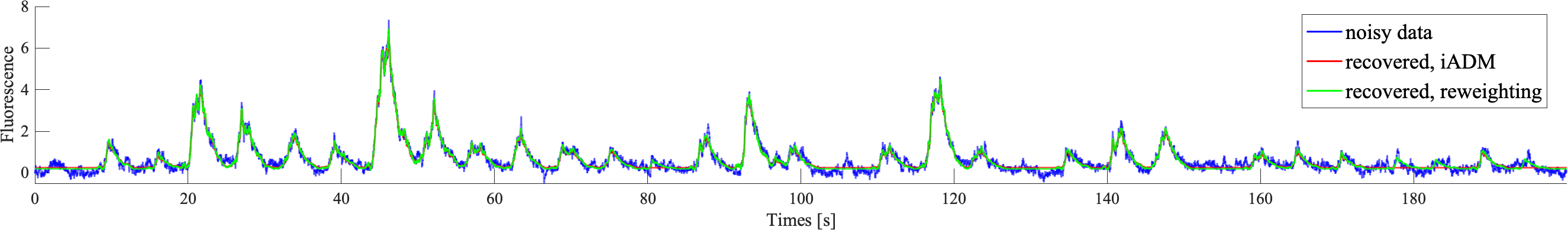}
\end{minipage}
\begin{minipage}[c]{\textwidth}
\subcaption{spike train, iADM algorithm}
\centering
	\includegraphics[width = \linewidth]{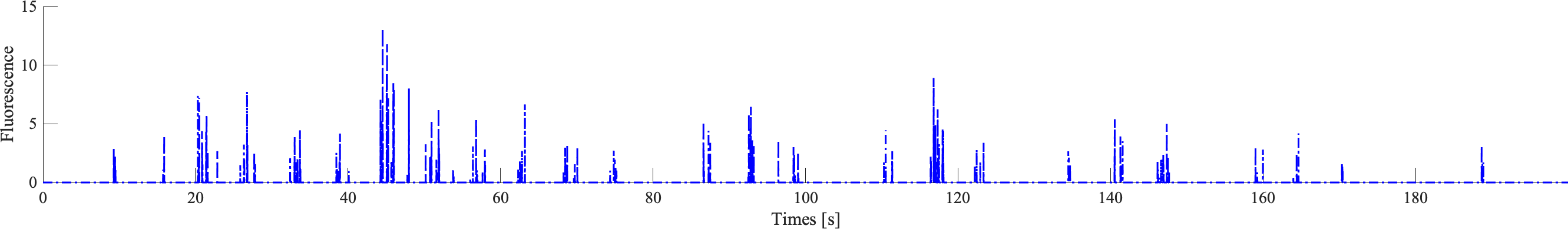}
\end{minipage}
\begin{minipage}[c]{\textwidth}
\subcaption{spike train, reweighted-iADM algorithm}
\centering
	\includegraphics[width = \linewidth]{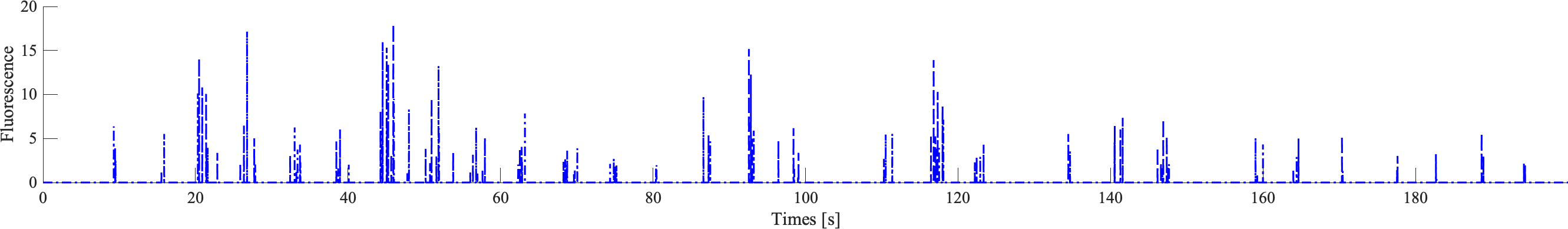}
\end{minipage}

\caption{\textbf{Estimation of spike train $\mb x_0$ for real calcium imaging dataset.} The first figure shows the estimation of calcium dynamics, the second figure shows the estimation of the spiking train $\mb x_0$ by iADM algorithm, and the third figure demonstrates the reweighting variant of iADM.}
\label{fig:calcium_spiketrain_real}
\end{figure*}

\paragraph{Real calcium imaging dataset.} Finally, we demonstrate the effectiveness of proposed methods on the real calcium imaging dataset\footnote{The data is obtain from the spikefinder website,\;\url{http://spikefinder.codeneuro.org/}.}. The data has been resampled to sampling rate $f=100Hz$, and linear drifting trends are removed from calcium traces using robust regression \cite{theis2016benchmarking}. Since these measurements are contaminated by large system noise, as is often the case in realistic settings, we choose a large sparsity penalty $\lambda = 6 \times 10^{-1}$ for \Cref{eqn:bilinear-lasso}. \Cref{fig:calcium_kernel_real} shows the recovered kernel by the proposed iADM and its reweighting variant.  \Cref{fig:calcium_spiketrain_real} shows the estimated spike train. By comparison, the reweighting method appears to produce better estimation of spiking activity.
%As we observe, the AR(2) kernel \eqref{eqn:kernel-ar2} indeed serves as a better model because it captures the short rising dynamics of the spiking activity.

\subsubsection{Spike sorting by convolutional dictionary learning}
\label{subsubsec:spike_sorting}
Electrophysiological activity recorded by electrodes usually record superpositions of waveforms generated from multiple neurons simultaneously \cite{rey2015past}. The goal of spike sorting is to estimate the spiking times from the measurement and decompose the spiking activities of the specific neurons. We refer interested readers to \cite{lewicki1998review,rey2015past} for a more detailed overview of this problem. Traditional spike sorting approaches \cite{quiroga2004unsupervised,chung2017fully,yger2018spike,chaure2018novel} are often time consuming, lack standardization, and involving manual intervention, which makes it difficult to maintain data provenance and assess the quality of scientific results.

In the following, we introduce a fully automated approach based on SaS-CDL and nonconvex optimization; this is similar to the approach taken by \cite{song2018spike}. Mathematically, the measured waveform can be modeled as a superposition of convolutions of individual neuron waveform templates and their corresponding spike trains,
\begin{align*}
   \underbrace{\mb y}_{\text{voltage signal}} \quad = \quad \sum_{k=1}^N \quad \underbrace{ \mb a_{0k} }_{\text{waveform template} } \conv \quad \underbrace{ \mb x_{0k} }_{\text{sparse spike train}} \quad  + \underbrace{\quad b \mb 1_m}_{\text{bias}}  \quad +\quad \underbrace{ \mb n }_{\text{noise}},
\end{align*}
where each waveform templates $\Brac{\mb a_{0k}}_{k=1}^N \in \bb R^{\di}$ correspond to different neurons, and therefore exhibit different kernel shapes. Given the signal $\mb y$, the task of spike sorting is to recover all $\Brac{\mb a_{0k}}_{k=1}^N$ and $\Brac{\mb x_{0k}}_{k=1}^N$; this is a classic example of the SaS-CDL problem as discussed in \Cref{subsec:cdl}.

The difficulty of spike sorting (or SaS-CDL) is not only captured by the shift-coherence of the individual waveforms $\mb a_{0k}$ individually, but also by the shift-coherence between different waveforms from $\Brac{ \mb a_{0k} }_{k=1}^N$. The problem increases with the cross-correlation of differing kernels. Let $\mb A_0 = \begin{bmatrix}
 \mb a_{01} & \cdots & \mb a_{0N}	\end{bmatrix}$. Quantitatively, we can define \emph{mutual incoherence} of $\mb A_0$ by
\begin{align*}
   \mu_m\paren{ \mb A_0 } = \max_{ 1\leq i <j \leq N} \norm{ \mb C_{\mb a_{0i}}^* \mb a_{0j} }{ \infty },
\end{align*}
which is essentially the largest shift-correlation between all kernels. The SaS-CDL problem becomes easy when $\mu_m\paren{ \mb A_0 }$ is small, and vice versa. In the following, we demonstrate the effectiveness of the proposed methods for spike sorting on one easy dataset (with small $\mu_m\paren{ \mb A_0 }$) and one difficult dataset (with large $\mu_m\paren{ \mb A_0 }$).

\begin{figure*}[!htbp]
\centering
%\captionsetup{font=normalsize,labelfont={bf,sf}}
\captionsetup[sub]{font=normalsize ,labelfont={bf,sf}}
\centering
\begin{minipage}[c]{0.32\textwidth}
\subcaption{Neuron 1}
\centering
	\includegraphics[width = \linewidth]{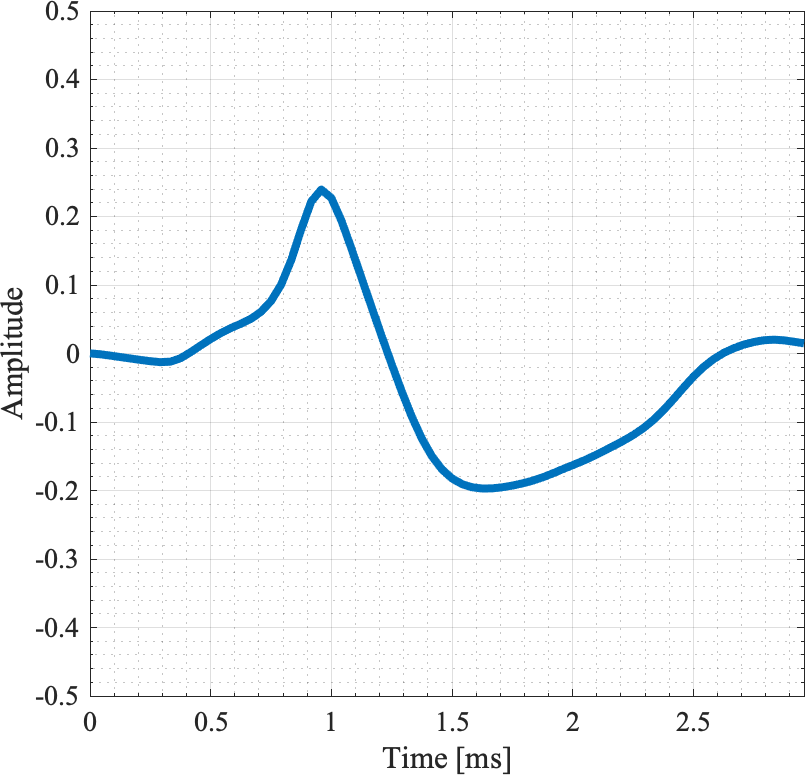}
\end{minipage}
\begin{minipage}[c]{0.32\textwidth}
\subcaption{Neuron 2}
\centering
	\includegraphics[width = \linewidth]{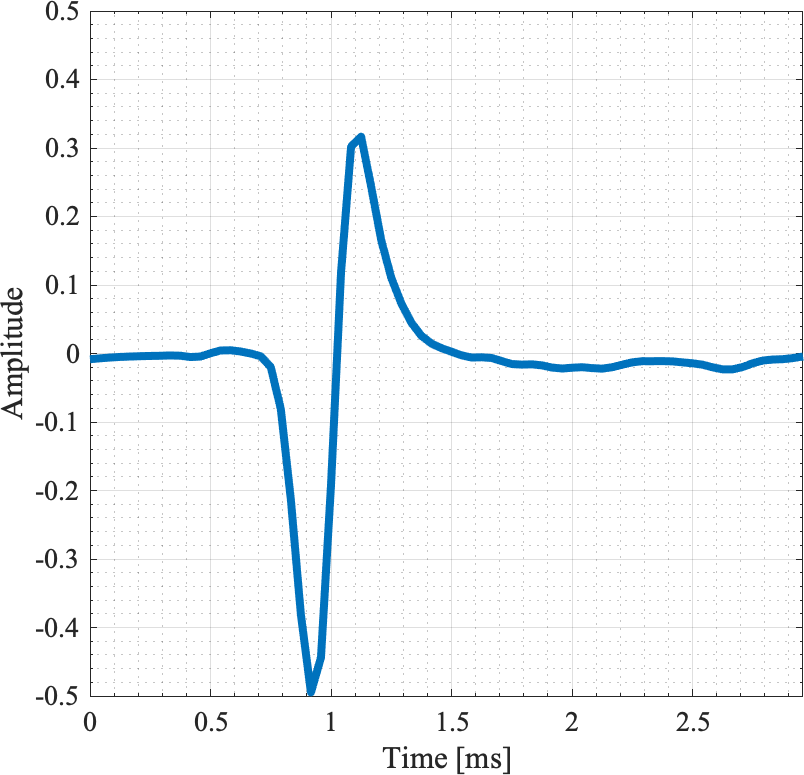}
\end{minipage}
\begin{minipage}[c]{0.32\textwidth}
\subcaption{Neuron 3}
\centering
	\includegraphics[width = \linewidth]{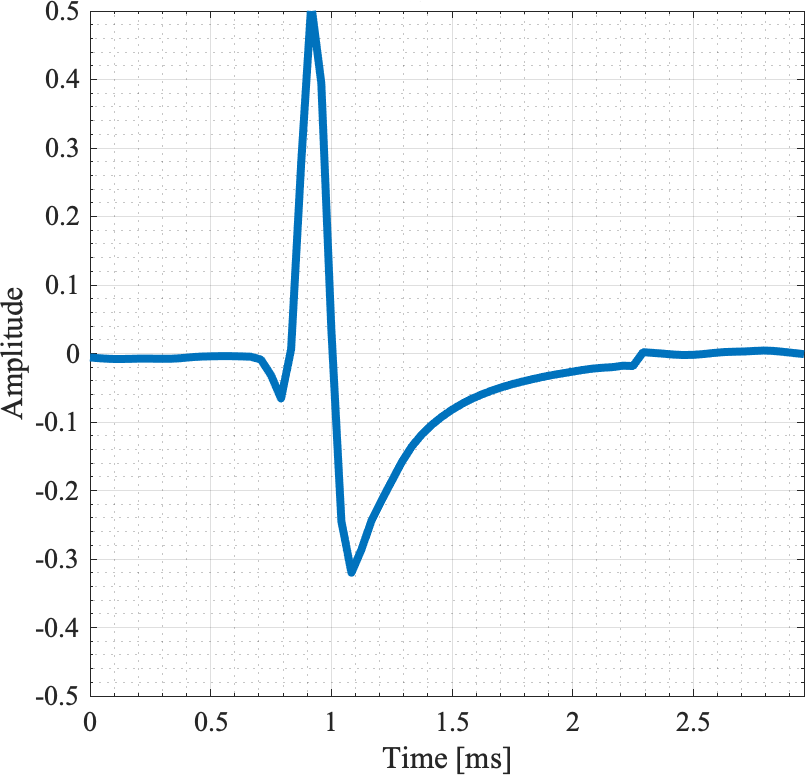}
\end{minipage}
\caption{\textbf{Recovered neuron waveform template of ``easy'' dataset.} The data contains three neurons of \emph{distinct} waveforms, with noise level $0.05$. Each subfigure corresponds to the recovered waveform template of one specific type of neuron.}
\label{fig:waveform-easy}
\end{figure*}

\begin{figure*}[!htbp]
\centering
%\captionsetup{font=normalsize,labelfont={bf,sf}}
\captionsetup[sub]{font=small,labelfont={bf,sf}}
\begin{minipage}[c]{\textwidth}
\subcaption{raw data vs. estimated sequence}
\centering
	\includegraphics[width = \linewidth]{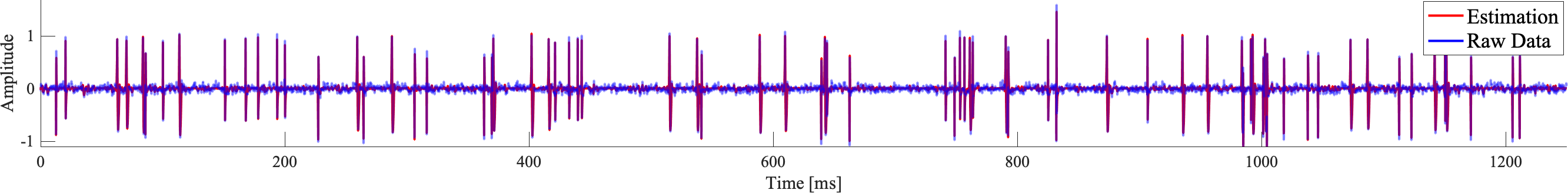}
\end{minipage}
\begin{minipage}[c]{\textwidth}
\subcaption{spike train for Neuron 1}
\centering
	\includegraphics[width = \linewidth]{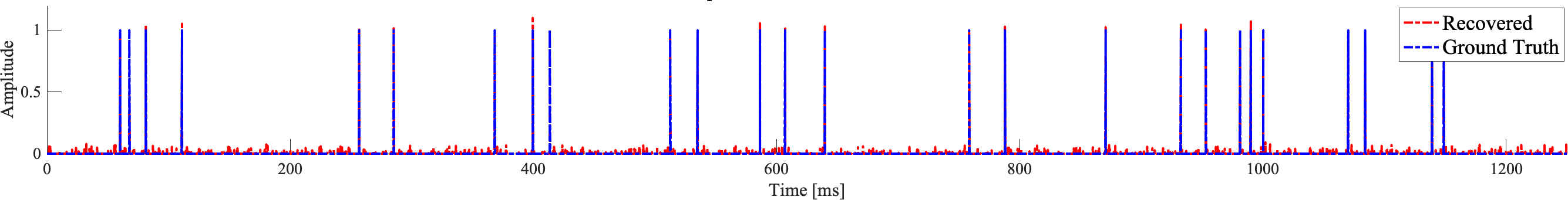}
\end{minipage}
\begin{minipage}[c]{\textwidth}
\subcaption{spike train for Neuron 2}
\centering
	\includegraphics[width = \linewidth]{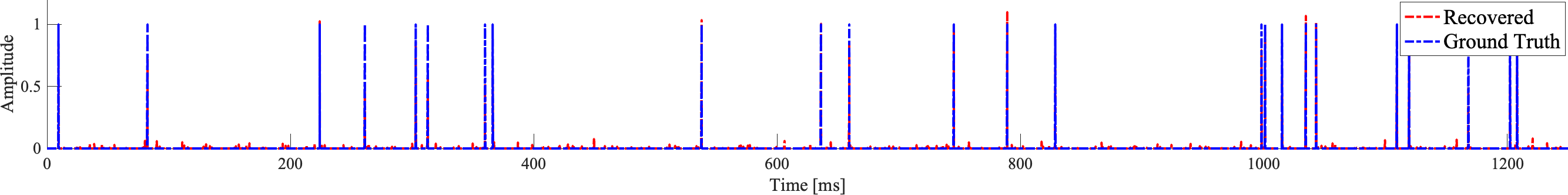}
\end{minipage}
\begin{minipage}[c]{\textwidth}
\subcaption{spike train for Neuron 3}
\centering
	\includegraphics[width = \linewidth]{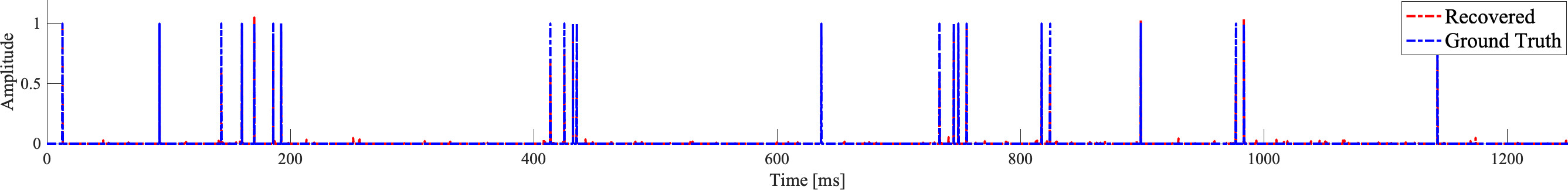}
\end{minipage}
\caption{\textbf{Detected spike train of ``easy'' dataset.} The data contains three neurons of \emph{distinct} waveforms, with noise level $0.05$. The first subfigure shows the estimation of the raw data sequence. The second to fourth subfigures show the predicted spike train for each neuron, respectively.}
\label{fig:spiketrain-easy}
\end{figure*}

We demonstrate the proposed reweighting variant of iADM algorithm on a classical spike-sorting dataset\footnote{It can be downloaded online at \url{https://vis.caltech.edu/~rodri/Wave_clus/Wave_clus_home.htm}.}. The signal is sampled at a frequency of $f = 24kHz$, and each time sequence records spiking activities of 3 different types of neurons. The waveform templates are constructed using a database of $594$ different average spike shapes compiled from recordings in the neocortex and basal ganglia. A more detailed description of dataset can be found in Section 4 of \cite{quiroga2004unsupervised}. We test the proposed method on two signal sequences of length $\sample = 10^5$, each measures the spiking activities of three different types of neurons with length $\di = 72$: one signal sequence is easy to deconvolve with low mutual coherence $\mu_m\paren{ \mb A_0 }$, and another is relatively more difficult with larger $\mu_m\paren{ \mb A_0 }$. The data is contaminated by random noise, with noise level $0.05$ (i.e., the standard deviation relative to the amplitude of the spike classes). The recovered waveform and sparse spike train for the ``easy'' case are shown in \Cref{fig:waveform-easy} and \Cref{fig:spiketrain-easy}, respectively. And the results for the ``difficult'' case are shown in \Cref{fig:waveform-easy} and \Cref{fig:spiketrain-easy}. As we observe, the proposed method successfully recovers the waveform templates and spiking locations for each type of neuron. As the latter ``difficult'' signal sequence contains neuron waveform of similar shapes, we observe slightly more false alarms in spike detection.

\begin{figure*}[!htbp]
%\captionsetup{font=normalsize,labelfont={bf,sf}}
\captionsetup[sub]{font=normalsize ,labelfont={bf,sf}}
\centering
\begin{minipage}[c]{0.32\textwidth}
\subcaption{Neuron 1}
\centering
	\includegraphics[width = \linewidth]{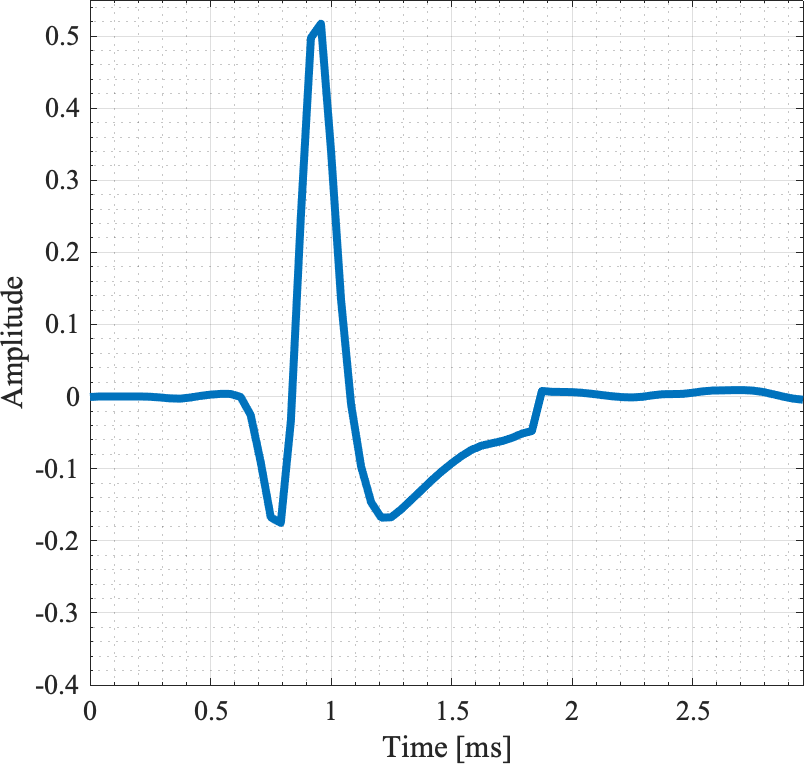}
\end{minipage}
\begin{minipage}[c]{0.32\textwidth}
\subcaption{Neuron 2}
\centering
	\includegraphics[width = \linewidth]{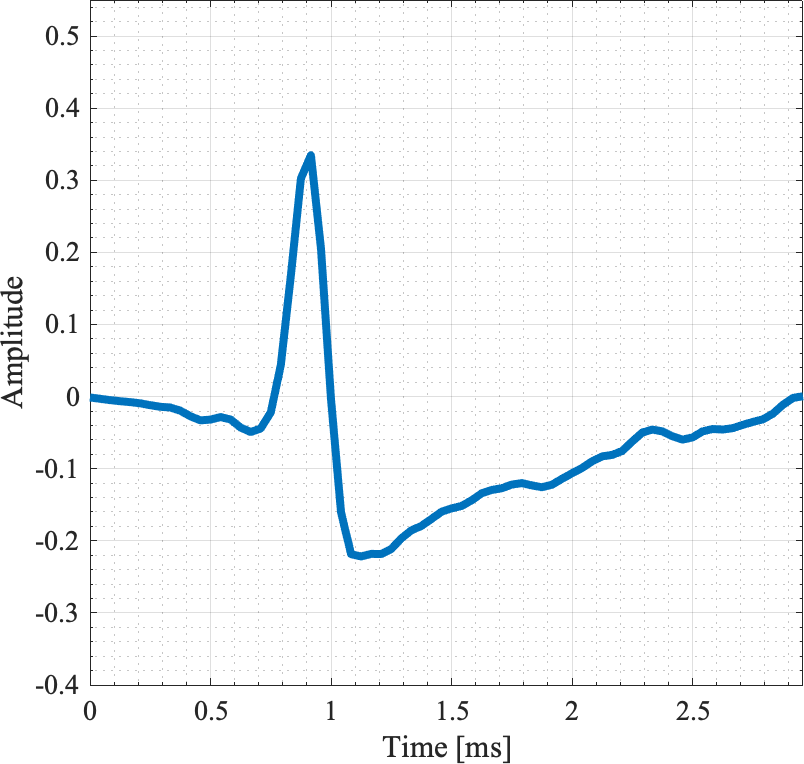}
\end{minipage}
\begin{minipage}[c]{0.32\textwidth}
\subcaption{Neuron 3}
\centering
	\includegraphics[width = \linewidth]{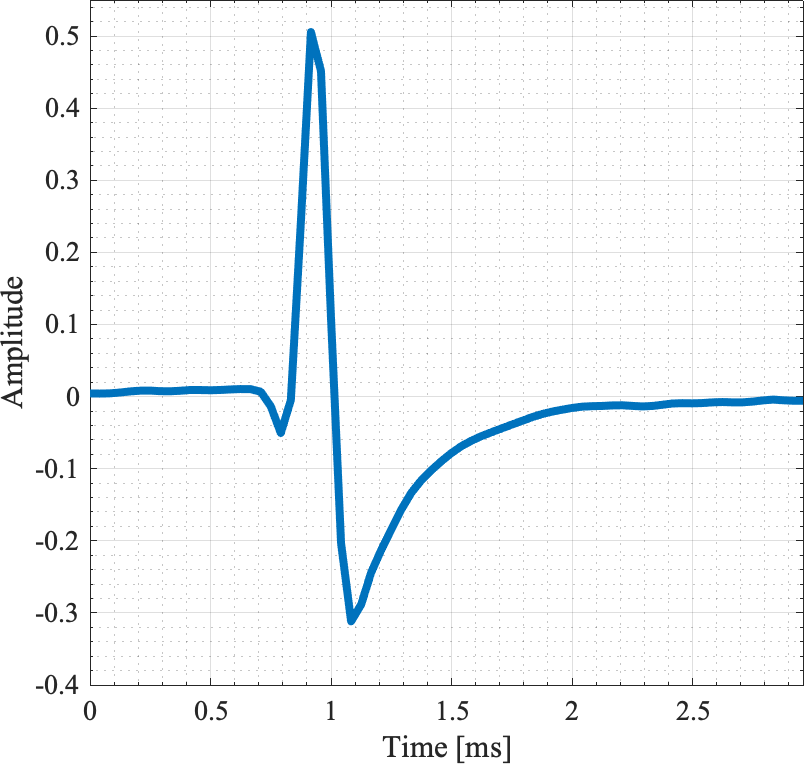}
\end{minipage}
\caption{\textbf{Recovered neuron waveform template of ``difficult'' dataset.} The data contains three neurons of \emph{similar} waveforms, with noise level $0.05$. Each subfigure corresponds to the recovered waveform template of one specific type of neuron.}
\label{fig:waveform-difficult}
\end{figure*}

\begin{figure*}[!htbp]
\centering
%\captionsetup{font=normalsize,labelfont={bf,sf}}
\captionsetup[sub]{font=small,labelfont={bf,sf}}
\begin{minipage}[c]{\textwidth}
\subcaption{raw data vs. estimated sequence}
\centering
	\includegraphics[width = \linewidth]{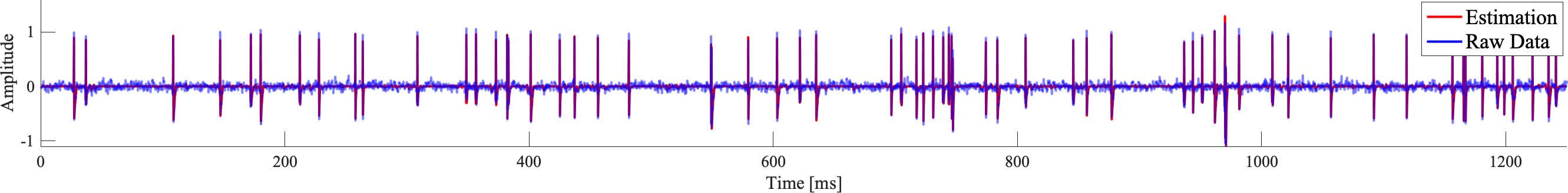}
\end{minipage}
\begin{minipage}[c]{\textwidth}
\subcaption{spike train for Neuron 1}
\centering
	\includegraphics[width = \linewidth]{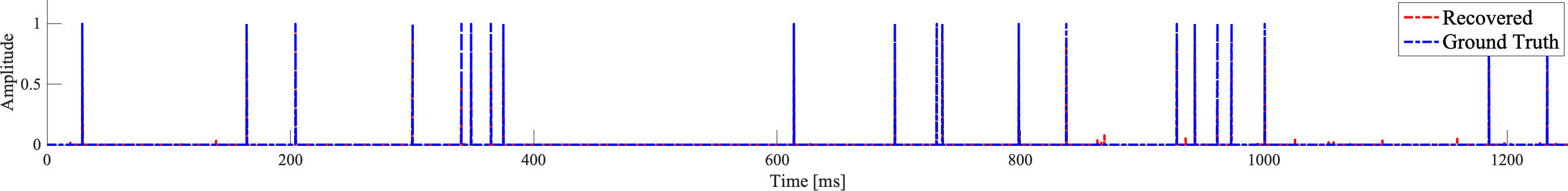}
\end{minipage}
\begin{minipage}[c]{\textwidth}
\subcaption{spike train for Neuron 2}
\centering
	\includegraphics[width = \linewidth]{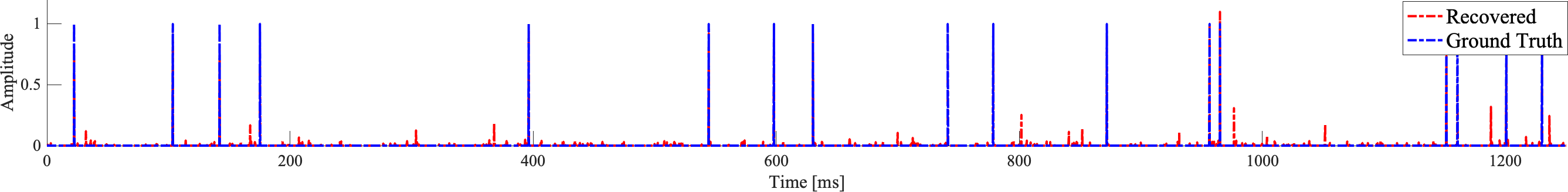}
\end{minipage}
\begin{minipage}[c]{\textwidth}
\subcaption{spike train for Neuron 3}
\centering
	\includegraphics[width = \linewidth]{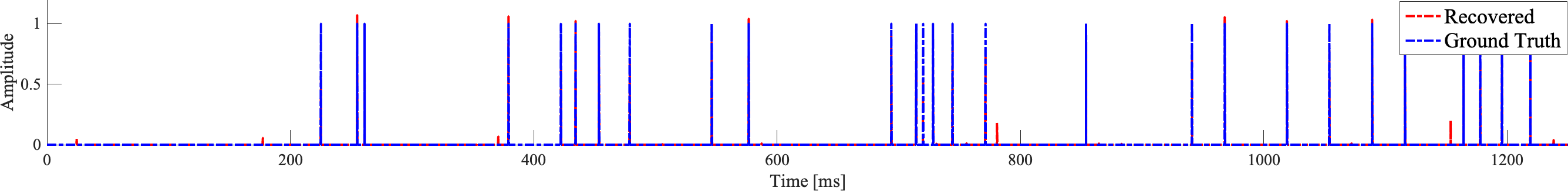}
\end{minipage}
\caption{\textbf{Detected spike train of ``difficult'' dataset.} The data contains three neurons of \emph{similar} waveforms, with noise level $0.05$. The first subfigure shows the estimation of the raw data sequence. The second to fourth subfigures show the predicted spike train for each neuron, respectively.}
\label{fig:spiketrain-difficult}
\end{figure*}

\subsection{Microscopy imaging and data analysis}
Finally, we apply our proposed method towards applications in microscopy, and demonstrate its effectiveness in image super-resolution and decomposition problem settings.
% !TEX root = ../main.tex

\subsubsection{Sparse blind deconvolution for super-resolution fluorescence microscopy}

Fluorescence microscopy is a widely used imaging method in biomedical research \cite{hell2007far,fernandez2008fluorescent}, and has enabled numerous breakthroughs in neuroscience \cite{grienberger2012imaging}, biology and biochemstry \cite{leung2011review,nienhaus2014fluorescent,boettiger2016super}. The spatial resolution of fluorescence microscopy is however limited by diffraction: the wavelength of the light (i.e., several hundred nanometers) is often larger than the typical molecular length scales in cells, preventing a detailed characterization of most subcellular structures.

\begin{figure*}[!htbp]
\captionsetup[sub]{font=small,labelfont={bf,sf}}
\centering
\begin{minipage}[c]{0.45\textwidth}
\subcaption{Frame 1, Time = 0s}
	\includegraphics[width = \linewidth]{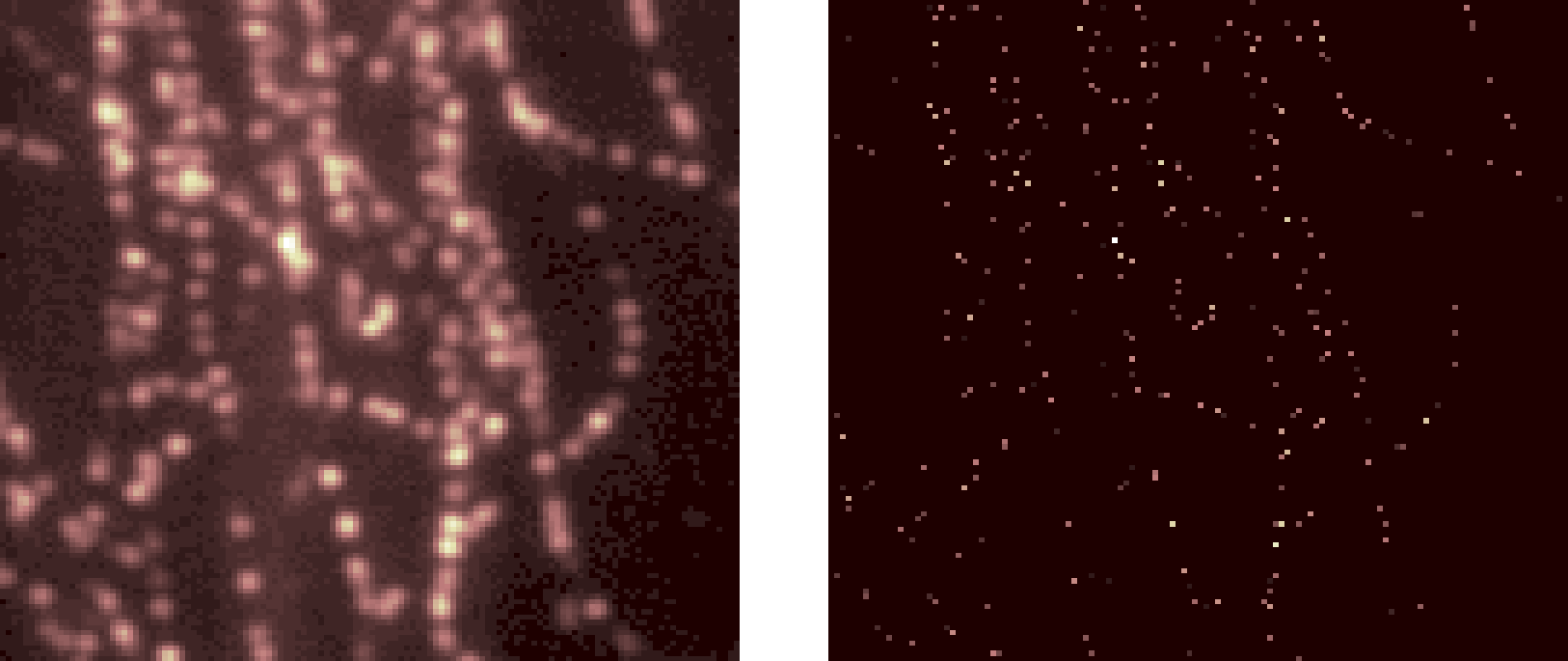}
\end{minipage}
\hspace{0.04\textwidth}
\begin{minipage}[c]{0.45\textwidth}
\subcaption{Frame 100, Time = 4s}
	\includegraphics[width = \linewidth]{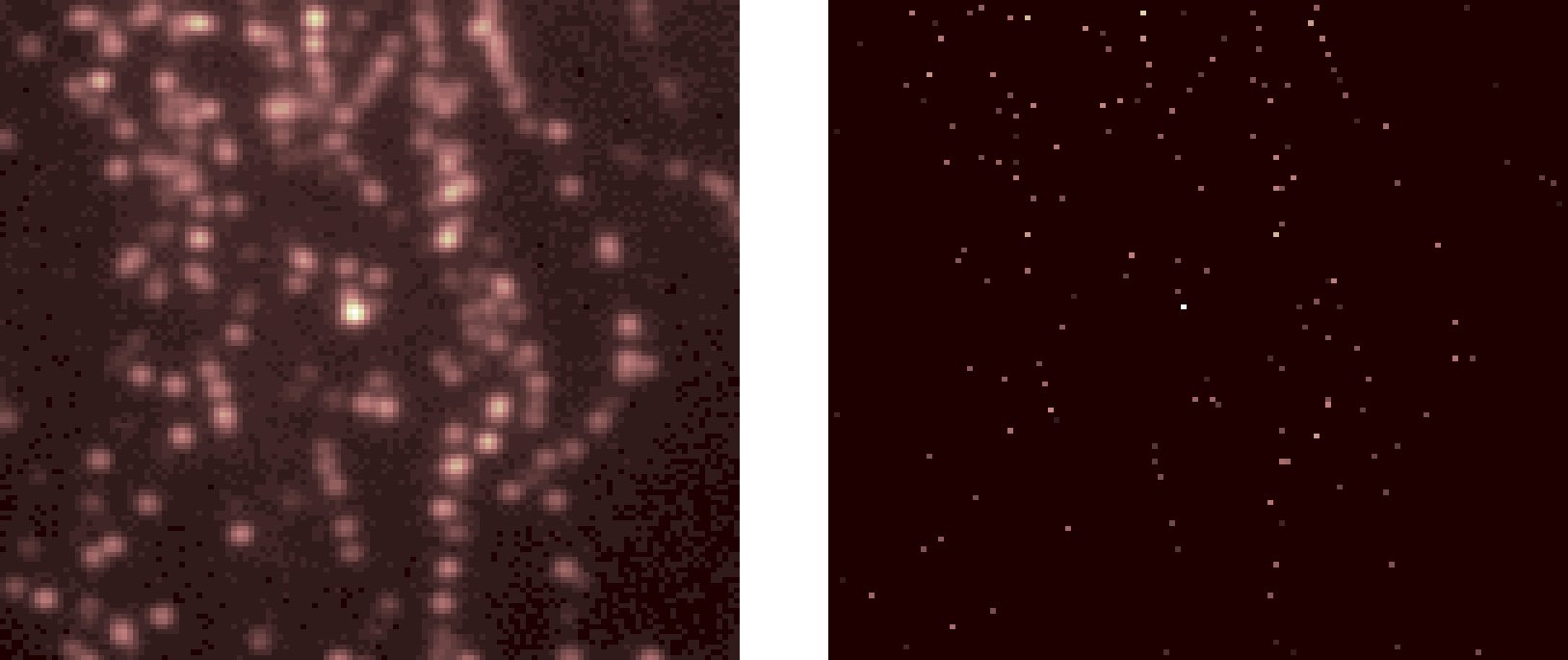}
\end{minipage}

\vspace{0.1in}
\begin{minipage}[c]{0.45\textwidth}
\subcaption{Frame 200, Time = 8s}
	\includegraphics[width = \linewidth]{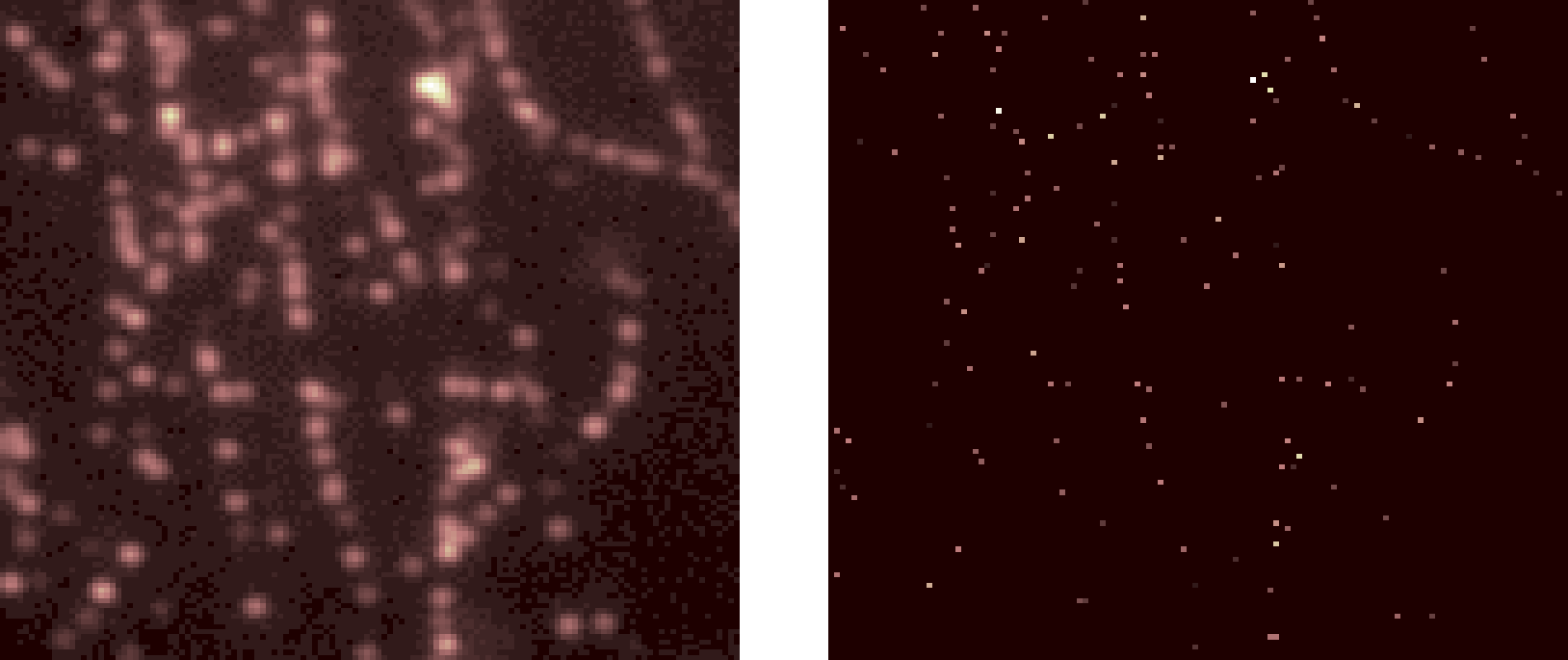}
\end{minipage}
\hspace{0.04\textwidth}
\begin{minipage}[c]{0.45\textwidth}
\subcaption{Frame 300, Time = 12s}
	\includegraphics[width = \linewidth]{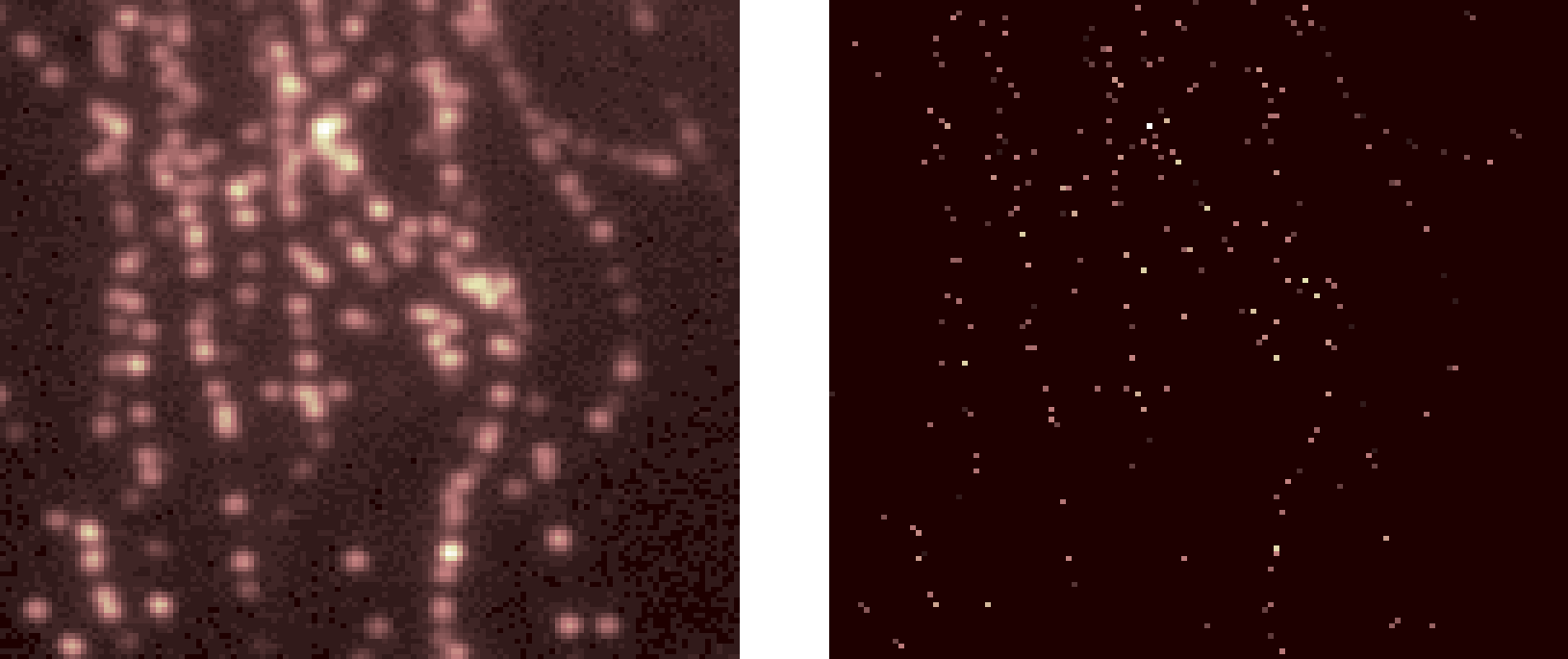}
\end{minipage}

\vspace{0.1in}
\begin{minipage}[c]{0.45\textwidth}
\subcaption{Frame 400, Time = 16s}
	\includegraphics[width = \linewidth]{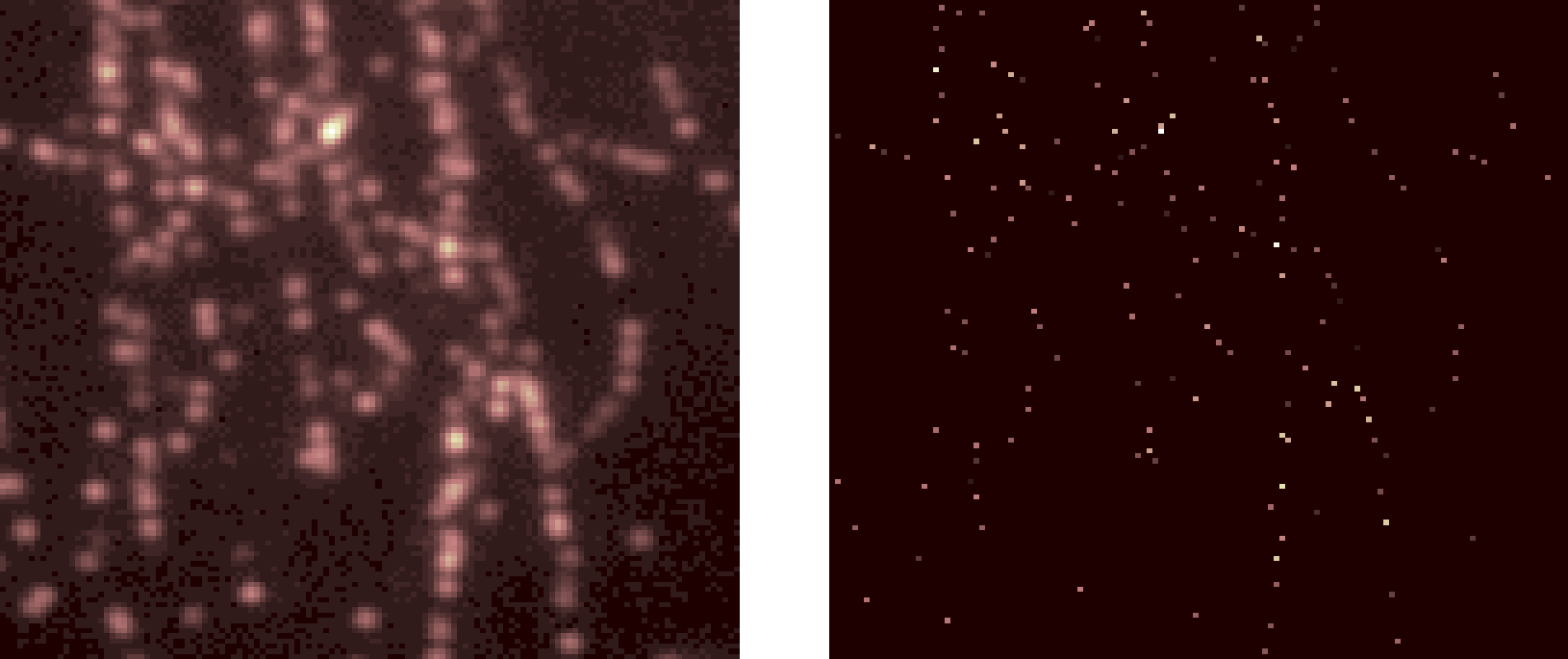}
\end{minipage}
\hspace{0.04\textwidth}
\begin{minipage}[c]{0.45\textwidth}
\subcaption{Frame 500, Time = 20s}
	\includegraphics[width = \linewidth]{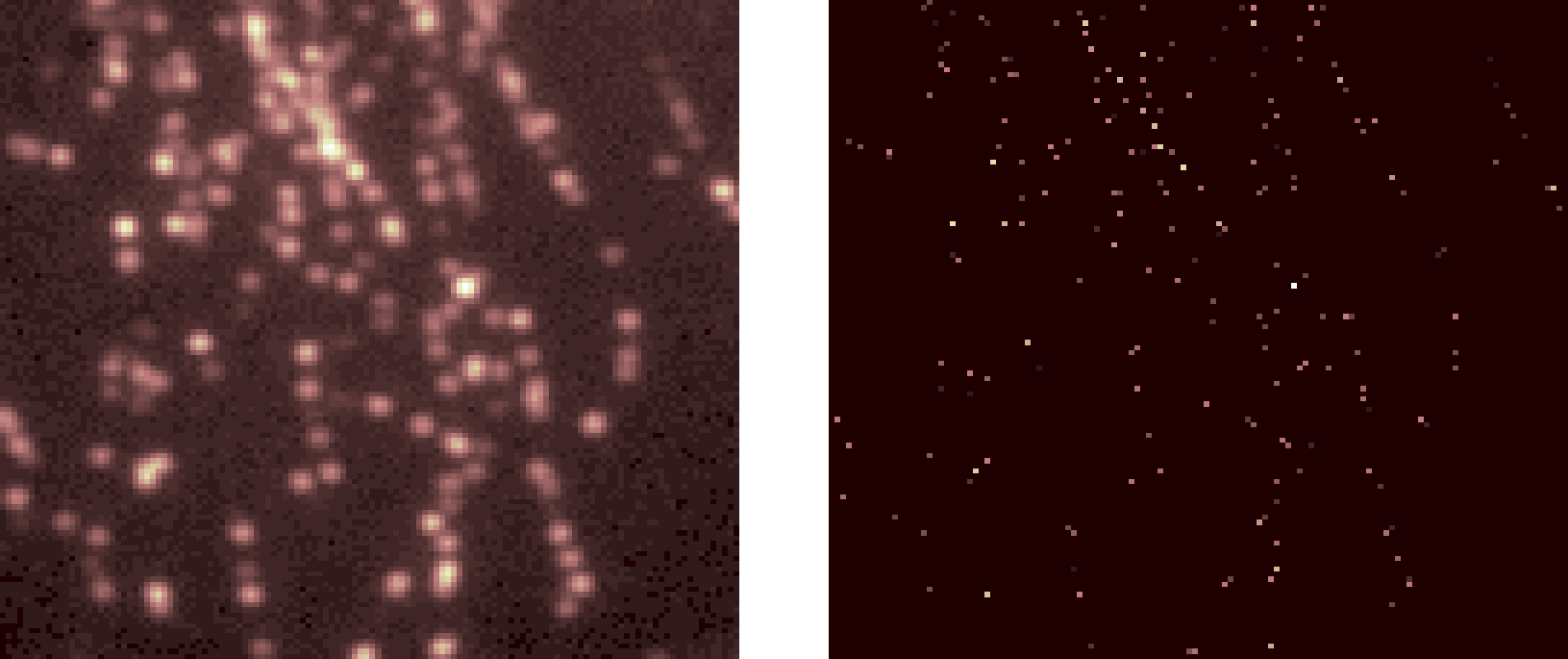}
\end{minipage}
\caption{\textbf{Predicted activation map for each individual frame.} For each subfigure, the left hand side shows the original video frame, and the right hand side presents the predicted activation map using our SaSD solver.}
\label{fig:STORM-activation-map}
\end{figure*}

\begin{figure*}[!htbp]
\captionsetup[sub]{font=large,labelfont={bf,sf}}
\centering
\begin{minipage}[c]{0.45\textwidth}
\subcaption{original image}
	\includegraphics[width = \linewidth]{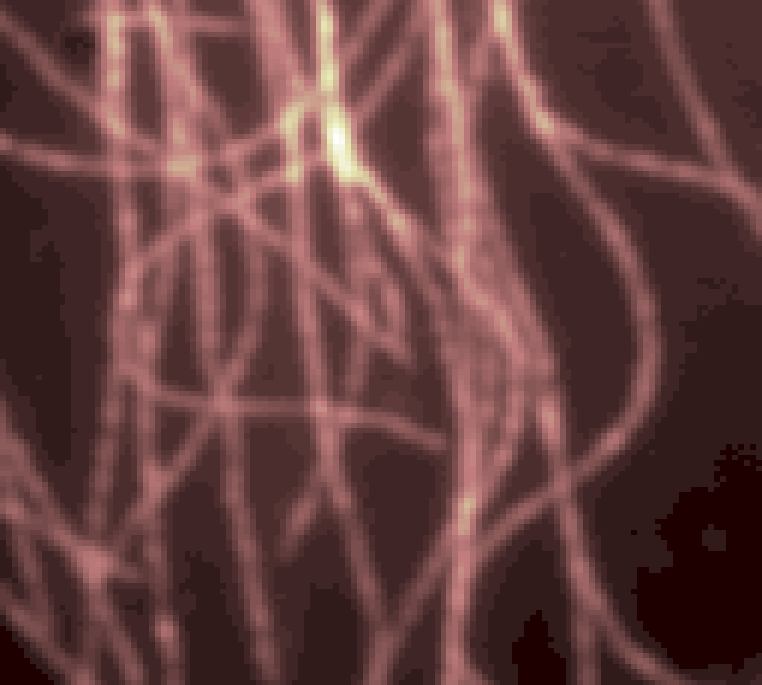}
\end{minipage}
\hspace{0.1in}
\begin{minipage}[c]{0.45\textwidth}
\subcaption{reconstructed image}
	\includegraphics[width = \linewidth]{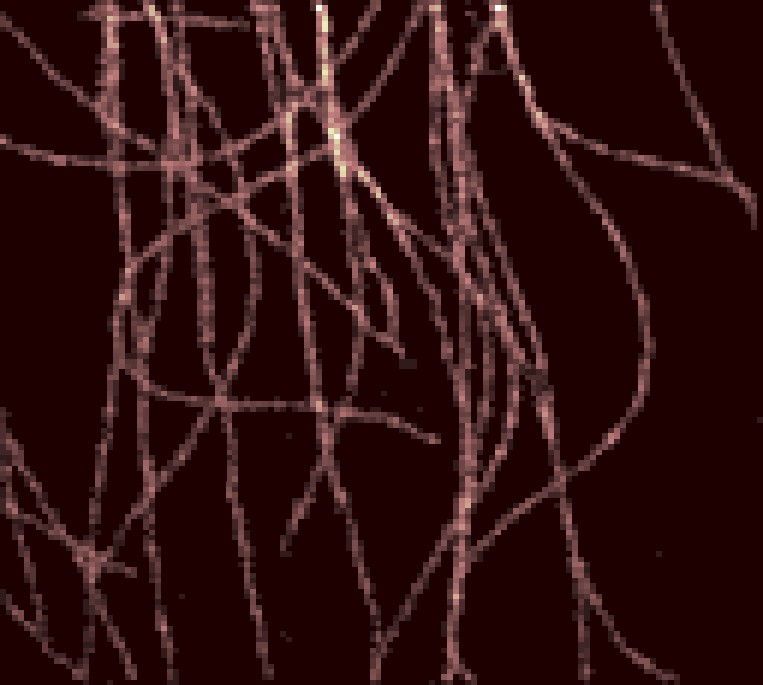}
\end{minipage}

\caption{\textbf{Aggregated result for STORM imaging.} The left hand side shows the original microscopy image, and the right hand side presents the super-resolution image obtained by our method. The pixel resolution is $100$ nm.}
\label{fig:STORM-result}
\end{figure*}

A computational imaging technique recently developed to overcome this resolution limit is \emph{stochastic optical reconstruction microscopy}\footnote{Similar methods with different names have been developed at the same time by using different fluorophores and microscopes, such as photoactivated localization microscopy (PALM) \cite{betzig2006imaging}, and fluorescence photoactivation localization microscopy (fPALM) \cite{hess2006ultra}.} (STORM) \cite{rust2006sub,huang2008three,huang2010breaking}. Instead of activating all the fluorophores at the same time, STORM randomly activates subsets of photoswitchable fluorescent probes to seperate the fluorophores present into multiple frames of sparsely activated molecules (see \Cref{fig:STORM-activation-map} and \Cref{fig:STORM-result}).
From the purspective of the sparsity-coherence tradeoff, this effectively reduces the sparsity of $\mb x_0$, making deconvolution easier to solve.
Therefore, if the location of these molecules can be precisely determined computationally for each frame, synthesizing all deconvolved frames produces a super-resolution microscopy image with near nanoscale resolutions.

For each frame, the localization task can be formulated as a sparse deconvolution problem, i.e.,
\begin{align*}
   \underbrace{\mb Y}_{ \text{frame} } \quad = \quad \underbrace{ \mb A_0}_{\text{point spread function}} \quad \cconv \quad \underbrace{ \mb X_0 }_{ \text{sparse point sources} } \quad + \quad \underbrace{\mb N}_{\text{noise}},
\end{align*}
where we want to recover $\mb X_0$ given $\mb Y$. The classical approaches solve the problem by fitting the blurred spots with Gaussian point-spread functions (PSFs) using either maximum-likelihood or Bayesian estimation techniques \cite{quan2010ultra,holden2011daostorm,zhu2012faster}. These approaches suffer from several limitations: (i) estimation is computationally expensive and poor in quality when dense clusters of fluorophores are activated; (ii) for 3D imaging, the PSF exhibits aberration across the focus plane \cite{sarder2006deconvolution}, making it almost impossible to directly estimate it from the data.

\begin{figure*}[!htbp]
%\captionsetup{font=normalsize,labelfont={bf,sf}}
\captionsetup[sub]{font=normalsize ,labelfont={bf,sf}}
\centering
\begin{minipage}[c]{0.3\textwidth}
\subcaption{PSF in 2D}
\centering
	\includegraphics[width = 0.6\linewidth]{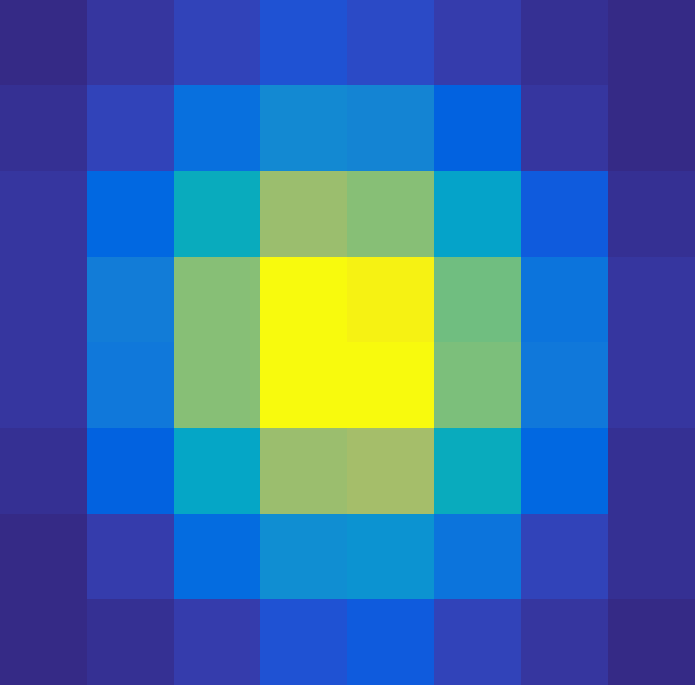}
\end{minipage}
\hspace{0.1in}
\begin{minipage}[c]{0.3\textwidth}
\subcaption{PSF in 3D}
\centering
	\includegraphics[width = 0.6\linewidth]{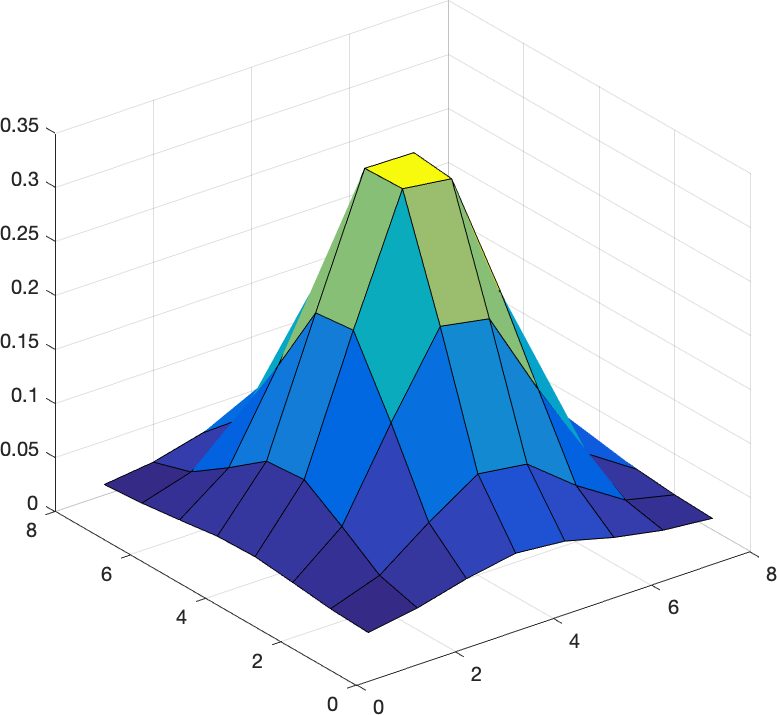}
\end{minipage}
\caption{\textbf{Estimated PSF for STORM imaging.} The left hand side shows the estimated $8\times 8$ PSF in 2D, the right hand side visualizes the PSF in 3D. }
\label{fig:STORM-PSF}
\end{figure*}

To deal with these challenges, we solve the single-molecule localization problem using our proposed method for SaSD to jointly estimate the PSF $\mb A_0$ and the point source map $\mb X_0$. Our frames come from the single-molecule localization microscopy (SMLM) benchmarking dataset\footnote{All the data can be downloaded at \url{http://bigwww.epfl.ch/smlm/datasets/index.html}.}. We apply the reweighted iADM algorithm on the 2D real video sequence ''Tubulin'', which contains $500$ high density frames. The fluorescence wavelength is $690$ nanometer (nm), the imaging frequency is $f = 25Hz$, and each frame is of size $128\times 128$. The single-molecule localization problem is solved on the same $128 \times 128$ pixel grid\footnote{Usually, the localization problem is solved on a finer grid (e.g., grid with $4-10$ times better resolution) so that the resulting resolution can reach $20-30$ nm. We will discuss potential methods to deal with this finer-grid SaSD problem in \Cref{subsec:future} as future work.}, where each pixel is of $100$ nm resolution. \Cref{fig:STORM-PSF} shows the recovered PSF, \Cref{fig:STORM-activation-map} presents the recovered activation map for each individual time frame, and \Cref{fig:STORM-result} presents the aggregated super-resolution image. These results show that our approach can automatically predict the PSF and the activation map for each video frame, producing higher resolution microscopy images without manual intervention.

%Here, we are estimating the point sources $\mb X_0$ on the same pixel grid as the original image. To obtain even higher resolution than the result we obtain here, people are usually estimating the points sources on a finer grid. This results in a simultaneous sparse deconvolution and super-resolution problem. We will discuss about possibilities of conquering this challenge in \Cref{sec:conclusion}.

\subsubsection{Convolutional dictionary learning for microscopy data analytics}

Recent advances in imaging and computational techniques have resulted in the ability to obtain microscopic data in unprecedented detail and volume. SaSD and its extensions are found to be well-suited for extracting motifs and location information from such datasets from neuroscience, material science and beyond, as we have seen from \Cref{subsubsec:spike_sorting}. In certain settings for microscopy, the observed image can also be decomposed as
\begin{align*}
   \underbrace{\mb Y}_{\text{microscopy image}}	\quad =\quad  \sum_{k=1}^K \quad \underbrace{\mb A_{0k}}_{\text{motif } k} \quad\cconv \quad \underbrace{ \mb X_{0k} }_{\text{activation map}} + \underbrace{\mb N}_{\text{noise}}.
\end{align*}
and useful information can be obtained by solving the resulting 2D SaS-CDL problem \cite{pnevmatikakis2016simultaneous,Cheung2018DictionaryLI}.
In this section, we demonstrate our proposed method for SaS-CDL on two different imaging modalities.

\begin{figure*}[!htbp]
\centering
\captionsetup[sub]{font=small,labelfont={bf,sf}}
\begin{minipage}[c]{0.47\textwidth}
\subcaption{\normalsize two-photon calcium image $\mb Y$}
\label{fig:calcium_2D_img}
	\includegraphics[width = \linewidth]{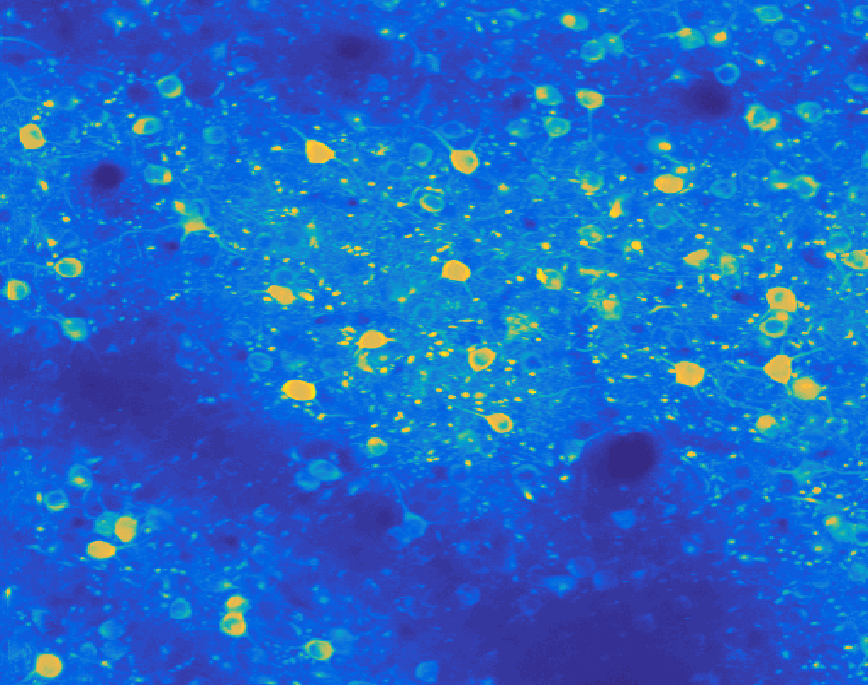}
\end{minipage}
\hspace{0.2in}
\begin{minipage}[c]{0.45\textwidth}
\subcaption{estimated kernel $\mb A_k\;(k=1,2)$}\label{fig:calcium_2D_kernel}
	\includegraphics[width = \linewidth]{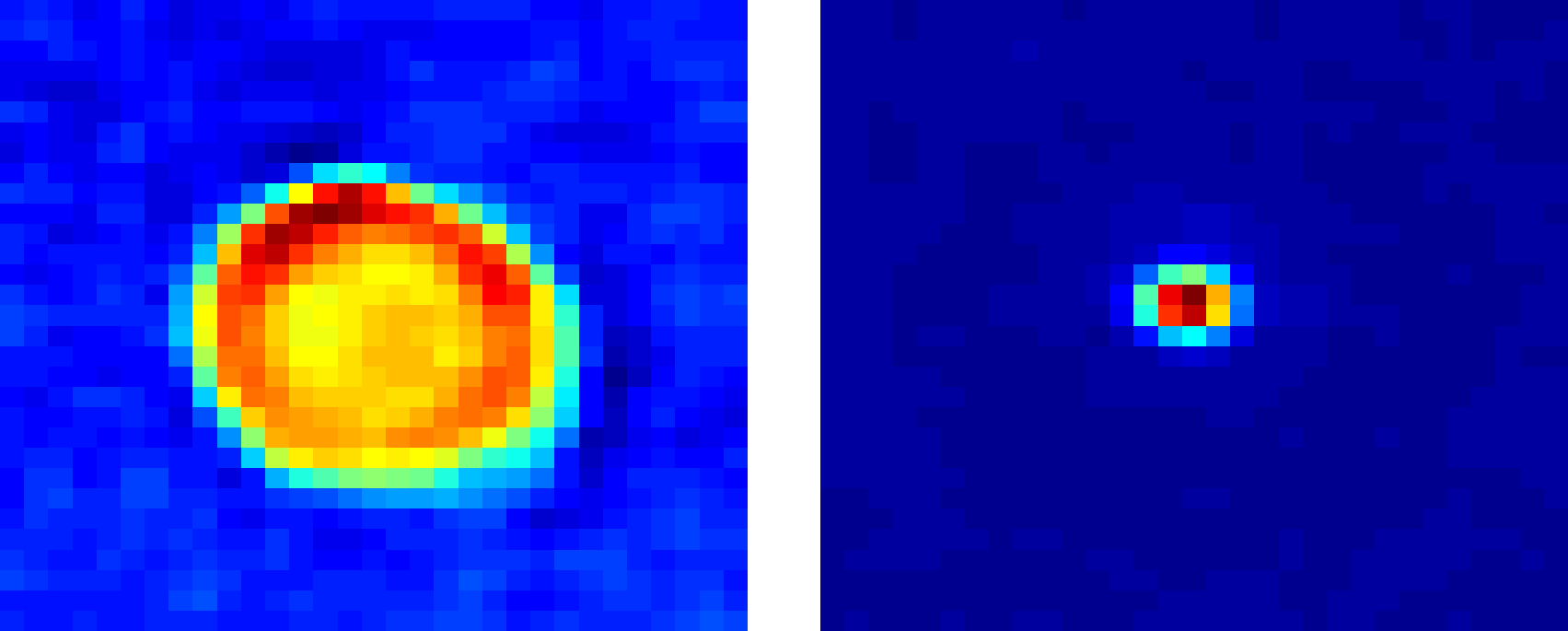}
	\subcaption{predicted activation map $\mb X_k\;(k=1,2)$}\label{fig:calcium_2D_map}
	\includegraphics[width = \linewidth]{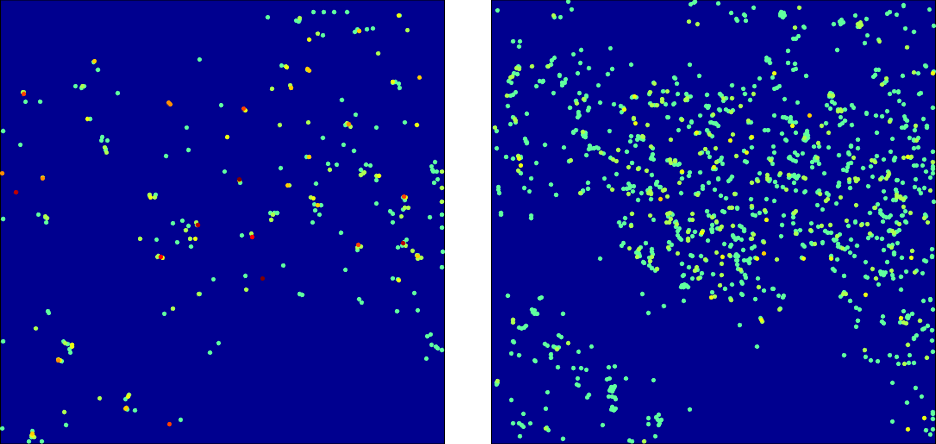}
	\subcaption{classified image $\mb Y_k= \mb A_k \cconv \mb X_k\;(k=1,2)$}\label{fig:calcium_2D_class}
	\includegraphics[width = \linewidth]{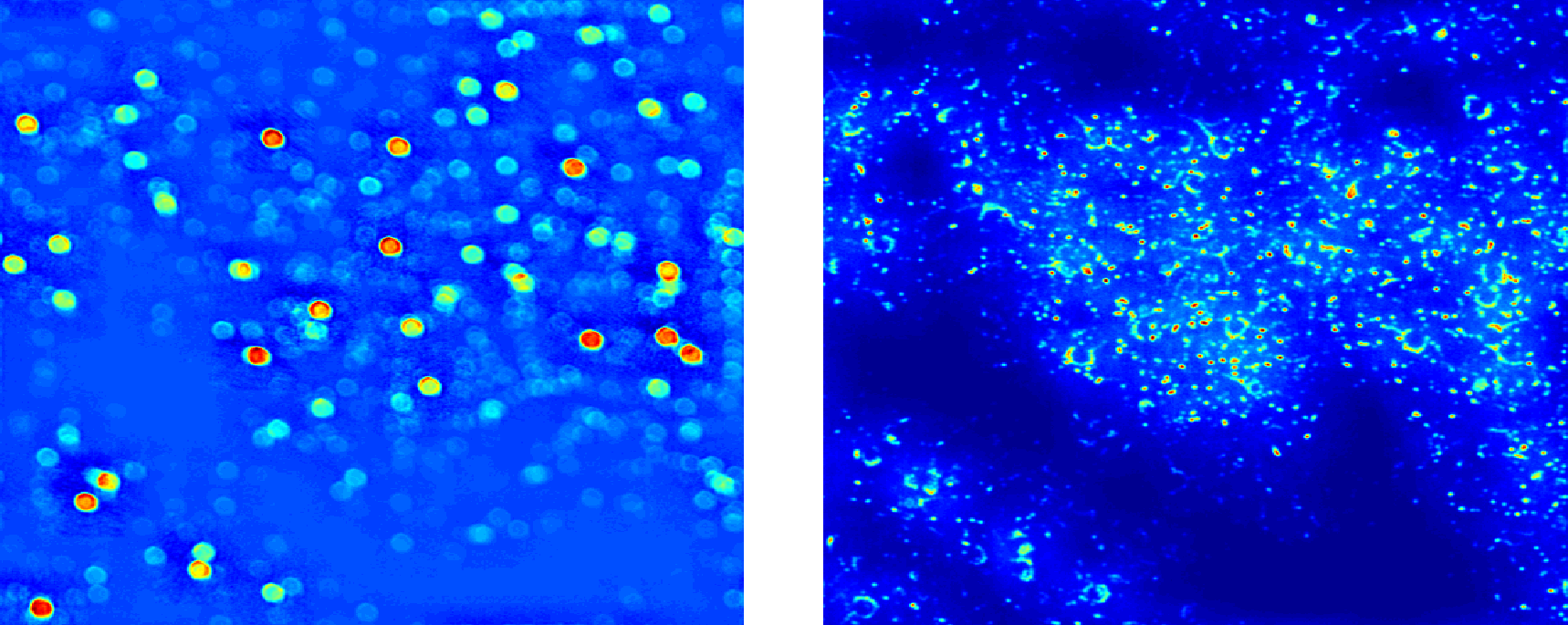}
\end{minipage}
\caption{\textbf{Localization and classification for calcium microscopy images.} (a) shows the original image; (b) shows the estimated kernel shape for the neuron (left) and dendrite (right); (c) presents the predicted activation map for the neuron (left) and dendrite (right); (d) presents the reconstructed image $\mb Y_k = \mb A_k \cconv \mb X_k\;(k=1,2)$ for the neuron (left) and dendrite (right).}
\label{fig:calcium_2D}
\end{figure*}

\paragraph{Neuronal localization for 2D calcium imaging.}

Tracking the spike locations of neurons in 2D calcium imaging video sequences is a challenging task due to the presence of (non)rigid motion, overlapping sources, and irregular background noise \cite{pnevmatikakis2016simultaneous,giovannucci2017onacid,giovannucci2019caiman}. Here we show how the SaS-CDL problem can serve as a basis for distinguishing between overlapping sources.
\Cref{fig:calcium_2D_img} shows a single $512\times 512$ frame from the two-photon fluorescence calcium microscopy dataset obtained by Allen Institute for Brain Science\footnote{The data can be found at \url{http://observatory.brain-map.org/visualcoding/search/overview}.}. The frame  shows the cross sections of two types of neuronal components, the somata and the denrdrites, whose fluorophores that are activated at the given time frame.
It is clear that these two components are primarily distinguished by their size. We decompose the frame into the somatic and dendritic components by solving SaS-CDL with the proposed method, giving us a rough estimate of the ``average'' somatic or dendritic motif (\Cref{fig:calcium_2D_kernel}), as well as the location of each component (\Cref{fig:calcium_2D_map}). This allows the image to be decomposed into images consisting of somata or dendrites exclusively (\Cref{fig:calcium_2D_class}).
Therefore SaS-CDL can either be the basis for a preprocessing step to remove undesired components, such as the dendrites, from a microscopy image. Furthermore, this deconvolution technique allows the individual activation map to be tracked for each video frame, opening a new way for nonrigid motion to be corrected across frames by synthesizing all activation maps. We left this as a promising future research direction.

\begin{figure*}[!htbp]
\centering
\captionsetup[sub]{font=small,labelfont={bf,sf}}
\begin{minipage}[c]{0.45\textwidth}
\subcaption{\normalsize STM image $\mb Y$}
\label{fig:em_2D_img}
	\includegraphics[width = \linewidth]{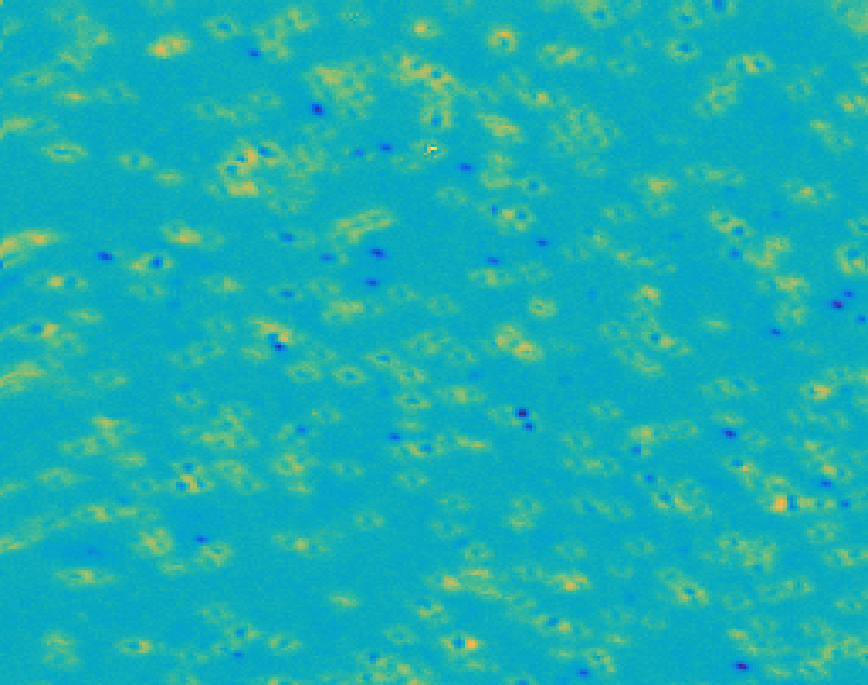}
\end{minipage}
\hspace{0.2in}
\begin{minipage}[c]{0.47\textwidth}
\subcaption{estimated kernel $\mb A_k\;(k=1,2)$}\label{fig:em_2D_kernel}
	\includegraphics[width = \linewidth]{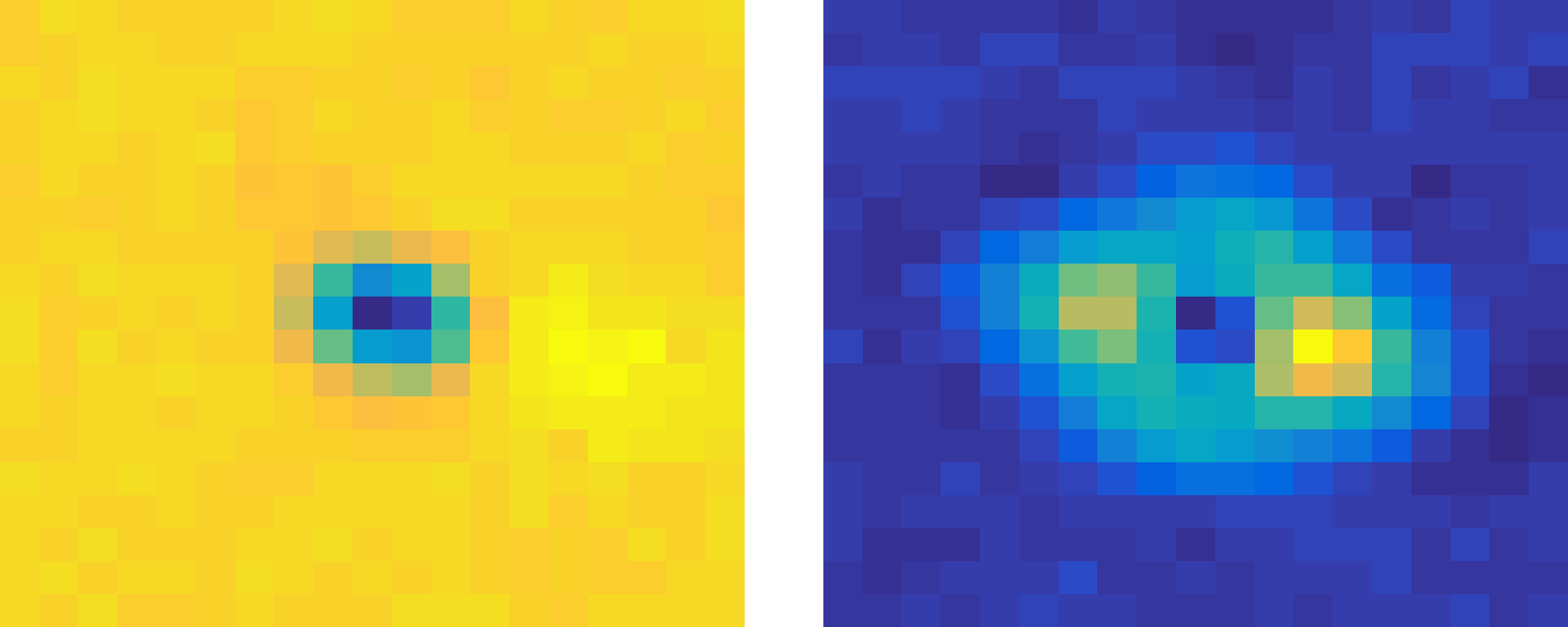}
	\subcaption{predicted activation map $\mb X_k\;(k=1,2)$}\label{fig:em_2D_map}
	\includegraphics[width = \linewidth]{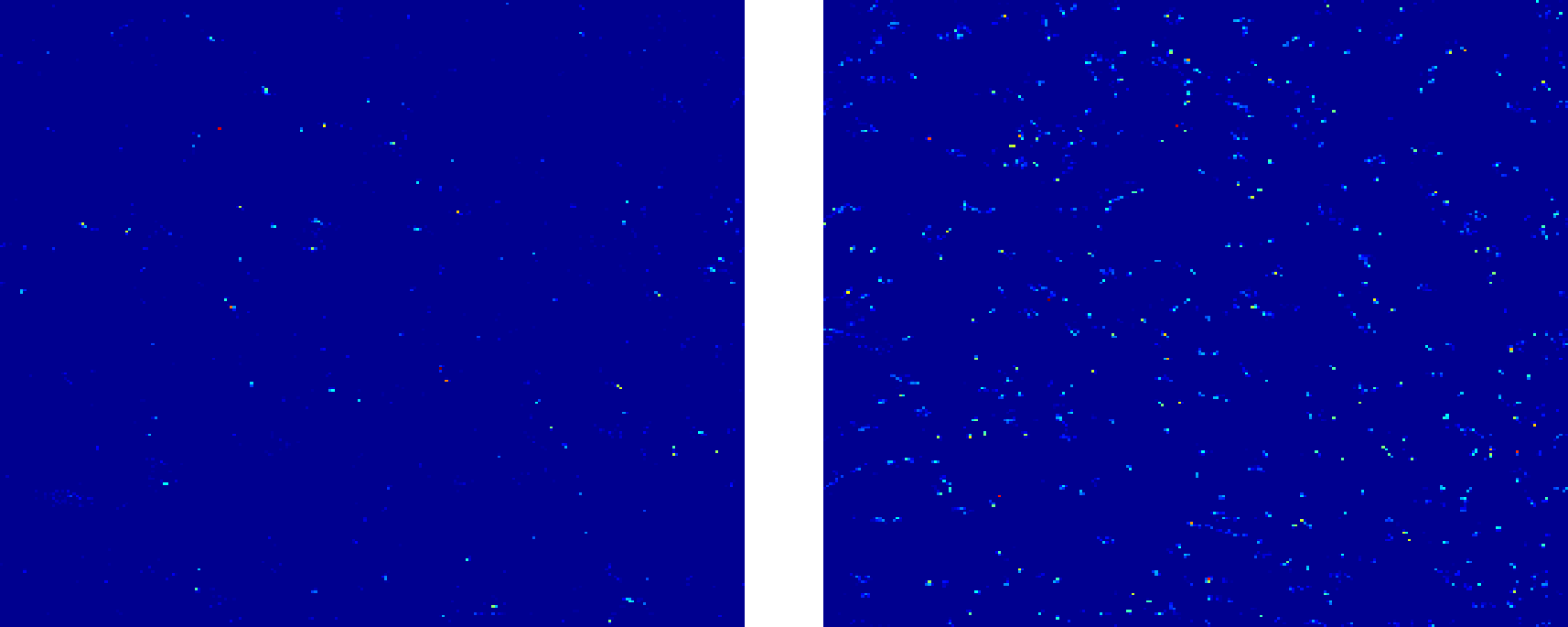}
	\subcaption{classified image $\mb Y_k= \mb A_k \cconv \mb X_k\;(k=1,2)$}\label{fig:em_2D_class}
	\includegraphics[width = \linewidth]{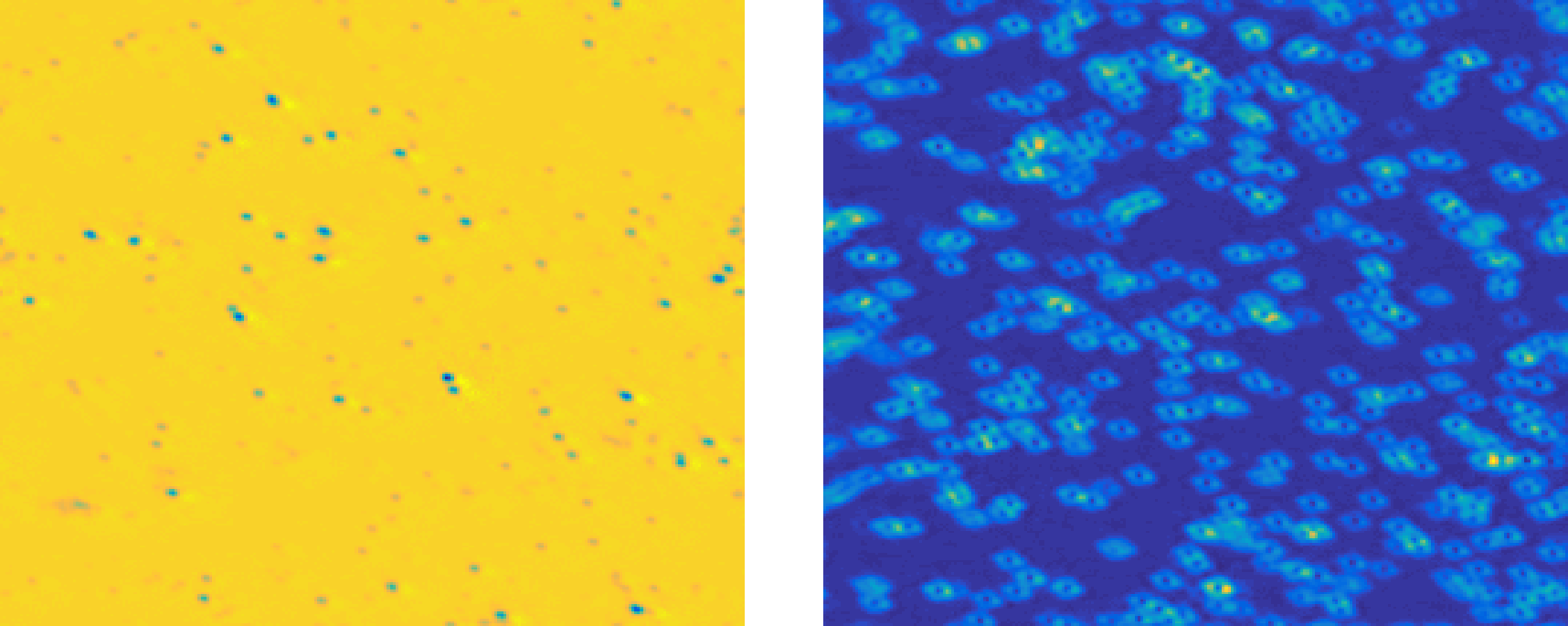}
\end{minipage}
\caption{\textbf{Defect detection for STM images.} (a) shows the original STM image; (b) shows the estimated kernel shape for the defects; (c) presents the predicted activation map for the defects; (d) presents the reconstructed image $\mb Y_k = \mb A_k \cconv \mb X_k\;(k=1,2)$ for the defect.}
\label{fig:STM-defect}
\end{figure*}

\paragraph{Defect detection in scan tunneling microscopy (STM) image.} Modern high-resolution microscopes, such as the scanning tunneling electron microscope, are commonly used to study specimens that have dense and aperiodic spatial structure \cite{crommie1993imaging,rutter2007scattering,roushan2009topological}. Extracting meaningful information from images obtained from such microscopes remains a formidable challenge \cite{kivelson2003detect}. For instance, \Cref{fig:em_2D_img} presents a STM NaFeAs sample image (with size $128\times 128$) of a Co-doped iron arsenide crystal lattice. A method for automatically acquiring the signatures of the defects (motifs) and their locations is highly desirable \cite{Cheung2018DictionaryLI}. Here we apply our proposed method to solve SaS-CDL and extract both the defect signatures (\Cref{fig:em_2D_kernel}) and their locations (see \Cref{fig:em_2D_map}), as well as decomposing the image into contributions based on the individual defects (\Cref{fig:em_2D_class}).

\section{Conclusion and Discussion}
\label{sec:conclusion}
% !TEX root = ../main.tex

\subsection{Relationship to the literature and conclusion}

\paragraph{Nonconvex optimization.} Unlike convex optimization problems, nonconvex functions usually have numerous spurious local minima, and one may also encounter ``flat'' saddle points that are very difficult to escape \cite{sun2015nonconvex}. In theory, even finding a local minimum of a general nonconvex function is NP-hard \cite{murty1987some} -- never mind the global minimum. However, recent advancements in nonconvex optimization \cite{sun2015nonconvex,ge2015escaping} showed that typical nonconvex problems in practice are often more structured, so that they often have much more benign geometric landscapes than the worst case: (i) all saddle points can be efficiently escaped by using negative curvature information; (ii) the equivalent ``good'' solutions (created by the intrinsic symmetry) are often the global optimizers of the nonconvex objective. This type of benign geometric structure has been discovered for many nonconvex problems in signal processing and machine learning, such as phase retrieval \cite{candes2015phase,sun2016geometric,qu2017convolutional}, dictionary learning \cite{qu2014finding,sun2016complete-i,sun2016complete-ii}, low rank matrix recovery \cite{ge2016matrix,chi2016guaranteed} (orthogonal) tensor decomposition \cite{ge2015escaping}, and phase synchronization problems \cite{boumal2016global}, etc. Inspired by similar benign geometric structure for a simplified nonconvex Dropped Quadratic formulation, this work provides an efficient and practical nonconvex optimization method for solving blind sparse deconvolution problems.

\paragraph{Blind deconvolution.} The blind deconvolution problem is an ill-posed problem in its most general form. Nonetheless, problems in practice often exhibits intrinsic low-dimensional structures, showing promises for efficient optimization. Motivated by a variety of applications, many low-dimensional models for (blind) deconvolution problems have been studied in the literature. \cite{ahmed2014blind,chi2016guaranteed,ling2015self,li2016identifiability,kech2017optimal,ahmed2018leveraging,li2018bilinear} studied the problem when the unknown signals $\mb a_0$ and $\mb x_0$ either live in known low-dimensional subspaces, or are sparse in some known dictionary. These results assumed that the subspace/dictionary are chosen at random, such that the problem does not exhibit the signed shift ambiguity and can be provably solved via convex relaxation\footnote{Some recent work \cite{li2018rapid,ma2017implicit} show this problem can also be provably solved via nonconvex approaches.}. However, the assumption of random subspace/dictionary model is often unrealistic in practice. Recently, \cite{wang2016blind,li2018global,qu2019blind} consider sparse blind deconvolution with multiple measurements, where they show the problem can be efficiently solved to global optimality when the kernel is invertible. In contrast, the SaS model studied in this work exhibits much broader applications.

Because of the shift symmetry, the SaS model does not appear to be amenable for convexification, and it exhibits a more complicated nonconvex geometry. To tackle this problem, Wipf et al. \cite{wipf2014revisiting} imposes $\ell_2$ regularization on $\mb a_0$ and provides an empirically reliable algorithm. Zhang et al. \cite{zhang2017global} studies the geometry of a simplified nonconvex objective over the sphere, and proves that in the dilute limit in which $\mb x_0$ is a single spike, all strict local minima are close to signed shift truncations of $\mb a_0$. Zhang et al. \cite{zhang2018structured} formulated the problem as an $\ell_4$ maximization problem over the sphere\footnote{A similar objective is considered for the multichannel sparse blind deconvolution problem \cite{li2018global}.}. They proved that on a restricted region of the sphere every local minimizer is near a truncated signed shift of $\mb a_0$, when $\mb a_0$ is well-conditioned and $\mb x_0$ is sparse. Kuo et al. \cite{kuo2019geometry} studies a Dropped Quadratic simplification of the Bilinear Lasso objective, which provably obtains exact recovery for an incoherent kernel $\mb a_0$ and sparse $\mb x_0$. However, both the $\ell_4$ maximization and Dropped Quadratic objectives are still quite far from practical formulations for solving SaSD. In contrast, as demonstrated in this work, optimizing the Bilinear Lasso formulation turns out to be much more effective in practice.

\paragraph{Geometry inspired optimization method for SaSD.} Inspired by the benign geometric structure of the nonconvex objective, we proposed efficient nonconvex optimization methods that directly optimizes the Bilinear Lasso. The new approach exploits the geometry by (i) using data driven initializations to avoid spurious local minimizers, (ii) adopting momentum accelerating for coherent kernels, and (iii) adaptively shrinking the penalty parameter $\lambda$ to achieve faster convergence and higher accuracy solutions. Our vanilla algorithm is a simple alternating descent method, which is inspired by the recent PALM methods \cite{bolte2014proximal,pock2016inertial}. In comparison with classical alternating minimization methods for sparse blind deconvolution \cite{chan2000convergence,sroubek2012robust,zhang2017global}, our approach does not require solving expensive Lasso subproblems, and the iterates make fast progress towards the optimal solution. On the other hand, as our method is first-order in nature, it is much more efficient than the second-order trust-region \cite{conn2000trust,boumal2016global} and curvilinear search \cite{goldfarb1980curvilinear} methods considered in \cite{zhang2017global,kuo2019geometry}.

\paragraph{Convolutional dictionary learning.} Furthermore, our approach has natural extensions for tackling the SaS-CDL problem when multiple unknown kernels present. By consider a similar nonconvex objective analogous to SaSD, our geometric inspired algorithm empirically solves the SaS-CDL problem to global optimality in a very efficient manner. The new method joins recent algorithmic development for solving CDL \cite{chalasani2013fast,huang2015convolutional,papyan2017convolutional,garcia2018convolutional,liu2018first,maggu2018convolutional,zisselman2018local}. Again, most\footnote{The recent work \cite{maggu2018convolutional} resembles some similarities to ours. However, the problem setting and formulation are still quite different.} of the previous approaches \cite{garcia2018convolutional} deploy an alternating minimization strategy, which exactly solves the expensive Lasso subproblem for each iteration. In contrast, our method is much more simple, efficient and effective, demonstrated by experiments on real datasets.

\subsection{Discussion and future work}\label{subsec:future}

Moving forward, we believe this work has opened up several future directions that could be of great empirical and theoretical interests.

\paragraph{Geometric analysis of Bilinear Lasso.}
The Bilinear Lasso formulation is one of most natural formulations for solving the SaSD problem. In light of our empirical success of solving the Bilinear Lasso, analyzing and understanding its global nonconvex landscapes is of great importance. As discuss in \Cref{sec:geometry}, the Dropped Quadratic formulation studied in \cite{kuo2019geometry} has commonalities with the Bilinear Lasso: both exhibit local minima at signed shifts, and both exhibit negative curvature in symmetry breaking directions. However, a major difference (and hence, major challenge) is that gradient methods for Bilinear Lasso do not retract to a subspaces  -- they retract to a more complicated, nonlinear set. As the empirical success we possessed here, better understandings of the geometric structure for the Bilinear Lasso in much needed. A better understanding will also shed light on SaS-CDL with multiple unknown kernels.

\paragraph{Parameterized sparse blind deconvolution.}

\begin{figure*}[!htbp]
\centering
\includegraphics[width = 1\linewidth]{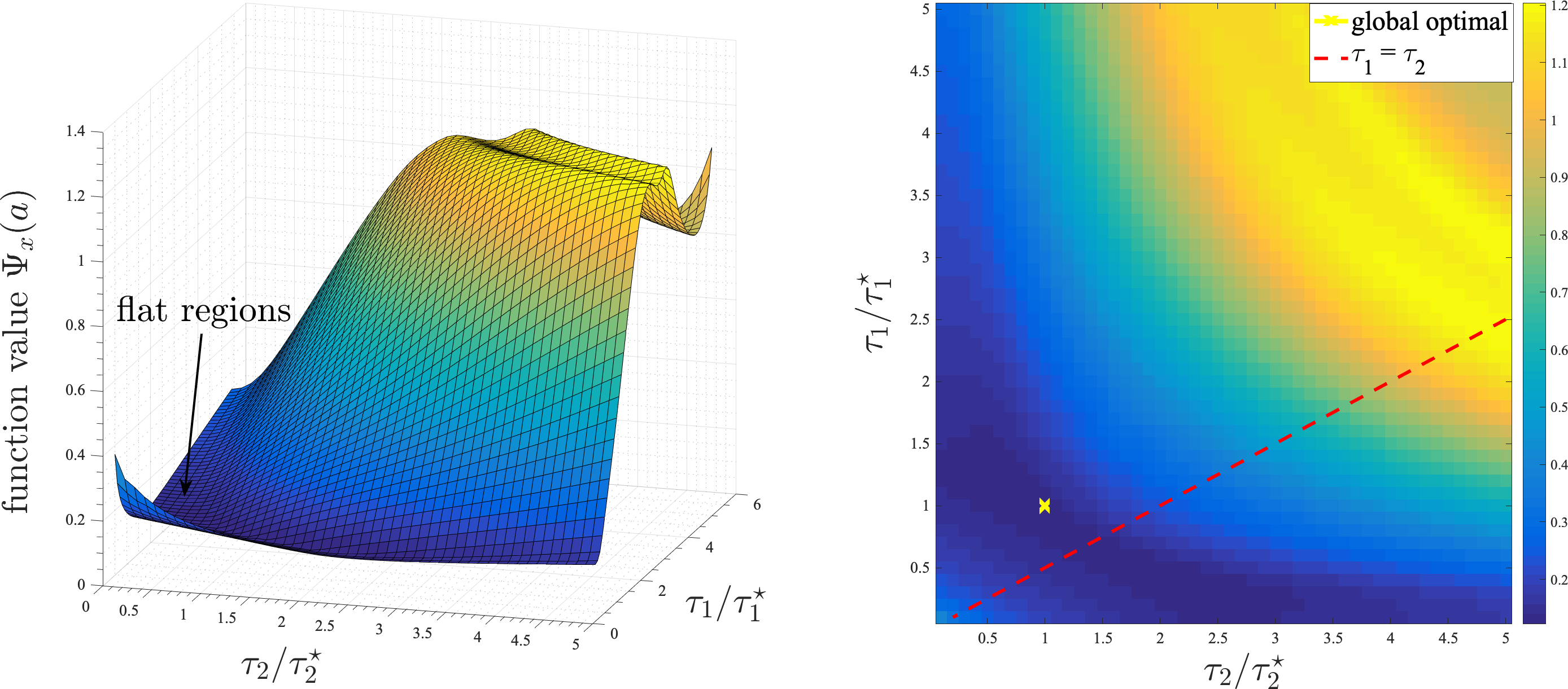}
\caption{ \textbf{Nonconvex landscape of parameterized SaSD, with AR(2) kernel and two unknown parameters.} The kernel $a_0(t) = \exp\paren{ -t/\tau_1^\star } - \exp\paren{ -t/\tau_2^\star } $ is parameterized by two parameters $\tau_1^\star = 0.2$ and $\tau_2^\star=0.1$. We generate the data $\mb y = \mb a_0(t) \conv \mb x_0$, where $\mb x_0 \sim_{i.i.d.}\mc B(\theta)$ with $\theta = 10^{-2}$. We plot the marginalized function landscape of $\Psi_x(a) = \min_{\mb x} \frac{1}{2} \norm{ \mb y - \mb a(\tau) \conv \mb x }{2}^2 + \lambda \norm{\mb x}{1} $ w.r.t. $\tau_1$ and $\tau_2$, where $\lambda = 10^{-3}$, $\di = 150$ and $\sample = 10^4$. The figures on the left and right hand sides are 3D and 2D plots of the function landscape, respectively. As we can see, the ground truth $(\tau_1^\star, \tau_2^\star)$ is the global minimizer to the nonconvex objective, but the landscape near region of the ground truth is very flat and therefore very difficult to make progress on minimizing the nonconvex objective.}
\label{fig:geometry-para-2d}
\end{figure*}

In this work, we studied the blind deconvolution problem with no prior knowledge of the kernel/motif $\mb a_0$. However, in many application, one can often obtain some side information, where the kernel is often determined by only a few parameters associated with the underlying physical processes. For example, in the calcium imaging problem we studied in \Cref{sec:exp}, an auto regression (AR) model is often used to characterize the spiking and decaying process of the kernel, which is only determined by one or two parameters \cite{vogelstein2010fast,friedrich2017fast}. Thus, how to estimate these kernel parameters raises a challenging but interesting question. Our preliminary investigation shows that nonconvex optimization landscapes of this parameterized ``semi-blind'' sparse deconvolution problem also possess benign geometric properties for certain types of kernels (see \Cref{fig:geometry-para-2d} for an illustration).

\paragraph{SaSD meets super-resolution.}
In many imaging applications, it is often desirable to solve blind deconvolution and super-resolution problems simultaneously. In other words, let $\mc D\brac{\cdot}$ be a downsampling operator, we want to recover the high-resolution kernel $\mb a_0$ and sparse activation map $\mb x_0$ from the low-resolution measurement of the form $\mb y = \mc D\brac{\mb a_0 \conv \mb x_0}$. This type of problem appears often due to the resolution/hardware limit of the imaging system, and therefore fine details of both $\mb a_0$ and $\mb x_0$ are missing due to downsampling. For instance, in \Cref{sec:exp} we show that the spatial resolution of fluorescent microscopy is constraint by the diffraction limit of the light \cite{huang2009super}. If we can solve this super-resolution SaSD problem, we can obtain much higher resolution image of living cells \emph{in vivo}. However, our preliminary investigations show that optimizing the natural nonconvex formulation
\begin{align*}
\min_{\mb a,\mb x}\quad\frac{1}{2}\norm{\mb y - \mc D \brac{\mb a\conv\mb x}}{2}^2+\lambda\norm{\mb x}{1},\quad \text{s.t.}\quad \mb a \;\in\; \bb S^{\Di-1}
\end{align*}
tends to produce downsampled $\mb a_0$ and $\mb x_0$. How to solve this problem is largely open and remains a very interesting question. One possibility is to enforce extra constraints on $\mb a_0$, such as penalizing $TV$-norm to promote smoothness.

\paragraph{Dealing with structured data.} Data in practice often possesses much richer structure than the basic SaS model we studied here. For instance, in calcium imaging, the signal we obtained often has drifting/motion issues across time frames, and it also exhibits low-rank background DC components \cite{pnevmatikakis2016simultaneous,giovannucci2017onacid}. In STORM optical microscopy, there are rich spatial and temporal correlations within and between video frames \cite{solomon2018sparsity}. Moreover, in many microscopy imaging data analysis problems, the motif we want to locate often exhibits unknown deformations and random rotations, and its shape is often asymmetric. How to deal with these extra structures raises a variety of challenging problems for future research.

\section*{Acknowledgement}
This work was funded by NSF 1343282, NSF CCF 1527809, and NSF IIS 1546411. We would like to thank Gongguo Tang, Shuyang Ling, Carlos Fernandez-Granda, Ruoxi Sun, Liam Paniski for fruitful discussions. QQ also acknowledges supports from Microsoft PhD fellowship and the Moore-Sloan Foundation.

\newpage

{\small
\bibliographystyle{alpha}
\bibliography{deconv,ncvx,cdl}

\newcommand{\etalchar}[1]{$^{#1}$}
\begin{thebibliography}{LGCWY18}

\bibitem[ABG07]{absil2007trust}
Pierre-Antoine. Absil, Christopher~G. Baker, and Kyle~A. Gallivan.
\newblock Trust-region methods on {R}iemannian manifolds.
\newblock {\em Foundations of Computational Mathematics}, 7(3):303--330, 2007.

\bibitem[AD18]{ahmed2018leveraging}
Ali Ahmed and Laurent Demanet.
\newblock Leveraging diversity and sparsity in blind deconvolution.
\newblock {\em IEEE Transactions on Information Theory}, 64(6):3975--4000,
  2018.

\bibitem[AMS09]{absil2009}
Pierre-Antoine. Absil, Robert Mahoney, and Rodolphe Sepulchre.
\newblock {\em Optimization Algorithms on Matrix Manifolds}.
\newblock Princeton University Press, 2009.

\bibitem[ANW10]{agarwal2010fast}
Alekh Agarwal, Sahand Negahban, and Martin~J Wainwright.
\newblock Fast global convergence rates of gradient methods for
  high-dimensional statistical recovery.
\newblock In {\em Advances in Neural Information Processing Systems}, pages
  37--45, 2010.

\bibitem[ARR14]{ahmed2014blind}
Ali Ahmed, Benjamin Recht, and Justin Romberg.
\newblock Blind deconvolution using convex programming.
\newblock {\em IEEE Transactions on Information Theory}, 60(3):1711--1732,
  2014.

\bibitem[B{\etalchar{+}}15]{bubeck2015convex}
S{\'e}bastien Bubeck et~al.
\newblock Convex optimization: Algorithms and complexity.
\newblock {\em Foundations and Trends{\textregistered} in Machine Learning},
  8(3-4):231--357, 2015.

\bibitem[BAC18]{boumal2016global}
Nicolas Boumal, Pierre-Antoine Absil, and Coralia Cartis.
\newblock Global rates of convergence for nonconvex optimization on manifolds.
\newblock {\em IMA Journal of Numerical Analysis}, 39(1):1--33, 2018.

\bibitem[BBM{\etalchar{+}}16]{boettiger2016super}
Alistair~N Boettiger, Bogdan Bintu, Jeffrey~R Moffitt, Siyuan Wang, Brian~J
  Beliveau, Geoffrey Fudenberg, Maxim Imakaev, Leonid~A Mirny, Chao-ting Wu,
  and Xiaowei Zhuang.
\newblock Super-resolution imaging reveals distinct chromatin folding for
  different epigenetic states.
\newblock {\em Nature}, 529(7586):418, 2016.

\bibitem[BDH{\etalchar{+}}13]{briers2013laser}
David Briers, Donald~D Duncan, Evan~R Hirst, Sean~J Kirkpatrick, Marcus
  Larsson, Wiendelt Steenbergen, Tomas Stromberg, and Oliver~B Thompson.
\newblock Laser speckle contrast imaging: theoretical and practical
  limitations.
\newblock {\em Journal of biomedical optics}, 18(6):066018, 2013.

\bibitem[Bec17]{beck2017first}
Amir Beck.
\newblock {\em First-order methods in optimization}, volume~25.
\newblock SIAM, 2017.

\bibitem[BK02]{baker2002limits}
Simon Baker and Takeo Kanade.
\newblock Limits on super-resolution and how to break them.
\newblock {\em IEEE Transactions on Pattern Analysis and Machine Intelligence},
  24(9):1167--1183, 2002.

\bibitem[BPC{\etalchar{+}}11]{boyd2011distributed}
Stephen Boyd, Neal Parikh, Eric Chu, Borja Peleato, Jonathan Eckstein, et~al.
\newblock Distributed optimization and statistical learning via the alternating
  direction method of multipliers.
\newblock {\em Foundations and Trends{\textregistered} in Machine learning},
  3(1):1--122, 2011.

\bibitem[BPS{\etalchar{+}}06]{betzig2006imaging}
Eric Betzig, George~H Patterson, Rachid Sougrat, O~Wolf Lindwasser, Scott
  Olenych, Juan~S Bonifacino, Michael~W Davidson, Jennifer Lippincott-Schwartz,
  and Harald~F Hess.
\newblock Imaging intracellular fluorescent proteins at nanometer resolution.
\newblock {\em Science}, 313(5793):1642--1645, 2006.

\bibitem[BST14]{bolte2014proximal}
J{\'e}r{\^o}me Bolte, Shoham Sabach, and Marc Teboulle.
\newblock Proximal alternating linearized minimization for nonconvex and
  nonsmooth problems.
\newblock {\em Mathematical Programming}, 146(1-2):459--494, 2014.

\bibitem[BT09]{beck2009fast}
Amir Beck and Marc Teboulle.
\newblock A fast iterative shrinkage-thresholding algorithm for linear inverse
  problems.
\newblock {\em SIAM Journal on Imaging Sciences}, 2(1):183--202, 2009.

\bibitem[BVG13]{benichoux2013fundamental}
Alexis Benichoux, Emmanuel Vincent, and R{\'e}mi Gribonval.
\newblock A fundamental pitfall in blind deconvolution with sparse and
  shift-invariant priors.
\newblock In {\em ICASSP-38th International Conference on Acoustics, Speech,
  and Signal Processing-2013}, 2013.

\bibitem[CE16]{campisi2016blind}
Patrizio Campisi and Karen Egiazarian.
\newblock {\em Blind image deconvolution: theory and applications}.
\newblock CRC press, 2016.

\bibitem[CF17]{chun2017convolutional}
Il~Yong Chun and Jeffrey~A Fessler.
\newblock Convolutional dictionary learning: Acceleration and convergence.
\newblock {\em IEEE Transactions on Image Processing}, 27(4):1697--1712, 2017.

\bibitem[CFG14]{candes2014towards}
Emmanuel~J Cand{\`e}s and Carlos Fernandez-Granda.
\newblock Towards a mathematical theory of super-resolution.
\newblock {\em Communications on pure and applied Mathematics}, 67(6):906--956,
  2014.

\bibitem[CGT00]{conn2000trust}
Andrew~R. Conn, Nicholas~I.M. Gould, and Philippe~L. Toint.
\newblock {\em Trust region methods}, volume~1.
\newblock SIAM, 2000.

\bibitem[Chi16]{chi2016guaranteed}
Yuejie Chi.
\newblock Guaranteed blind sparse spikes deconvolution via lifting and convex
  optimization.
\newblock {\em IEEE Journal of Selected Topics in Signal Processing},
  10(4):782--794, 2016.

\bibitem[CLE93]{crommie1993imaging}
MF~Crommie, CP~Lutz, and DM~Eigler.
\newblock Imaging standing waves in a two-dimensional electron gas.
\newblock {\em Nature}, 363(6429):524, 1993.

\bibitem[CLS15]{candes2015phase}
Emmanuel~J. Cand{\`e}s, Xiaodong Li, and Mahdi Soltanolkotabi.
\newblock Phase retrieval via wirtinger flow: Theory and algorithms.
\newblock {\em IEEE Transactions on Information Theory}, 61(4):1985--2007,
  April 2015.

\bibitem[CMB{\etalchar{+}}17]{chung2017fully}
Jason~E Chung, Jeremy~F Magland, Alex~H Barnett, Vanessa~M Tolosa, Angela~C
  Tooker, Kye~Y Lee, Kedar~G Shah, Sarah~H Felix, Loren~M Frank, and Leslie~F
  Greengard.
\newblock A fully automated approach to spike sorting.
\newblock {\em Neuron}, 95(6):1381--1394, 2017.

\bibitem[CPR13]{chalasani2013fast}
Rakesh Chalasani, Jose~C Principe, and Naveen Ramakrishnan.
\newblock A fast proximal method for convolutional sparse coding.
\newblock In {\em The 2013 International Joint Conference on Neural Networks
  (IJCNN)}, pages 1--5. IEEE, 2013.

\bibitem[CRQ18]{chaure2018novel}
Fernando Chaure, Hernan~Gonzalo Rey, and Rodrigo~Quian Quiroga.
\newblock A novel and fully automatic spike sorting implementation with
  variable number of features.
\newblock {\em Journal of Neurophysiology}, 2018.

\bibitem[CSL{\etalchar{+}}18]{Cheung2018DictionaryLI}
Sky~C Cheung, John~Y Shin, Yenson Lau, Zhengyu Chen, Ju~Sun, Yuqian Zhang,
  John~N Wright, and Abhay~N Pasupathy.
\newblock Dictionary learning in fourier transform scanning tunneling
  spectroscopy.
\newblock {\em arXiv preprint arXiv:1807.10752}, 2018.

\bibitem[CW98]{chan1998total}
Tony~F Chan and Chiu-Kwong Wong.
\newblock Total variation blind deconvolution.
\newblock {\em IEEE transactions on Image Processing}, 7(3):370--375, 1998.

\bibitem[CW00]{chan2000convergence}
Tony~F Chan and Chiu-Kwong Wong.
\newblock Convergence of the alternating minimization algorithm for blind
  deconvolution.
\newblock {\em Linear Algebra and its Applications}, 316(1-3):259--285, 2000.

\bibitem[CWB08]{candes2008enhancing}
Emmanuel~J Candes, Michael~B Wakin, and Stephen~P Boyd.
\newblock Enhancing sparsity by reweighted $\ell_1$ minimization.
\newblock {\em Journal of Fourier analysis and applications}, 14(5-6):877--905,
  2008.

\bibitem[ETS11]{ekanadham2011blind}
Chaitanya Ekanadham, Daniel Tranchina, and Eero~P Simoncelli.
\newblock A blind sparse deconvolution method for neural spike identification.
\newblock In {\em Advances in Neural Information Processing Systems}, pages
  1440--1448, 2011.

\bibitem[FST08]{fernandez2008fluorescent}
Marta Fern{\'a}ndez-Su{\'a}rez and Alice~Y Ting.
\newblock Fluorescent probes for super-resolution imaging in living cells.
\newblock {\em Nature Reviews Molecular cell Biology}, 9(12):929, 2008.

\bibitem[FZP17]{friedrich2017fast}
Johannes Friedrich, Pengcheng Zhou, and Liam Paninski.
\newblock Fast online deconvolution of calcium imaging data.
\newblock {\em PLoS Computational Biology}, 13(3):e1005423, 2017.

\bibitem[GBW18]{gilboa2018efficient}
Dar Gilboa, Sam Buchanan, and John Wright.
\newblock Efficient dictionary learning with gradient descent.
\newblock {\em arXiv preprint arXiv:1809.10313}, 2018.

\bibitem[GCW18]{garcia2018convolutional}
Cristina Garcia-Cardona and Brendt Wohlberg.
\newblock Convolutional dictionary learning: A comparative review and new
  algorithms.
\newblock {\em IEEE Transactions on Computational Imaging}, 4(3):366--381,
  2018.

\bibitem[GFG{\etalchar{+}}19]{giovannucci2019caiman}
Andrea Giovannucci, Johannes Friedrich, Pat Gunn, Jeremie Kalfon, Brandon~L
  Brown, Sue~Ann Koay, Jiannis Taxidis, Farzaneh Najafi, Jeffrey~L Gauthier,
  Pengcheng Zhou, et~al.
\newblock Caiman an open source tool for scalable calcium imaging data
  analysis.
\newblock {\em Elife}, 8:e38173, 2019.

\bibitem[GFK{\etalchar{+}}17]{giovannucci2017onacid}
Andrea Giovannucci, Johannes Friedrich, Matt Kaufman, Anne Churchland, Dmitri
  Chklovskii, Liam Paninski, and Eftychios~A Pnevmatikakis.
\newblock Onacid: Online analysis of calcium imaging data in real time.
\newblock In {\em Advances in Neural Information Processing Systems}, pages
  2381--2391, 2017.

\bibitem[GHJY15]{ge2015escaping}
Rong Ge, Furong Huang, Chi Jin, and Yang Yuan.
\newblock Escaping from saddle points---online stochastic gradient for tensor
  decomposition.
\newblock In {\em Proceedings of The 28th Conference on Learning Theory}, pages
  797--842, 2015.

\bibitem[GK12]{grienberger2012imaging}
Christine Grienberger and Arthur Konnerth.
\newblock Imaging calcium in neurons.
\newblock {\em Neuron}, 73(5):862--885, 2012.

\bibitem[GLM16]{ge2016matrix}
Rong Ge, Jason~D Lee, and Tengyu Ma.
\newblock Matrix completion has no spurious local minimum.
\newblock In {\em Advances in Neural Information Processing Systems}, pages
  2973--2981, 2016.

\bibitem[GMWZ17]{goldfarb2017using}
Donald Goldfarb, Cun Mu, John Wright, and Chaoxu Zhou.
\newblock Using negative curvature in solving nonlinear programs.
\newblock {\em Computational Optimization and Applications}, 68(3):479--502,
  2017.

\bibitem[Gol80]{goldfarb1980curvilinear}
Donald Goldfarb.
\newblock Curvilinear path steplength algorithms for minimization which use
  directions of negative curvature.
\newblock {\em Mathematical Programming}, 18(1):31--40, 1980.

\bibitem[HA15]{huang2015convolutional}
Furong Huang and Animashree Anandkumar.
\newblock Convolutional dictionary learning through tensor factorization.
\newblock In {\em Feature Extraction: Modern Questions and Challenges}, pages
  116--129, 2015.

\bibitem[Hay94]{haykin1994blind}
Simon~S Haykin.
\newblock {\em Blind deconvolution}.
\newblock Prentice Hall, 1994.

\bibitem[HBZ09]{huang2009super}
Bo~Huang, Mark Bates, and Xiaowei Zhuang.
\newblock Super-resolution fluorescence microscopy.
\newblock {\em Annual Review of Biochemistry}, 78:993--1016, 2009.

\bibitem[HBZ10]{huang2010breaking}
Bo~Huang, Hazen Babcock, and Xiaowei Zhuang.
\newblock Breaking the diffraction barrier: super-resolution imaging of cells.
\newblock {\em Cell}, 143(7):1047--1058, 2010.

\bibitem[Hel07]{hell2007far}
Stefan~W Hell.
\newblock Far-field optical nanoscopy.
\newblock {\em science}, 316(5828):1153--1158, 2007.

\bibitem[HGM06]{hess2006ultra}
Samuel~T Hess, Thanu~PK Girirajan, and Michael~D Mason.
\newblock Ultra-high resolution imaging by fluorescence photoactivation
  localization microscopy.
\newblock {\em Biophysical journal}, 91(11):4258--4272, 2006.

\bibitem[HUK11]{holden2011daostorm}
Seamus~J Holden, Stephan Uphoff, and Achillefs~N Kapanidis.
\newblock Daostorm: an algorithm for high-density super-resolution microscopy.
\newblock {\em Nature Methods}, 8(4):279, 2011.

\bibitem[HWBZ08]{huang2008three}
Bo~Huang, Wenqin Wang, Mark Bates, and Xiaowei Zhuang.
\newblock Three-dimensional super-resolution imaging by stochastic optical
  reconstruction microscopy.
\newblock {\em Science}, 319(5864):810--813, 2008.

\bibitem[HYZ08]{hale2008fixed}
Elaine~T Hale, Wotao Yin, and Yin Zhang.
\newblock Fixed-point continuation for $\backslash$ell\_1-minimization:
  Methodology and convergence.
\newblock {\em SIAM Journal on Optimization}, 19(3):1107--1130, 2008.

\bibitem[JGN{\etalchar{+}}17]{jin2017escape}
Chi Jin, Rong Ge, Praneeth Netrapalli, Sham~M Kakade, and Michael~I Jordan.
\newblock How to escape saddle points efficiently.
\newblock In {\em Proceedings of the 34th International Conference on Machine
  Learning}, pages 1724--1732, 2017.

\bibitem[JNJ18]{jin2017accelerated}
Chi Jin, Praneeth Netrapalli, and Michael~I Jordan.
\newblock Accelerated gradient descent escapes saddle points faster than
  gradient descent.
\newblock In {\em Conference On Learning Theory}, pages 1042--1085, 2018.

\bibitem[KBF{\etalchar{+}}03]{kivelson2003detect}
Steven~A Kivelson, Ian~P Bindloss, Eduardo Fradkin, Vadim Oganesyan,
  JM~Tranquada, Aharon Kapitulnik, and Craig Howald.
\newblock How to detect fluctuating stripes in the high-temperature
  superconductors.
\newblock {\em Reviews of Modern Physics}, 75(4):1201, 2003.

\bibitem[KH96]{kundur1996blind}
Deepa Kundur and Dimitrios Hatzinakos.
\newblock Blind image deconvolution.
\newblock {\em IEEE Signal Processing Magazine}, 13(3):43--64, 1996.

\bibitem[KK17]{kech2017optimal}
Michael Kech and Felix Krahmer.
\newblock Optimal injectivity conditions for bilinear inverse problems with
  applications to identifiability of deconvolution problems.
\newblock {\em SIAM Journal on Applied Algebra and Geometry}, 1(1):20--37,
  2017.

\bibitem[KZLW19]{kuo2019geometry}
Han-Wen Kuo, Yuqian Zhang, Yenson Lau, and John Wright.
\newblock Geometry and symmetry in short-and-sparse deconvolution.
\newblock In {\em International Conference on Machine Learning (ICML)}, June
  2019.

\bibitem[LB95]{loke1995least}
MH~Loke and RD~Barker.
\newblock Least-squares deconvolution of apparent resistivity pseudosections.
\newblock {\em Geophysics}, 60(6):1682--1690, 1995.

\bibitem[LB18]{li2018global}
Yanjun Li and Yoram Bresler.
\newblock Global geometry of multichannel sparse blind deconvolution on the
  sphere.
\newblock In {\em Advances in Neural Information Processing Systems}, pages
  1132--1143, 2018.

\bibitem[LC11]{leung2011review}
Bonnie~O Leung and Keng~C Chou.
\newblock Review of super-resolution fluorescence microscopy for biology.
\newblock {\em Applied Spectroscopy}, 65(9):967--980, 2011.

\bibitem[Lew98]{lewicki1998review}
Michael~S Lewicki.
\newblock A review of methods for spike sorting: the detection and
  classification of neural action potentials.
\newblock {\em Network: Computation in Neural Systems}, 9(4):R53--R78, 1998.

\bibitem[LGCWY18]{liu2018first}
Jialin Liu, Cristina Garcia-Cardona, Brendt Wohlberg, and Wotao Yin.
\newblock First-and second-order methods for online convolutional dictionary
  learning.
\newblock {\em SIAM Journal on Imaging Sciences}, 11(2):1589--1628, 2018.

\bibitem[Li18]{li2018bilinear}
Yanjun Li.
\newblock {\em Bilinear inverse problems with sparsity: optimal identifiability
  conditions and efficient recovery}.
\newblock PhD thesis, University of Illinois at Urbana-Champaign, 2018.

\bibitem[LLB16]{li2016identifiability}
Yanjun Li, Kiryung Lee, and Yoram Bresler.
\newblock Identifiability in blind deconvolution with subspace or sparsity
  constraints.
\newblock {\em IEEE Transactions on Information Theory}, 62(7):4266--4275,
  2016.

\bibitem[LLSW18]{li2018rapid}
Xiaodong Li, Shuyang Ling, Thomas Strohmer, and Ke~Wei.
\newblock Rapid, robust, and reliable blind deconvolution via nonconvex
  optimization.
\newblock {\em Applied and Computational Harmonic Analysis}, 2018.

\bibitem[LPP{\etalchar{+}}17]{lee2017first}
Jason~D Lee, Ioannis Panageas, Georgios Piliouras, Max Simchowitz, Michael~I
  Jordan, and Benjamin Recht.
\newblock First-order methods almost always avoid strict saddle points.
\newblock {\em Mathematical Programming}, pages 1--27, 2017.

\bibitem[LS15]{ling2015self}
Shuyang Ling and Thomas Strohmer.
\newblock Self-calibration and biconvex compressive sensing.
\newblock {\em Inverse Problems}, 31(11):115002, 2015.

\bibitem[LWDF11a]{levin2011understanding}
Anat Levin, Yair Weiss, Fredo Durand, and William~T Freeman.
\newblock Understanding blind deconvolution algorithms.
\newblock {\em IEEE Transactions on Pattern Analysis and Machine Intelligence},
  33(12):2354--2367, 2011.

\bibitem[LWDF11b]{levin2009understanding}
Anat Levin, Yair Weiss, Fredo Durand, and William~T Freeman.
\newblock Understanding blind deconvolution algorithms.
\newblock {\em IEEE Transactions on Pattern Analysis and Machine Intelligence},
  33(12):2354--2367, 2011.

\bibitem[MCCM18]{maggu2018convolutional}
Jyoti Maggu, Emilie Chouzenoux, Giovanni Chierchia, and Angshul Majumdar.
\newblock Convolutional transform learning.
\newblock In {\em International Conference on Neural Information Processing},
  pages 391--398, 2018.

\bibitem[MK87]{murty1987some}
Katta~G. Murty and Santosh~N. Kabadi.
\newblock Some {NP}-complete problems in quadratic and nonlinear programming.
\newblock {\em Mathematical programming}, 39(2):117--129, 1987.

\bibitem[MWCC17]{ma2017implicit}
Cong Ma, Kaizheng Wang, Yuejie Chi, and Yuxin Chen.
\newblock Implicit regularization in nonconvex statistical estimation: Gradient
  descent converges linearly for phase retrieval, matrix completion and blind
  deconvolution.
\newblock {\em arXiv preprint arXiv:1711.10467}, 2017.

\bibitem[Nes13a]{nesterov2007gradient}
Yu~Nesterov.
\newblock Gradient methods for minimizing composite functions.
\newblock {\em Mathematical Programming}, 140(1):125--161, 2013.

\bibitem[Nes13b]{nesterov2013introductory}
Yurii Nesterov.
\newblock {\em Introductory lectures on convex optimization: A basic course},
  volume~87.
\newblock Springer Science \& Business Media, 2013.

\bibitem[NN14]{nienhaus2014fluorescent}
Karin Nienhaus and G~Ulrich Nienhaus.
\newblock Fluorescent proteins for live-cell imaging with super-resolution.
\newblock {\em Chemical Society Reviews}, 43(4):1088--1106, 2014.

\bibitem[NP06]{nesterov2006cubic}
Yurii Nesterov and Boris~T Polyak.
\newblock Cubic regularization of newton method and its global performance.
\newblock {\em Mathematical Programming}, 108(1):177--205, 2006.

\bibitem[NW06]{nocedal2006numerical}
Jorge Nocedal and Stephen Wright.
\newblock {\em Numerical optimization}.
\newblock Springer Science \& Business Media, 2006.

\bibitem[PB{\etalchar{+}}14]{parikh2014proximal}
Neal Parikh, Stephen Boyd, et~al.
\newblock Proximal algorithms.
\newblock {\em Foundations and Trends{\textregistered} in Optimization},
  1(3):127--239, 2014.

\bibitem[Pol64]{polyak1964some}
Boris~T Polyak.
\newblock Some methods of speeding up the convergence of iteration methods.
\newblock {\em USSR Computational Mathematics and Mathematical Physics},
  4(5):1--17, 1964.

\bibitem[PRSE17]{papyan2017convolutional}
Vardan Papyan, Yaniv Romano, Jeremias Sulam, and Michael Elad.
\newblock Convolutional dictionary learning via local processing.
\newblock In {\em Proceedings of the IEEE International Conference on Computer
  Vision}, pages 5296--5304, 2017.

\bibitem[PS16]{pock2016inertial}
Thomas Pock and Shoham Sabach.
\newblock Inertial proximal alternating linearized minimization (ipalm) for
  nonconvex and nonsmooth problems.
\newblock {\em SIAM Journal on Imaging Sciences}, 9(4):1756--1787, 2016.

\bibitem[PSG{\etalchar{+}}16]{pnevmatikakis2016simultaneous}
Eftychios~A Pnevmatikakis, Daniel Soudry, Yuanjun Gao, Timothy~A Machado, Josh
  Merel, David Pfau, Thomas Reardon, Yu~Mu, Clay Lacefield, Weijian Yang,
  et~al.
\newblock Simultaneous denoising, deconvolution, and demixing of calcium
  imaging data.
\newblock {\em Neuron}, 89(2):285--299, 2016.

\bibitem[QLL{\etalchar{+}}10]{quan2010ultra}
Tingwei Quan, Pengcheng Li, Fan Long, Shaoqun Zeng, Qingming Luo, Per~Niklas
  Hedde, Gerd~Ulrich Nienhaus, and Zhen-Li Huang.
\newblock Ultra-fast, high-precision image analysis for localization-based
  super resolution microscopy.
\newblock {\em Optics Express}, 18(11):11867--11876, 2010.

\bibitem[QLZ19]{qu2019blind}
Qing Qu, Xiao Li, and Zhihui Zhu.
\newblock A nonconvex approach for exact and efficient multichannel sparse
  blind deconvolution.
\newblock {\em arXiv preprint arXiv:1908.10776}, 2019.

\bibitem[QNBS04]{quiroga2004unsupervised}
R~Quian Quiroga, Zoltan Nadasdy, and Yoram Ben-Shaul.
\newblock Unsupervised spike detection and sorting with wavelets and
  superparamagnetic clustering.
\newblock {\em Neural Computation}, 16(8):1661--1687, 2004.

\bibitem[QSW14]{qu2014finding}
Qing Qu, Ju~Sun, and John Wright.
\newblock Finding a sparse vector in a subspace: Linear sparsity using
  alternating directions.
\newblock In {\em Advances in Neural Information Processing Systems}, pages
  3401--3409, 2014.

\bibitem[QZEW17]{qu2017convolutional}
Qing Qu, Yuqian Zhang, Yonina Eldar, and John Wright.
\newblock Convolutional phase retrieval.
\newblock In {\em Advances in Neural Information Processing Systems}, pages
  6086--6096, 2017.

\bibitem[RBZ06a]{rust2006stochastic}
Michael~J Rust, Mark Bates, and Xiaowei Zhuang.
\newblock Stochastic optical reconstruction microscopy (storm) provides
  sub-diffraction-limit image resolution.
\newblock {\em Nature Methods}, 3(10):793, 2006.

\bibitem[RBZ06b]{rust2006sub}
Michael~J Rust, Mark Bates, and Xiaowei Zhuang.
\newblock Sub-diffraction-limit imaging by stochastic optical reconstruction
  microscopy (storm).
\newblock {\em Nature Methods}, 3(10):793, 2006.

\bibitem[RCG{\etalchar{+}}07]{rutter2007scattering}
Gregory~M Rutter, JN~Crain, NP~Guisinger, T~Li, PN~First, and JA~Stroscio.
\newblock Scattering and interference in epitaxial graphene.
\newblock {\em Science}, 317(5835):219--222, 2007.

\bibitem[RPQ15]{rey2015past}
Hernan~Gonzalo Rey, Carlos Pedreira, and Rodrigo~Quian Quiroga.
\newblock Past, present and future of spike sorting techniques.
\newblock {\em Brain Research Bulletin}, 119:106--117, 2015.

\bibitem[RSP{\etalchar{+}}09]{roushan2009topological}
Pedram Roushan, Jungpil Seo, Colin~V Parker, Yew~San Hor, David Hsieh, Dong
  Qian, Anthony Richardella, M~Zahid Hasan, Robert~Joseph Cava, and Ali
  Yazdani.
\newblock Topological surface states protected from backscattering by chiral
  spin texture.
\newblock {\em Nature}, 460(7259):1106, 2009.

\bibitem[SFB18]{song2018spike}
Andrew~H Song, Francisco Flores, and Demba Ba.
\newblock Spike sorting by convolutional dictionary learning.
\newblock {\em arXiv preprint arXiv:1806.01979}, 2018.

\bibitem[SGG{\etalchar{+}}09]{shtengel2009interferometric}
Gleb Shtengel, James~A Galbraith, Catherine~G Galbraith, Jennifer
  Lippincott-Schwartz, Jennifer~M Gillette, Suliana Manley, Rachid Sougrat,
  Clare~M Waterman, Pakorn Kanchanawong, Michael~W Davidson, et~al.
\newblock Interferometric fluorescent super-resolution microscopy resolves 3d
  cellular ultrastructure.
\newblock {\em Proceedings of the National Academy of Sciences},
  106(9):3125--3130, 2009.

\bibitem[SGHK03]{stosiek2003vivo}
Christoph Stosiek, Olga Garaschuk, Knut Holthoff, and Arthur Konnerth.
\newblock In vivo two-photon calcium imaging of neuronal networks.
\newblock {\em Proceedings of the National Academy of Sciences},
  100(12):7319--7324, 2003.

\bibitem[SM12]{sroubek2012robust}
Filip Sroubek and Peyman Milanfar.
\newblock Robust multichannel blind deconvolution via fast alternating
  minimization.
\newblock {\em IEEE Transactions on Image processing}, 21(4):1687--1700, 2012.

\bibitem[SMSE18]{solomon2018sparsity}
Oren Solomon, Maor Mutzafi, Mordechai Segev, and Yonina~C Eldar.
\newblock Sparsity-based super-resolution microscopy from correlation
  information.
\newblock {\em Optics Express}, 26(14):18238--18269, 2018.

\bibitem[SN06]{sarder2006deconvolution}
Pinaki Sarder and Arye Nehorai.
\newblock Deconvolution methods for 3-d fluorescence microscopy images.
\newblock {\em IEEE Signal Processing Magazine}, 23(3):32--45, 2006.

\bibitem[SQW15]{sun2015nonconvex}
Ju~Sun, Qing Qu, and John Wright.
\newblock When are nonconvex problems not scary?
\newblock {\em arXiv preprint arXiv:1510.06096}, 2015.

\bibitem[SQW16a]{sun2016complete-i}
Ju~Sun, Qing Qu, and John Wright.
\newblock Complete dictionary recovery over the sphere i: Overview and the
  geometric picture.
\newblock {\em IEEE Transactions on Information Theory}, 63(2):853--884, 2016.

\bibitem[SQW16b]{sun2016complete-ii}
Ju~Sun, Qing Qu, and John Wright.
\newblock Complete dictionary recovery over the sphere ii: Recovery by
  riemannian trust-region method.
\newblock {\em IEEE Transactions on Information Theory}, 63(2):885--914, 2016.

\bibitem[SQW18]{sun2016geometric}
Ju~Sun, Qing Qu, and John Wright.
\newblock A geometric analysis of phase retrieval.
\newblock {\em Foundations of Computational Mathematics}, 18(5):1131--1198,
  2018.

\bibitem[STP17]{stella2017forward}
Lorenzo Stella, Andreas Themelis, and Panagiotis Patrinos.
\newblock Forward--backward quasi-newton methods for nonsmooth optimization
  problems.
\newblock {\em Computational Optimization and Applications}, 67(3):443--487,
  2017.

\bibitem[TBF{\etalchar{+}}16]{theis2016benchmarking}
Lucas Theis, Philipp Berens, Emmanouil Froudarakis, Jacob Reimer,
  Miroslav~Rom{\'a}n Ros{\'o}n, Tom Baden, Thomas Euler, Andreas~S Tolias, and
  Matthias Bethge.
\newblock Benchmarking spike rate inference in population calcium imaging.
\newblock {\em Neuron}, 90(3):471--482, 2016.

\bibitem[Tib96]{tibshirani1996regression}
Robert Tibshirani.
\newblock Regression shrinkage and selection via the lasso.
\newblock {\em Journal of the Royal Statistical Society. Series B
  (Methodological)}, pages 267--288, 1996.

\bibitem[VPM{\etalchar{+}}10]{vogelstein2010fast}
Joshua~T Vogelstein, Adam~M Packer, Timothy~A Machado, Tanya Sippy, Baktash
  Babadi, Rafael Yuste, and Liam Paninski.
\newblock Fast nonnegative deconvolution for spike train inference from
  population calcium imaging.
\newblock {\em Journal of Neurophysiology}, 104(6):3691--3704, 2010.

\bibitem[WC16]{wang2016blind}
Liming Wang and Yuejie Chi.
\newblock Blind deconvolution from multiple sparse inputs.
\newblock {\em IEEE Signal Processing Letters}, 23(10):1384--1388, 2016.

\bibitem[WNF09]{wright2009sparse}
Stephen~J Wright, Robert~D Nowak, and M{\'a}rio~AT Figueiredo.
\newblock Sparse reconstruction by separable approximation.
\newblock {\em IEEE Transactions on Signal Processing}, 57(7):2479--2493, 2009.

\bibitem[WZ14]{wipf2014revisiting}
David Wipf and Haichao Zhang.
\newblock Revisiting bayesian blind deconvolution.
\newblock {\em The Journal of Machine Learning Research}, 15(1):3595--3634,
  2014.

\bibitem[XZ13]{xiao2013proximal}
Lin Xiao and Tong Zhang.
\newblock A proximal-gradient homotopy method for the sparse least-squares
  problem.
\newblock {\em SIAM Journal on Optimization}, 23(2):1062--1091, 2013.

\bibitem[YHV17]{yellin2017blood}
Florence Yellin, Benjamin~D Haeffele, and Ren{\'e} Vidal.
\newblock Blood cell detection and counting in holographic lens-free imaging by
  convolutional sparse dictionary learning and coding.
\newblock In {\em IEEE 14th International Symposium on Biomedical Imaging},
  pages 650--653. IEEE, 2017.

\bibitem[YSE{\etalchar{+}}18]{yger2018spike}
Pierre Yger, Giulia~LB Spampinato, Elric Esposito, Baptiste Lefebvre,
  St{\'e}phane Deny, Christophe Gardella, Marcel Stimberg, Florian Jetter,
  Guenther Zeck, Serge Picaud, et~al.
\newblock A spike sorting toolbox for up to thousands of electrodes validated
  with ground truth recordings in vitro and in vivo.
\newblock {\em Elife}, 7:e34518, 2018.

\bibitem[YWHM10]{yang2010image}
Jianchao Yang, John Wright, Thomas~S Huang, and Yi~Ma.
\newblock Image super-resolution via sparse representation.
\newblock {\em IEEE Transactions on Image Processing}, 19(11):2861--2873, 2010.

\bibitem[ZCB{\etalchar{+}}14]{zhou2014classification}
Yin Zhou, Hang Chang, Kenneth Barner, Paul Spellman, and Bahram Parvin.
\newblock Classification of histology sections via multispectral convolutional
  sparse coding.
\newblock In {\em Proceedings of the IEEE Conference on Computer Vision and
  Pattern Recognition}, pages 3081--3088, 2014.

\bibitem[ZKW18]{zhang2018structured}
Yuqian Zhang, Han-wen Kuo, and John Wright.
\newblock Structured local minima in sparse blind deconvolution.
\newblock In {\em Advances in Neural Information Processing Systems}, pages
  2328--2337, 2018.

\bibitem[ZLK{\etalchar{+}}17]{zhang2017global}
Yuqian Zhang, Yenson Lau, Han-Wen Kuo, Sky Cheung, Abhay Pasupathy, and John
  Wright.
\newblock On the global geometry of sphere-constrained sparse blind
  deconvolution.
\newblock In {\em Computer Vision and Pattern Recognition (CVPR), 2017 IEEE
  Conference on}, pages 4381--4389. IEEE, 2017.

\bibitem[ZSE19]{zisselman2018local}
Ev~Zisselman, Jeremias Sulam, and Michael Elad.
\newblock A local block coordinate descent algorithm for the csc model.
\newblock In {\em Proceedings of the IEEE Conference on Computer Vision and
  Pattern Recognition}, pages 8208--8217, 2019.

\bibitem[ZZEH12]{zhu2012faster}
Lei Zhu, Wei Zhang, Daniel Elnatan, and Bo~Huang.
\newblock Faster storm using compressed sensing.
\newblock {\em Nature Methods}, 9(7):721, 2012.

\end{thebibliography}
}

\newpage
\appendices

\edited{The appendix is organized as follows. In Appendix \ref{app:basic}, we introduce the basic notations and terms that are used throughout the draft. In Appendix \ref{app:algorithm}, we describe the proposed algorithmic pipeline for solving sparse deconvolution problems in very detail. Finally, Appendix \ref{app:details} provides all the missing complementary details of implementing the proposed algorithm for solving SaS-BD and SaS-CDL.}

\section{Basic notations}\label{app:basic}
Throughout this paper, all vectors/matrices are written in bold font $\mb a$/$\mb A$; indexed values are written as $a_i, A_{ij}$. Vectors with all zero and all one entries are denoted as $\mb 0_m$ and $\mb 1_m$, respectively, with $m$ denoting its length. The $i$-th canonical basis vector is denoted by $\mb e_i$. We use $\bb S^{n-1}$ to denote an $n$-dimensional unit sphere in the Euclidean space $\bb R^n$. We use $\mb z^{(k)}$ to denote the optimization variable $\mb z$ at $k$th iteration. We let $[m] =\Brac{1,2,\cdots,m}$. For a multivariate function $\Psi(\mb a,\mb x)$, we use $\Psi_{\mb a}(\mb x)$ and $\Psi_{\mb x}(\mb a)$ to denote marginal functions of $\Psi(\mb a,\mb x)$ with one variable fixed, respectively. Next, we define several useful operators appear throughout the paper and the appendices.

\paragraph{Some basic operators.} We use $ \res{n}{m} $ to denote a zero-padding operator $	\res{n}{m}\mb v = \begin{bmatrix}
 \mb v \\ \mb 0_{n-m}\end{bmatrix}$, which zero-pads a length $n$ vector $\mb v \in \bb R^n$ to length $m$ ($n \leq m$). Correspondingly, its adjoint operator $\res{n}{m}^*$ denotes the restriction of a vector of length-$m$ to its first $n$ coordinate (and $\res{n}{m}^* = \res{m}{n} $). Similarly, given a subset $\mc I \subseteq [m]$ and a vector $\mb v \in \bb R^{ \abs{\mc I} }$, we use $\res{ \mc I }{m}: \bb R^{ \abs{\mc I} } \mapsto \bb R^m $ to denote an operator that maps $\mb v$ to a zero-padded vector whose entries in $\mc I$ corresponding to those of $\mb v$.

We use $\mc P_{\mb v}$ and $\mc P_{\mb v^\perp}$ to denote projections onto $\mb v$ and its orthogonal complement, respectively. We let $\mc P_{\bb S^{n-1}}(\cdot)$ to be the $\ell_2$-normalization operator. To sum up, for any two vectors $\mb v$ and $\mb u \in \bb R^n$, we have
\begin{align*}
	\mc P_{\mb v^\perp} \mb u \;=\; \mb u - \frac{\mb v \mb v^\top }{\norm{\mb v}{2}^2} \mb u,\quad  \mc P_{\mb v} \mb u \;=\; \frac{\mb v \mb v^\top }{\norm{\mb v}{2}^2} \mb u,\quad \mc P_{\bb S^{n-1}} \mb u \;=\; \frac{\mb u}{\norm{\mb u}{2}}.
\end{align*}

\paragraph{Circular convolution and circulant matrices.}
The convolution operator $\conv$ is \emph{circular} with modulo-$m$: $\paren{\mb a\conv \mb x}_i = \sum_{j=0}^{m-1} a_j x_{i-j}$, and we use $\cconv$ to specify the \emph{circular} convolution in 2D. For a vector $\mb v \in \bb R^m$, let $\mathrm{s}_\ell[\mb v]$ denote the cyclic shift of $\mb v$ with length $\ell$. In addition, we use $\Shift{\mb v}{\ell}$ to denote a $3m-2$ length zero-pad shift, i.e.,
\begin{align*}
   	\Shift{\mb v}{\ell}\; =\;  \mathrm{s}_\ell [\begin{bmatrix}
 \mb 0_{m-1} \\
 \mb v \\
 	\mb 0_{m-1}
\end{bmatrix}].
\end{align*}
We introduce the circulant matrix $\mb C_{\mb v}\in \bb R^{m \times m}$ generated through $\mb v \in \bb R^m$,
\begin{align*}
   \mb C_{\mb v} \;=\; \begin{bmatrix}
   v_1 & v_m & \cdots & v_3 & v_2 \\
   v_2 & v_1 & v_m  & & v_3 \\
   \vdots & v_2 & v_1 & \ddots &\vdots \\
   v_{m-1} &  & \ddots  & \ddots  &v_m\\
   v_m &  v_{m-1} & \cdots  & v_2 &v_1
 \end{bmatrix} \;=\; \begin{bmatrix}
 	\shift{ \mb v }{0} & \shift{ \mb v }{1} & \cdots & \shift{ \mb v }{m-1}
 \end{bmatrix}.
\end{align*}
Now the circulant convolution can also be written in a simpler matrix-vector product form. For instance, for any $\mb u \in \bb R^m$ and $\mb v \in \bb R^n$ ($n \leq m$),
\begin{align*}
   \mb u \conv \mb v \;=\; \mb C_{\mb u}  \cdot \res{n}{m} \mb v \;=\; \mb C_{\res{n}{m} \mb v} \cdot \mb u \; =\; \mb v \conv \mb u.
\end{align*}
In addition, the correlation between $\mb u$ and $\mb v$ can be also written in a similar form of convolution operator which reverses one vector before convolution. Let $\wc{\mb v}$ denote a \emph{cyclic reversal} of $\mb v \in \bb R^m$, i.e., $\wc{\mb v} = \brac{ v_1, v_m, v_{m-1},\cdots,v_2 }^\top$, and define two correlation matrices $\mb C_{\mb v}^* \mb e_j = \mathrm{s}_j[\mb v]$ and $\wc{\mb C}_{\mb v} \mb e_j = \mathrm{s}_{-j}[\mb v]$. The two operators satisfy
\begin{align*}
   \mb C_{\res{n}{m}\mb v}^* \mb u \;=\; \wc{\mb v} \conv \mb u,\quad \wc{\mb C}_{\res{n}{m}\mb v} \mb u \;=\; \mb v \conv \wc{\mb u}.	
\end{align*}

\paragraph{Notation for several distributions.} We use $i.i.d.$ to denote \emph{identically} and \emph{independently distributed} random variables. In addition, we introduce and denote several distributions as follows.
\begin{itemize}[leftmargin=*]
	\item We use $\mc N(\mu,\sigma^2)$ to denote the Gaussian distribution with mean $\mu$ and variance $\sigma^2$, \edited{and use $\mc U(\bb S^{n-1})$ to denote a uniform distribution over the sphere $\bb S^{n-1}$};
	\item we use $\mc B(\theta)$ to denote the Bernoulli distribution with parameter $\theta$ controling the nonzero probability;
	\item we use $\mc {BG}(\theta)$ to denote Bernoulli-Gaussian distribution, i.e., if $u \sim \mc {BG}(\theta) $, then $u = b \cdot g$ with $b \sim \mc B(\theta)$ and $g \sim \mc N(0,1)$;
	\item we use $\mc {BR}(\theta)$ to denote Bernoulli-Rademacher distribution, i.e., if $u \sim \mc {BR}(\theta) $, then $u = b \cdot r$ with $b \sim \mc B(\theta)$ and $r$ follows Rademacher distribution. 
\end{itemize}

%Let us define a reversal operator
%\begin{align*}
%   \mc R_m \doteq \begin{bmatrix}
%    1 & 0&\cdots &\cdots & 0 \\
%    0 & & & \iddots &1\\
%    0 & &0 &1 &0 \\
%    0&  \iddots & \iddots  &  \iddots&  \vdots \\
%    0 &1 & 0 & \cdots & 0
% \end{bmatrix},
%\end{align*}
%we use $\wc{\mb b}$ to denote a cyclic reversal of a vector $\mb b = \brac{ b_1,b_2,\cdots, b_m}^\top $, i.e.,
%\begin{align*}
%   \wc{\mb b} = \brac{ b_1, b_n, b_{m-1},\cdots,b_2 }^\top.
%\end{align*}
%We use $\wc{\mb C}_{\mb b}$ to denote the reversal matrix
%\begin{align*}
%   \wc{\mb C}_{\mb b} = \mb C_{\mb b} \mc R_m,\qquad \wc{\mb C}_{\mb u \conv \mb v} = \mb C_{\mb u} \mb C_{ \res{n}{m} \mb v} \mc R_m.
%\end{align*}
%A nice property of the reversal matrix $\wc{\mb C}_{\mb b}$ is that it is a symmetric matrix, 
%where we have $\mc R_m \mb b = \wc{\mb b}$.

% !TEX root = ../main.tex

\section{Algorithmic Pipeline}\label{app:algorithm}
\edited{
In this part of appendix, we introduce a general algorithmic pipeline for solving sparse deconvolution problems, including SaSD and SaS-CDL. We describe the optimization problem in a more general form here. Namely, we consider the following problem
\begin{align}\label{eqn:problem-general}
   \min_{\mb a,\mb x} \; \Psi(\mb a,\mb x) \;\;=\;\; \psi(\mb a,\mb x) \;+\; \lambda \cdot g(\mb x), \qquad \mathrm{s.t.}\quad \mb a \;\in\; \mc M,
\end{align}
where $\psi(\mb a,\mb x)$ is a data fidelity term that we to be twice continuously differentiable, $g(\mb x)$ is a convex (possibly nonsmooth) sparse promoting penalty, and $\mc M$ is a smooth Riemannian manifold. Again, the penalty $\lambda>0$ balances the weights of two terms $\psi(\mb a,\mb x)$ and $g(\mb x)$. The objective in \Cref{eqn:problem-general} generalizes the Bilinear Lasso formulation for SaSD and SaS-CDL problems:
\begin{itemize}[leftmargin=*]
\item \textbf{SaSD.} Recall from \Cref{eqn:bilinear-lasso}, we have
\begin{align*}
   \psi(\mb a,\mb x)\;=\; \frac{1}{2} \norm{ \mb y - \mb a \conv \mb x }{2}^2,\quad g(\mb x) \;=\; \norm{\mb x}{1},\quad \mc M \;=\; \bb S^{\Di-1}.
\end{align*}
\item \textbf{SaS-CDL.} Let $\mb A = \begin{bmatrix} \mb a_1 & \cdots & \mb a_N \end{bmatrix}$ and $\mb X = \begin{bmatrix} \mb x_1 & \cdots & \mb x_N \end{bmatrix}$, by \Cref{eqn:bilinear-lasso-CDL}, 
\begin{align*}
   	\psi(\mb A,\mb X)\;=\; \frac{1}{2} \norm{ \mb y - \sum_{k=1}^N \mb a_k \conv \mb x_k }{2}^2,\quad g(\mb X) \;=\; \norm{\mb X}{1},\quad \mc M \;=\; \Brac{ \mb A \in \bb R^{n \times N} \;\mid\; \mb a_k \in \bb S^{\Di-1}, \;1\leq k\leq N }.
\end{align*}
\end{itemize}
For the rest of this appendix, we introduce our algorithms based on the general formulation in \Cref{eqn:problem-general} for the ease of exposition. We defer more implementation details for SaSD and SaS-CDL to Appendix \ref{app:details}.
% which can be cast as an optimization problem where one minimizes a data fidelity term $\psi(\mb a,\mb x),$ plus a sparsity penalty $g(\mb x)$ over a Riemannian manifold $\mc M$,
%\begin{align}\label{eqn:problem-general}
%   \min_{\mb a,\mb x} \Psi(\mb a,\mb x) = \psi(\mb a,\mb x) + \lambda \cdot g(\mb x), \qquad \mathrm{s.t.}\quad \mb a \in \mc M.
%\end{align}
%Here $\lambda$ is a parameter balancing the two terms, and we assume that $\psi(\mb a,\mb x)$ is twice continuously differentiable and convex w.r.t.\ one variable when the other is fixed, and $g(\mb x)$ is convex but possibly nonsmooth.

%For ease of exposition, most of our discussion will be based on \eqref{eqn:problem-general}, with the intention that the appropriate quantities from the Bilinear Lasso can be plugged in. Readers interested in implementation may refer to \Cref{subsec:momentum} for an accelerated first-order solver for \Cref{eqn:problem-general}, and to \Cref{subsubsec:iADM-BL} for details regarding SaSD. Details for SaS-CDL can be found in Appendix \ref{app:details}.
}

\subsection{Alternating descent method}\label{subsec:adm}

\subsubsection{Vanilla ADM}

\edited{We begin this part of appendix by introducing a \emph{vanilla} first-order method for solving \Cref{eqn:problem-general} based on alternating descent method (ADM). The method minimizes the objective by alternating between taking descent steps on one variable with the other fixed. The basic algorithm pipeline is summarized in \Cref{alg:ADM}.}

\begin{algorithm}
\begin{algorithmic}
\caption{Alternating Descent Method (ADM)}
\renewcommand{\algorithmicrequire}{\textbf{Input:}}
\renewcommand{\algorithmicensure}{\textbf{Output:}}
\Require~~\
Measurement $\mb y\in \bb R^\sample$; stepsizes $t_0$ and $\tau_0$; penalty $\lambda>0$.
\Ensure~~\
Final iterate $\mb a_\star$, $\mb x_\star$.
\State Initialize $\mb a\iter0$ using \Cref{eqn:initialization}, $\mb x\iter0 \leftarrow \mb 0_\sample$, and $k \leftarrow 0$.
\While { not converged }
\State Fix $\mb a\iter k$ and take a proximal gradient step on $\mb x$ with stepsize $t_k$
\begin{align*}
   \mb x^{(k+1)} \;\leftarrow \; \mb x^{(k)} - t_k \mc G_{t_k} \paren{ \mb x^{(k)} } ,
 % \mb x\iter{k+1} \;\leftarrow\;
  %  \text{prox\,}_{t_k\cdot\lambda\norm{\cdot}1} \paren{ \mb x\iter k -
 %   t_k\cdot\nabla_{\mb x\,} \psi\paren{\mb a\iter k, \mb x\iter k}}.
\end{align*}

\State Fix $\mb x\iter{k+1}$ and take a Riemannian gradient step on $\mb a$ with stepsize $\tau_k$
\begin{align*}
  \mb a\iter{k+1} \;\leftarrow \;  \mc R_{\mb a^{(k)}}^{\mc M} \paren{ -\tau \cdot \grad \psi_{\mb x\iter{k+1}}(\mb a^{(k)}) },
\end{align*}
\State Update $k \leftarrow k+1$.
\EndWhile \label{alg:ADM}
\end{algorithmic}
\end{algorithm}

\paragraph{Fix $\mb a$ and take a proximal gradient step on $\mb x$.} Given $\mb a$ being fixed, the marginal function of $\Psi(\mb a,\mb x)$,
\begin{align*}
    \Psi_{\mb a}(\mb x) = \psi_{\mb a}(\mb x) +  \lambda \cdot g(\mb x),
\end{align*}
is \emph{nonsmooth} w.r.t. $\mb x$. A classical way to deal with nonsmoothness is by considering its \emph{smooth envelope} \cite{stella2017forward}, and take a proximal gradient step on the smooth variant \cite{parikh2014proximal},
\begin{align}\label{eqn:prox-grad}
   \mb x^{(k+1)} \; = \; \mc P_t(\mb x^{(k)}) \;=\;   \mb x^{(k)} - t \mc G_t(\mb x^{(k)}),\qquad \mc G_t(\mb x) \;=\; t^{-1} \paren{ \mb x - \mc P_t(\mb x)  },
\end{align}
where $\mc G_t(\mb x)$ is termed as \emph{composite gradient mapping}, and $\mc P_t(\mb x^{(k)})$ is a proximal mapping that we introduce in the following. For any $t>0$, consider a quadratic approximation of $\Psi_{\mb a}(\mb x)$ at a given point $\ol{\mb x}$,
\begin{align*}
   Q_{\mb a}^t\paren{\mb x,\ol{\mb x}} \;=\; \psi_{\mb a}\paren{ \ol{\mb x} }  + \innerprod{ \mb x - \ol{\mb x} }{ \nabla \psi_{\mb a}(\ol{ \mb x} ) } + \frac{1}{2t} \norm{ \mb x - \ol{\mb x} }{2}^2 + \lambda \cdot g(\mb x).
\end{align*}
For a convex $g(\cdot)$, $Q_{\mb a}^t\paren{\mb x,\ol{\mb x}}$ admits a unique minimizer via the proximal mapping
\begin{align*}
   \mc P_t \paren{\ol{\mb x} } \;=\; \arg \min_{\mb x} Q_{\mb a}^t\; \paren{\mb x, \ol{\mb x} } \;=\; \mathrm{prox}_g^{ \lambda t} \paren{ \ol{\mb x} - t \nabla \psi_{\mb a}(\ol{\mb x})  },
\end{align*}
where we denote the \emph{proximal operator} of $g(\cdot)$ by
\begin{align*}
	\mathrm{prox}_g^{\rho} (\mb x) \;\doteq\; \arg\min_{\mb z} \Brac{ \rho \cdot g(\mb z) + \frac{1}{2} \norm{ \mb x - \mb z }{2}^2  }.
\end{align*}
By plugging $\mc P_t \paren{\ol{\mb x} }$ back into $Q_{\mb a}^t\paren{\mb x,\ol{\mb x}}$, it gives the so-called \emph{forward-backward envelope} \cite{stella2017forward} of $\Psi_{\mb a}(\mb x)$ as
\begin{align*}
   	F_{\mb a}^t(\ol{\mb x}) \;=\; \min_{\mb x} \Brac{ Q_{\mb a}^t\paren{\mb x,\ol{\mb x}} } \;=\; Q_{\mb a}^t\paren{ \mc P_t \paren{\ol{\mb x}},\ol{\mb x}},
\end{align*}
which serves as a smooth upper bound (envelope), i.e.,
\begin{align}\label{eqn:prox-envelope}
	F_{\mb a}^t(\mb x) \;=\; Q_{\mb a}^t\paren{ \mc P_t \paren{\mb x}, \mb x} \;\geq\; \Psi_{\mb a}( \mc P_t \paren{\mb x})
\end{align}
for any $t \in \paren{0, 1/L_\psi}$, where $L_{\psi}$ is the Lipschitz constant of $\nabla \psi_{\mb a}(\mb x)$ \cite{nesterov2013introductory,beck2017first}. Indeed, the composite gradient mapping $\mc G_t(\mb x)$ in \Cref{eqn:prox-grad} can be interpreted as the gradient on the smooth envelope $F_{\mb a}^t(\mb x)$, so that the proximal step in \Cref{eqn:prox-grad} can be viewed as a gradient descent method. Additionally, we can show that the function value produced by the proximal gradient in \Cref{eqn:prox-grad} is nonincreasing $\Psi_a(\mb x^{(k+1)}) \leq \Psi_a(\mb x^{(k)})$, when $ t \in \paren{0, 1/L_{\psi_{\mb a}}  }$ \cite{beck2009fast,beck2017first}.

The parameter $t$ is usually set to be $1/L_{\psi_{\mb a}} $ for fast convergence. However, computing $L_{\psi_{\mb u}} $ for each iteration can be expensive. Instead, we use a backtracking rule (see \Cref{alg:backtracking-t}) to adaptively choose $t$ based on the inequality in \Cref{eqn:prox-envelope}.

\begin{algorithm}
\begin{algorithmic}
\renewcommand{\algorithmicrequire}{\textbf{Input:}}
\renewcommand{\algorithmicensure}{\textbf{Output:}}
\Require~~\
$\mb a$, $\mb x$, $t_0$, $\beta \in (0,1)$
\Ensure~~\
$t$, $\mc P_t(\mb x)$,
\State Set $t \leftarrow t_0$ and compute $\mc P_t(\mb x)$.
\While { $ \Psi_{\mb a}\paren{ \mc P_t(\mb x) } \;\geq\; Q_{\mb a}^t \paren{ \mc P_t(\mb x), \mb x } $ }
\State Set $t \; \leftarrow \; \beta t $ and update $\mc P_t(\mb x)$.
\EndWhile
\end{algorithmic}
\caption{Backtracking rule for stepsize $t$}
\label{alg:backtracking-t}
\end{algorithm}

\begin{figure*}[!htbp]
\centering
\includegraphics[width = 0.4\linewidth]{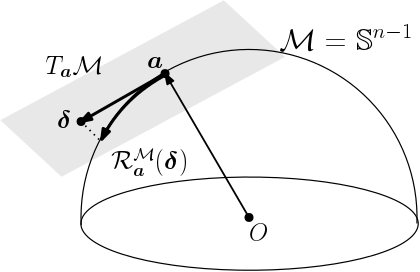}
\caption{\textbf{An illustration of manifold optimization with $\mc M = \bb S^{\Di-1}$.} $T_{\mb a}\mc M$ denote the tangent space of $\mc M$ at the point $\mb a$, and $\mc R_{\mb a}^{\mc M}(\mb \delta)$ denote the retraction operator at the point $\mb a \in \mc M$.}
\label{fig:exp-map}
\end{figure*}

\paragraph{Fix $\mb x$ and take a Riemannian gradient step on $\mb a$.} As our optimization variable $\mb a$ in constraint over the Riemannian manifold $\mc M$, we consider the Riemannian derivative on $\psi_{\mb x}(\mb a)$ \cite{absil2009}. Starting from an iterate $\mb a^{(k)}$, we take a Riemannian gradient step on $\psi_{\mb x}(\mb a)$ by
\begin{align}\label{eqn:riemannian-grad}
   \mb a^{(k+1)} = \mc R_{\mb a^{(k)}}^{\mc M} \paren{ -\tau \cdot \grad \psi_{\mb x}(\mb a^{(k)}) },
\end{align}
where $\tau$ is the stepsize, which can be adaptively chosen by the \emph{Riemannian linesearch} based on \emph{Armijo condition} (see \Cref{alg:backtracking-tau}). We use $\grad \psi_{\mb x}(\mb a)$ to denote the Riemannian gradient of $\psi_{\mb x}(\mb a)$, which is defined over the \emph{tangent space} of $\mc M$ at the point $\mb a$,
\begin{align*}
   \grad \psi_{\mb x}(\mb a) \;=\;  	\mb P_{ T_{\mb a}\mc M} \nabla \psi_{\mb x}(\mb a),
\end{align*}
where $\mb P_{ T_{\mb a}\mc M}$ is the projection operator onto the tangent space $T_{\mb a}\mc M$.
On the other hand, $\mc R_{\mb a}^{\mc M}(\mb \delta)$ denotes the retraction operator, which pulls a vector $\mb \delta$ from the tangent space $T_{\mb a}\mc M$ to its closest point on the Riemannian manifold $\mc M$. \Cref{fig:exp-map} provides an illustration of the tangent space $T_{\mb a}\mc M$ and the retraction operator $\mc R_{\mb a}^{\mc M}(\mb \delta)$ when $\mc M = \bb S^{n-1}$, we refer readers to Chapter 3 and 4 of \cite{absil2009} for more detailed definitions.

\begin{algorithm}
\begin{algorithmic}
\renewcommand{\algorithmicrequire}{\textbf{Input:}}
\renewcommand{\algorithmicensure}{\textbf{Output:}}
\Require~~\
$\mb a$, $\mb x$, $\tau_0$, $\eta \in (0.5,1)$, $\beta \in (0,1)$,
\Ensure~~\
$\tau$, $\mc R_{\mb a}^{\mc M} \paren{ -\tau  \mb P_{T_{\mc M} } \nabla \psi_{\mb x}(\mb a)  }$
\State Initialize $\tau \leftarrow \tau_0$.
\While { $  \psi_{\mb x}( \mc R_{\mb a}^{\mc M}(-\tau \cdot \grad \psi_{\mb x}(\mb a)  ) ) \;\geq\;  \psi_{\mb x} (\mb a ) - \tau \cdot \eta \cdot \norm{ \grad \psi_{\mb x}(\mb a) }{2}^2	 $ }
\State $\tau \leftarrow \beta \tau $.
\EndWhile
\end{algorithmic}
\caption{Riemannian linesearch for stepsize $\tau$}
\label{alg:backtracking-tau}
\end{algorithm}

\subsubsection{Accelerated ADM via momentum method}\label{subsec:momentum}

As aforementioned in \Cref{subsec:acceleration}, problems in practice often raise additional challenges. The kernel $\mb a_0$ we encounter in practice is often smooth, so that the underlying kernel $\mb a_0$ is of large incoherence $\mu_s(\mb a_0)$. This results in ill-conditioned problems and slow convergence of first-order methods \cite{nesterov2013introductory,bubeck2015convex,beck2017first}. As we have discussed in \Cref{subsubsec:main-momentum}, a natural idea to improve solution accuracy and convergence speed is to employ a momentum acceleration strategy, which can be traced back to the \emph{heavy ball} method of Polyak \cite{polyak1964some}.

\begin{algorithm}
\caption{Inertial Alternating Descent Method (iADM)}
\label{alg:iADM}
\begin{algorithmic}
\renewcommand{\algorithmicrequire}{\textbf{Input:}}
\renewcommand{\algorithmicensure}{\textbf{Output:}}
\Require~~\
measurement $\mb y$; initial values $\mb a\iter 0$, $\mb x\iter 0$;
\ penalty $\lambda > 0$;
\ momentum parameter $\beta \in [0,1)$.
\Ensure~~\
Final iterate $\mb a_\star$, $\mb x_\star$.
\State Initialize $k \leftarrow 0$, set $\mb a\iter{-1} = \mb a\iter 0$, $\mb x\iter{-1} = \mb x\iter 0$.
\While { not converged }
\State Fix $\mb a^{(k)}$, and update $\mb x$ using proximal gradient descent with momentum
\begin{align*}
    \mb w\iter k \;&=\; \mb x\iter k + \beta \cdot \paren{ \mb x\iter k - \mb x\iter{k-1} }, \\
  \mb x\iter{k+1} \;&=\; \mathrm{prox}_{g}^{\lambda t}  \paren{
    \mb w\iter k - t_k \cdot \nabla \psi_{\mb a\iter k}\paren{ \mb w\iter k} }.
\end{align*}

\State Fix $\mb x^{(k+1)}$, and update $\mb a$ by using the Riemannian gradient descent with momentum
\begin{align*}
  \mb z\iter k \;&=\; \mc R_{\mb a\iter k}^{\mc M} \paren{
    \beta \cdot \paren{\mc R_{\mb a\iter{k-1}}^{\mc M} }^{-1}\paren{\mb a\iter k} }
  , \\
  \mb a\iter{k+1} \;&=\; \mc R_{\mb z\iter k}^{\mc M} \paren{
    - \tau_k \cdot \grad  \psi_{\mb x\iter{k+1}} \paren{\mb z\iter k} },
\end{align*}

\State Set $k \leftarrow k +1$.
\EndWhile
\end{algorithmic}
\end{algorithm}

For our particular problem in \Cref{eqn:problem-general}, we apply momentum acceleration to sub-iterations of ADM on both $\mb a$ and $\mb x$. When updating $\mb x$ with $\mb a$ fixed, recall from \Cref{eqn:main-xiter-momentum} in \Cref{subsubsec:main-momentum}, we modify the original iteration by adding an inertial term $\mb w\iter k$, which incorporates information from previous updates. Similarly, when we update $\mb a$ with $\mb x$ fixed, we modify the Riemannian gradient step in \Cref{eqn:riemannian-grad} by
\begin{align}
   	  \mb z\iter k \;&=\; \mc R_{\mb a\iter k}^{\mc M} \Big(
    \beta \cdot \underbrace{\paren{\mc R_{\mb a\iter{k-1}}^{\mc M}}^{-1}\paren{\mb a\iter k} }_{\text{inertial term}} \Big), \label{eqn:inertia-a} \\
  \mb a\iter{k+1} \;&=\; \mc R_{\mb z\iter k}^{\mc M} \paren{
    - \tau_k \cdot \grad \psi_{ \mb x\iter{k+1} } \paren{\mb z\iter k} }. \nonumber
\end{align}
Here, $\paren{\mc R_{\mb a}^{\mc M} }^{-1}\paren{ \mb b } : \mc M \rightarrow T_{\mb a}\mc M$ denotes the \emph{inverse retraction operator}, i.e., \ $\mc R_{\mb a}^{\mc M}\paren{ \paren{\mc R_{\mb a}^{\mc M} }^{-1}(\mb b) } \equiv \mb b$. It maps a point $\mb b \in\mc M$ to the tangent space $T_{\mb a}\mc M$ of $\mb a$. Intuitively, when $\beta$ is small, and $\mb a\iter{k-1}$ and $\mb a\iter k$ are close, we approximately have
\begin{align*}
    \mb z\iter k \;\approx\; \mb a \iter k \; + \beta \cdot 	\mb \delta \iter k,\quad \mb \delta \iter k \;=\; \paren{\mc R_{\mb a\iter{k-1}}^{\mc M}}^{-1}\paren{\mb a\iter k} \;\approx\;\mb a\iter k - \mb a\iter{k-1},
\end{align*}
which reduces to the standard update in the Euclidean space. The overall algorithmic pipeline is summarized in \Cref{alg:iADM}, and we term the algorithm inertial ADM (iADM). Similarly to the ADM, we can either set the stepsizes $t_k$ and $\tau_k$ to be constants, or choose them via backtracking (linesearch) (\Cref{alg:backtracking-t,alg:backtracking-tau}). The parameter $\beta \in [0,1)$ controls the weight of inertial term. Empirically, good choices for $\beta$ lie somewhere between $0.8$ to $0.9$, and iADM reverts to ADM when $\beta$ is set to zero. An iteration-dependent schedule for $\beta$ is also discussed in \cite{pock2016inertial}. Unlike the ADM algorithm which decreases the function value $\Psi(\mb a,\mb x)$ monotonically, the iterates of iADM exhibit some oscillation effects and they can diverge when $\beta$ is chosen too large.

\subsection{Adaptive update of the penalty $\lambda$ through the solution path}\label{subsec:homotopy}

For sparse deconvolution problems, the parameter $\lambda$ controls the sparsity of the solution $\mb x$: the larger $\lambda$ is, the sparser $\mb x$ is produced, and vice versa. \cite{kuo2019geometry} suggests that a good choice could be $\lambda= \mc O\paren{ 1/ \paren{ \theta \di } } $,  where $\theta$ is the parameter of Bernoulli distribution controlling the sparsity level. However, the sparsity level $\theta$ is often not known ahead of time for many real applications, but the choice of $\lambda$ is crucial for convergence speed and recovery accuracy. In this subsection, we introduce two schemes to adaptively update $\lambda$.
\begin{enumerate}[label=(\roman*)]
\item \textbf{Homotopy continuation method,} which improves both algorithmic convergence speed and recovery accuracy by shrinking the $\lambda$ through the solution path;
\item \textbf{Reweighting method}, which improves robustness against noise by adaptively enforcing different penalties $\lambda$ on different entries of $\mb x$.
\end{enumerate}

\begin{algorithm}
\caption{Homotopy continuation method}\label{alg:homotopy}
\begin{algorithmic}
\renewcommand{\algorithmicrequire}{\textbf{Input:}}
\renewcommand{\algorithmicensure}{\textbf{Output:}}
\Require~~\
Measurement $\mb y\in\bb R^m$; initial and final sparse penalties $\lambda_0$, $\lambda_\star$ ($\lambda_0>\lambda_\star$); initialization $(\mb a \iter 0, \mb x \iter 0)$; decay penalty parameter $\eta \in (0,1)$; precision factor $\delta \in (0,1)$ and tolerance $\eps_\star$.
\Ensure~~\
final solution $(\mb a_\star,\mb x_\star)$.
\State Initialize $k \leftarrow 0$, $\lambda \iter 0 \leftarrow \lambda_0$, $\eps\iter 0 \leftarrow \delta  \lambda\iter 0$.
\State Set $K \; \leftarrow \; \left\lfloor \log\paren{ \lambda_\star /\lambda_0 } / \log\paren{ \eta }   \right\rfloor   $.
\While { $k \;\leq\;  K  $ }
\State \emph{Solve} \Cref{eqn:problem-general} with $\lambda\iter k$ to $\paren{\mb a\iter{k+1},\mb x\iter{k+1} }$ of precision $\eps\iter k$, using $\paren{\mb a\iter k,\mb x\iter k }$ as warm restart.
\State \emph{Update the parameters}: $\lambda\iter{k+1} \leftarrow \eta \lambda\iter k $, and $\eps\iter{k+1} \leftarrow \delta \lambda\iter{k+1} $.
\State \emph{Update} $k \;\leftarrow\; k +1$.
\EndWhile
\State \emph{Final round}: from $\paren{\mb a^{(K+1)},\mb x^{(K+1)} }$, solve \Cref{eqn:problem-general} with penalty $\lambda_\star$ to $\paren{\mb a_\star,\mb x_\star}$ of precision $\eps_\star$.
\end{algorithmic}
\end{algorithm}

\paragraph{Homotopy continuation method.} As discussed in \Cref{subsec:main-homotopy}, the geometric intuition (see \Cref{fig:homotopy-lambda}) suggests a homotopy continuation approach \cite{hale2008fixed,wright2009sparse,xiao2013proximal}, which chooses a solution path for $(\mb a,\mb x)$ by adaptively decreasing $\lambda$. The overall algorithmic pipeline is summarized in \Cref{alg:homotopy}. More concretely, we start by solving  \Cref{eqn:problem-general} with a large penalty $\lambda_0$ (e.g., $\lambda_0 = \norm{  \wc{\mb a}^{(0)} \conv \mb y }{\infty} $), and correspondingly choose a large solution tolerance $\eps = \delta \lambda_0$. The problem in \Cref{eqn:problem-general} can be solved using any local descent methods (e.g., ADM and iADM described in the previous section). Once \Cref{eqn:problem-general} is solved with given $\lambda$ and $\eps$, we sequentially decrease the penalty $\lambda$ by $\eta$ and the solution tolerance $\eps$. We use an approximate solution for $(\mb a,\mb x)$ at the end of each stage to warm restart the next stage, and repeatedly solves \Cref{eqn:problem-general} until the target penalty $\lambda_\star$ and precision $\eps_\star$ reached.

In practice, we usually set the parameters $\eta = 0.9 $ and $\delta = 10^{-1}$. As we show in \Cref{sec:exp_synthetic}, we observe linear convergence for the \emph{homotopy} continuation method works for SaSD. For SaS-CDL problem, we observe that the homotopy continuation method could occasionally produce duplicated kernels, because a large penalty $\lambda$ in the beginning stage could attract multiple different kernels to the same solution initially.

\begin{algorithm}
\caption{Reweighting method}
\label{alg:reweighted}
\begin{algorithmic}
\renewcommand{\algorithmicrequire}{\textbf{Input:}}
\renewcommand{\algorithmicensure}{\textbf{Output:}}
\Require~~\
Measurement $\mb y\in\bb R^m$; penalty $\lambda>0$; initialization $(\mb a \iter 0, \mb x \iter 0)$.
\Ensure~~\
final solution $(\mb a_\star$, $\mb x_\star)$.
\State Initialize $k \leftarrow 0$, $\mb w\iter 0 = \mb 1_m$.
\While { not converged }
\State \emph{Solve} a weighted subproblem 
\begin{align}\label{eqn:weighted-subproblem}
   \min_{\mb a,\mb x}\; \Psi^{\mb w\iter k}(\mb a,\mb x) \;=\; \psi(\mb a,\mb x) + \lambda \cdot g\paren{\mb w\iter k \odot \mb x}, \qquad \mb a \in \mc M
\end{align}
to a solution $\paren{\mb a\iter{k+1},\mb x\iter{k+1} }$, by using a warm restart $\paren{\mb a \iter k, \mb x \iter k}$.
\State \emph{Update the weights}: Compute $\eps\iter k$ using \Cref{eqn:weights}. Update the weight $\mb w\iter{k+1}$ by
\begin{align}\label{eqn:weights-1}
	w_i \iter{k+1} \;= \; \frac{1}{  \abs{ x_i\iter{k} } + \eps\iter k },\qquad 1\leq i\leq m.
\end{align}
\State \emph{Update} $k \;\leftarrow\; k +1$.
\EndWhile
\end{algorithmic}
\end{algorithm}

\paragraph{Reweighting.} Real data is often contaminated by noise, it is preferred to set large $\lambda$ on zero entries of $\mb x_0$ to suppress the noise, and set small $\lambda$ on nonzero entries of $\mb x_0$ to promote sparse solutions. This inspires us to introduce the reweighing scheme \cite{candes2008enhancing} (see \Cref{alg:reweighted}), the basic idea is to adaptively adjust the penalty $\lambda$ for each entry of $\mb x$ by considering a weighted variant of the problem \eqref{eqn:problem-general},
\begin{align}\label{eqn:problem-weighted}
   \min_{\mb a,\mb x}\; \Psi^{\mb w}(\mb a,\mb x) \;=\; \psi(\mb a,\mb x) + \lambda \cdot g\paren{\mb w \odot \mb x}, \qquad \mb a \in \mc M.
\end{align}
where $ \mb w \in \bb R_+^\sample$ is the weight and $\odot$ denotes Hadamard products. Taking SaSD problem for instance, when $g(\cdot) = \norm{\cdot}{1}$, the desired choice of the weight $\mb w$ is expected to be inversely proportional to the magnitude of the true signal $\mb x_0$,
\begin{align}\label{eqn:weights}
   w_i \;=\; \begin{cases}
 	\frac{1}{ \abs{x_{0,i}}  }, & \quad x_{0,i} \not = 0, \\
 	+\infty, & \quad x_{0,i} = 0,
 \end{cases}
\end{align}
which makes $\norm{ \mb w \odot \mb x_0  }{1} = \norm{\mb x_0}{0} $ 
%\yl{This is completely incorrect; applying reweighting using \Cref{eqn:weights} converges to $\log(\abs(x)+\eps)$}. 
The large (actually infinite) entries in $w_i$ force the solution $\mb x$ to concentrate on the indices where $w_i$ is small (actually finite), and by construction these correspond precisely to the indices where $x_0$ is nonzero. This suggests more generally that large weights could be used to discourage nonzero entries in the recovered signal, while small weights could be used to encourage nonzero entries. Although it is impossible to construct the precise weights in \Cref{eqn:weights} without knowing the signal $\mb x_0$ itself, we consider an iterative procedure (as shown in \Cref{alg:reweighted}) that alternates between estimating $\mb x_0$ and refining the weights $\mb w$.

In the following, we take $g(\cdot) = \norm{\cdot}{1}$ for an example, and provide more details of solving the weighted subproblem in \Cref{eqn:weighted-subproblem} in \Cref{alg:reweighted} with a given $\mb w$. This subproblem can be solved by either ADM or iADM without much modification. The new objective does not affect the update for $\mb a$ when $\mb x$ is fixed. When we update $\mb x$ with $\mb a$ fixed. Notice that the $\ell_1$-penalty is separable, so that we have
\begin{align*}
   \Psi^{\mb w}(\mb a,\mb x) \;=\; \psi(\mb a,\mb x) + \lambda \cdot \norm{\mb w \odot \mb x}{1} \;=\; 	\psi(\mb a,\mb x) + \sum_{i=1}^m \underbrace{\lambda w_i}_{\lambda_i} \abs{x_i}.
\end{align*}
The separability of $\ell_1$-penalty implies that we can just update each entry $x_i$ with different penalty $\lambda_i$ using proximal gradient.

For weight refinement in \Cref{eqn:weights-1}, we introduce a scalar $\eps>0$ in order to provide algorithmic stability, ensuring that a zero-valued component in $\mb x$ does not strictly prohibit a nonzero estimate in the next step. Let $\Brac{ \abs{x}_{(i)} }$ denote a descent reordering of $\Brac{\abs{x_i}}$, we empirically set
\begin{align*}
  \eps = \max\Brac{ \abs{ x }_{(i_0)} , 10^{-3}},
\end{align*}
where $i_0 = \left\lceil \Di / \log(\sample/\Di) \right\rceil$. In general, the reweighing method tends to be reasonably robust to the choice of small $\eps$.

\subsection{Miscellaneous}\label{subsec:misc}
For the remaining of this part of the appendix, we discuss about various aspects of practical issues in solving \Cref{eqn:problem-general}. We first discuss about the initialization strategy for SaSD and SaS-CDL. Second, problems in practice often possess extra structures beyond the general form we considered here (e.g., nonnegativity, bias), and our solutions often require post-processing. We discuss about these issues in more technical details.

\begin{figure}[t!]
	\centering
	\input{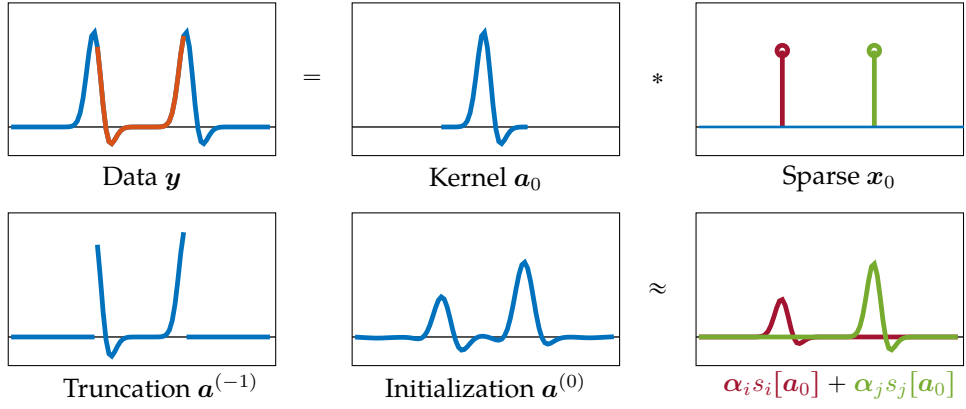}
	\caption{ \textbf{Illustration of data-driven initialization for $\mb a$:} using a piece of the observed data $\mb y$ to generate a good initial point $\mb a^{(0)}$.  Top: data $\mb y = \mb a_0 \conv \mb x_0$ is a superposition of shifts of the true kernel $\mb a_0$. Bottom: a length-$\di$ window contains pieces of just a few shifts. Bottom middle: one step of the generalized power method approximately fills in the missing pieces, yielding an initialization that is close to a linear combination of shifts of $\mb a_0$ (right).
	}
	\label{fig:init}
\end{figure}

\paragraph{Data-driven initialization.} For the SaSD problem, we usually initialize $\mb x$ by $\mb x^{(0)} = \mb 0$, so that our initialization is sparse. For the optimization variable $\mb a \in \bb R^n$, recall from \Cref{subsec:geometry} that it is desirable to obtain an initialization $\mb a^{(0)}$ which is close to $\mc S_{\mc I}$ spanned by a few shifts of $\mb a_0$ (as described in \Cref{eqn:geometry-subspace}). When $\mb x_0$ is sparse, \Cref{eqn:linear-combine-y} implies that our measurement $\mb y $ is a linear combination of a few shifts of $\mb a_0$. Therefore, intuitively an arbitrary consecutive length-$\di$ truncation $\wh{\mb y}$ of the data $\mb y$ should be not far away from such a subsphere $\mc S_{\mc I}$. As illustrated in \Cref{fig:init}, one step of the generalized power method \cite{kuo2019geometry}
\begin{align}\label{eqn:DQ-initialization}
   \mb a^{(0)} = \mc P_{\bb S^{\Di-1}}	 \nabla \paren{ - \nabla \phit{DQ}\paren{ \begin{bmatrix}
 \mb 0_{\Di-1} \\ \wh{\mb y} \\ \mb 0_{\Di-1}
 \end{bmatrix}
 } }
\end{align}
produces a refined initialization that is very close to $\mc S_{\mc I}$ with $\abs{\mc I} \leq \mc O(\theta \di)$. Moreover, in practice, we find that even a simpler initialization $\mb a^{(0)}$ (without the power iteration) in \Cref{eqn:initialization} works stably well for solving SaSD.

When dealing with multiple kernel SaS-CDL problems, we take several independent random truncations of $\mb y$, and repeat the procedure \Cref{eqn:initialization} to initialize different kernels.

\paragraph{Nonnegativity constraint.} When the signal $\mb x_0$ is nonnegative, we add an extra nonnegativity constraint to the original problem
\begin{align}\label{eqn:problem-general-pos}
   \min_{\mb a,\mb x} \Psi(\mb a,\mb x) \;=\; \psi(\mb a,\mb x) \;+\; \lambda \cdot g(\mb x), \qquad \mathrm{s.t.}\quad \mb a \in \mc M,\quad \mb x \geq 0.
\end{align}
To enforce the nonnegativity constraint on $\mb x$ in ADM or iADM, we simply modify the proximal gradient step for updating $\mb x$ in \Cref{alg:ADM,alg:iADM} by
\begin{align*}
    \mb x^{(k+1)} \;=\; \max \Brac{\mathrm{prox}_g^{ \lambda t} \paren{ \mb x^{(k)} - t \nabla \psi_{\mb a}(\mb x^{(k)})  }, \; 0},
\end{align*}
which projects the solution to the nonnegative orthant. 

\paragraph{Removing bias components.} In practice, the measurement $\mb y$ often contains a constant direct current (DC) component. Taking SaSD as an example, we often have the measurement
\begin{align*}
   \mb y \;=\; \mb a_0 \conv \mb x_0	 \;+\; b_0 \mb 1_\sample,
\end{align*}
where $b_0$ describes the magnitude of the DC component. To deal with this issue, it is natural to reformulate the Bilinear Lasso problem in \Cref{eqn:bilinear-lasso} as
\begin{align*}
   \min_{\mb a,\; \mb x, b} \Psi(\mb a,\mb x,b) \;=\;  \frac{1}{2} \norm{ \mb y - \mb a \conv \mb x - b \mb 1_\sample }{2}^2  \;+\; \lambda \norm{\mb x}{1} ,\quad \text{s.t.} \quad \mb a \in \bb S^{\Di-1},
\end{align*}
and modify the optimization methods accordingly. For ADM and iADM in \Cref{alg:ADM,alg:iADM}, we initialize $b$ as the mean value of the sequence $\mb y$, and update $\mb a$ and $\mb x$ in the original way with $b$ fixed. For optimizing variable $b$, we simply add an extra step after updating $\mb a$ and $\mb x$ by
\begin{align*}
   b^{(k)} = \frac{1}{\sample} \innerprod{ \mb 1_\sample }{ \mb y - \mb a^{(k)} \conv \mb x^{(k)} }.
\end{align*}

\paragraph{Shift correction.} The shift symmetry implies that we can only solve sparse deconvolution problems up to a shift ambiguity. However, as predicting the precise activation locations $\mb x_0$ could be mission critical in many applications, post-processing is often needed to correct the shift ambiguity by exploiting the structure of the data $\mb y$.

As aforementioned in \Cref{sec:geometry}, our optimization space $\Di = 3\di-2$ for the kernel $\mb a_0$ is larger the original dimension of $\mb a_0$, due to shift truncations. We need to truncation the solution $\mb a$ produced by our algorithm to obtain an approximation of the original kernel. A natural idea is to find a length-$\di$ subvector (submatrix) of the produced solution $\mb a_\star$ that maximizes the Frobenius norm across all length-$\di$ subvectors. Therefore, we simply shift $\mb a_\star$ so that the chosen length-$\di$ window is in the top left, and remove other zero-padding entries if needed. Correspondingly, the solution $\mb x_\star$ can be corrected by shifting the same amount of length in the opposite direction. However, using Frobenius norm could be unstable when large noise presents. \Cref{alg:shift_correction} presents an alternative approach based on the reconstruction error, which turns out to be more reliable in some cases.

\begin{algorithm}
\caption{Shift correction}
\label{alg:shift_correction}
\begin{algorithmic}
\renewcommand{\algorithmicrequire}{\textbf{Input:}}
\renewcommand{\algorithmicensure}{\textbf{Output:}}
\Require~~\
observation $\mb y$, optimal solution $(\mb a_\star, \mb x_\star)\in \bb R^\Di \times \bb R^\sample $.
\Ensure~~\
Solution $(\mb a$, $\mb x)\in \bb R^{\di} \times \bb R^\sample  $ after shift correction.

\For{ $i\;=\;1\;:\;2\di+1$}
\State Set $\wh{\mb a} = \res{\Di}{\di}\mathrm{s}_{-i+1}\brac{\mb a_\star}$, $\wh{\mb x} = \mathrm{s}_{i-1}\brac{\mb x_\star}$;
\State Compute $\wh{\mb y}_i = \wh{\mb a} \conv \wh{\mb x}$;
\EndFor
\State Find $i_\star = \arg\min_i \Brac{ \norm{\wh{\mb y}_i - \mb y}{2}  } $;
\State Set $\mb a = \res{\Di}{\di}\mathrm{s}_{-i_\star+1}\brac{\mb a_\star}$, $\mb x = \mathrm{s}_{i_\star-1}\brac{\mb x_\star}$.
\end{algorithmic}
\end{algorithm}

\section{Implementation details for SaSD and SaS-CDL}\label{app:details}

Finally, we provide missing implementation details of the proposed ADM and iADM algorithms for both SaSD and SaS-CDL. We show how to solve these problems for both cases when the observation is 1-dimensional (1D) and 2-dimensional (2D).

\subsection{Technical details for solving 1D problems}

\subsubsection{Implementations details for SaSD} 
We optimize \Cref{eqn:problem-general} in Appendix \ref{app:algorithm} for the SaSD problem, 
\begin{align*}
   \min_{\mb a,\mb x}\; \Psi(\mb a,\mb x)\;= \; \underbrace{\frac{1}{2} \norm{ \mb y - \mb a \conv \mb x }{2}^2 }_{\psi(\mb a,\mb x) } \;+\; \lambda \cdot \underbrace{ \norm{\mb x}{1} }_{g(\mb x)} ,\qquad \text{s.t.}\quad \mb a \in \underbrace{\bb S^{n-1}}_{\mc M},
\end{align*}
where $\psi(\mb a,\mb x) = \frac{1}{2} \norm{ \mb y - \mb a \conv \mb x }{2}^2$, $g(\mb x)=\norm{\mb x}{1}$, and $\mc M = \bb S^{n-1}$. Next, we provide missing implementation details (e.g., exact forms of the gradients) of SaSD for ADM and iADM in \Cref{alg:ADM} and  \Cref{alg:iADM}.

\paragraph{Update $\mb x$ with $\mb a$ fixed.} For the proximal gradient step on $\mb x$ in \Cref{eqn:prox-grad}, the proximal operator of $g(\cdot) = \norm{ \cdot }{1} $ is the soft thresholding operator
\begin{align*}
   	\mathrm{prox}_{\norm{ \cdot }{1}}^{\lambda t}(z) \;=\; \mc S_{\lambda t}(z),\qquad \mc S_{\lambda t}(z) \;=\; \sign(z) \cdot \max\Brac{ \abs{z} - \lambda t , 0}.
\end{align*}
The gradient of $\psi_{\mb a}(\mb x)$ is 
\begin{align*}
   \nabla \psi_{\mb a}(\mb x)\;=\; \wc{\mb a} \conv \paren{ \mb a \conv \mb x - \mb y }.
\end{align*}

\paragraph{Update $\mb a$ with $\mb x$ fixed.} For the Riemannian manifold $\mc M = \bb S^{\Di-1}$, its tangent space $T_{\mb a}\bb S^{\Di-1}$ and the projection onto $T_{\mb a}\bb S^{\Di-1}$ are
\begin{align*}
   	T_{\mb a}\bb S^{\Di-1} \;=\; \Brac{ \mb z \in \bb R^\Di \;\mid\; \mb a^\top \mb z = 0 },\qquad \mc P_{T_{\mb a}\bb S^{\Di-1}} = \mc P_{\mb a^\perp} = \mb I_\Di - \frac{1  }{ \norm{\mb a}{2}^2 }  \mb a \mb a^\top.
\end{align*}
For the Riemannian gradient step on $\mb a$ presented in \Cref{eqn:riemannian-grad}, the Riemannian gradient of $\psi_{\mb x}(\mb a)$ over $\bb S^{\Di-1}$ is
\begin{align*}
  \grad \psi_{\mb x}(\mb a) \;=\; \mc P_{\mb a^\perp} \nabla \psi_{\mb x}(\mb a), \qquad \nabla \psi_{\mb x}(\mb a) \;=\; \res{\Di}{\sample}^* \wc{\mb x} \conv \paren{ \mb a \conv \mb x - \mb y }.
\end{align*}
In addition, for $\mc M = \bb S^{\Di-1}$, the retraction operator $\mc R_{\mb a}^{\bb S^{\Di-1}}(\mb \delta)$ for $\mb \delta \in T_{\mb a}\bb S^{\Di-1}$, and its inversion $\paren{\mc R_{\mb a}^{\bb S^{\Di-1}}}^{-1}(\mb b)$ for $\mb b \in \bb S^{\Di-1}$, can be specified as 
\begin{align}\label{eqn:retract-sphere}
   \mc R_{\mb a}^{\bb S^{n-1}}(\mb \delta) \;=\; \mb a \cdot \cos\paren{ \norm{\mb \delta}{2} } + \frac{ \mb \delta }{ \norm{\mb \delta}{2} } \sin \paren{ \norm{\mb \delta}{2} },\quad \paren{\mc R_{\mb a}^{\bb S^{n-1}}	}^{-1} (\mb \delta) \;=\; \frac{ \alpha }{ \sin \alpha } \mc P_{\mb a^\perp} \mb \delta,
\end{align}
where $\alpha = \arccos\paren{ \mb a^\top \mb \delta}$.

\subsubsection{Implementations details for SaS-CDL} 

Since the SaSD problem can be considered a special case of SaS-CDL with $N = 1$, the derivations we obtained for SaSD can be easily extended to SaS-CDL. Recall from \Cref{eqn:bilinear-lasso-CDL}, the general objective in \Cref{eqn:problem-general} can specified as
\begin{align*}
   \min_{\mb A,\; \mb X}\; \Psi(\mb A,\mb X) = \underbrace{ \frac{1}{2} \norm{ \mb y - \sum_{k=1}^N \mb a_k \conv \mb x_k }{2}^2 }_{ \psi(\mb A,\mb X)  } \;+\; \lambda \underbrace{ \norm{\mb X}{1} }_{g(\mb X)},\qquad \text{s.t.} \quad \mb A \;\in\; \underbrace{ \mc {OB}(\Di,N) }_{\mc M},
\end{align*}
where $\mc {OB}(n,N)$ is called the \emph{oblique} manifold with
\begin{align*}
	\mc {OB}(\Di,N) = \Brac{ \mb Z
 \in \bb R^{\Di \times N} \;\mid\; \mb Z = \begin{bmatrix}
   \mb z_1 & \cdots & \mb z_N	
 \end{bmatrix}, \; \mb z_i \in \bb S^{\Di-1}, \; 1\leq i \leq N } \;=\; \underbrace{\bb S^{\Di-1} \times \cdots \times \bb S^{\Di-1}}_{N},
\end{align*}
which is essentially a product of $N$ spheres. Next, we provide missing implementation details of solving SaS-CDL by ADM and iADM in  \Cref{alg:ADM} and \Cref{alg:iADM}.

\paragraph{Update $\mb X$ with $\mb A$ fixed.} First, for the proximal gradient step on $\mb X$ presented in \Cref{eqn:prox-grad}, similarly we have  $\mathrm{prox}_{\norm{ \cdot }{1}}^{\lambda t}(z) = \mc S_{\lambda t}(z)$ and 
\begin{align*}
	\nabla \psi_{\mb A}\paren{ \mb X } &\;=\; \begin{bmatrix} \nabla_{\mb x_1} \psi_{\mb A}(\mb X) & \nabla_{\mb x_2} \psi_{\mb A}(\mb X) & \cdots & \nabla_{\mb x_N} \psi_{\mb A}(\mb X) \end{bmatrix}, \\
   	\nabla_{\mb x_i} \psi_{\mb A}(\mb X) &\;=\;\wc{\mb a}_i \conv \paren{ \sum_{j=1}^N \mb a_j \conv \mb x_j - \mb y} ,\quad 1 \leq i \leq N.
\end{align*}
Second, for the Riemannian manifold $\mc M = \mc {OB}(\Di,N)$, its tangent space $T_{\mb A}\mc {OB}(\Di,N)$ and the projection onto $T_{\mb A}\mc {OB}(\Di,N)$ are
\begin{align*}
  	T_{\mb A}{\mc {OB}}(\Di,N) \;=\; T_{\mb a_1}\bb S^{\Di-1} \times \cdots \times  T_{\mb a_N}\bb S^{\Di-1},\qquad \mc P_{ T_{\mb A}{\mc {OB}} }(\mb Z) = \begin{bmatrix}
 	\mb P_{\mb a_1^\perp} \mb z_1 & \mb P_{\mb a_2^\perp} \mb z_1 & \cdots & \mb P_{\mb a_N^\perp} \mb z_N
 \end{bmatrix}.
\end{align*}

\paragraph{Update $\mb A$ with $\mb X$ fixed.} For the Riemannian gradient step on $\mb A$ presented in \Cref{eqn:riemannian-grad}, we have the Riemannian gradient of $\psi_{\mb X}(\mb A)$ over $\mc {OB}(\Di,N)$ as
\begin{align*}
   \grad \psi_{\mb X}(\mb A) \;&=\; \begin{bmatrix}
 	 \grad_{\mb a_1} \psi_{\mb X}(\mb A), \grad_{\mb a_2} \psi_{\mb X}(\mb A), \cdots, \grad_{\mb a_N} \psi_{\mb X}(\mb A)
 \end{bmatrix},\\
   \grad_{\mb a_i} \psi_{\mb X}(\mb A)	\;&=\; \mc P_{\mb a_i^\perp} \nabla_{\mb a_i} \psi_{\mb X}(\mb A) = \mc P_{\mb a_i^\perp} \res{n}{m}^* \wc{\mb x}_i \conv \paren{ \sum_{j=1}^N \mb a_j \conv \mb x_j - \mb y },\; 1\leq i \leq N,
\end{align*}
and the retraction operator $\mc R_{\mb A}^{\mc {OB}(\Di,N)}(\mb \Delta)$ for $\mb \Delta = \begin{bmatrix} \mb \delta_1 & \mb \delta_2 & \cdots & \mb \delta_N \end{bmatrix} \in T_{\mb A}{\mc {OB}}(\Di,N)$ can be specified as 
\begin{align*}
   \mc R_{\mb A}^{\mc {OB}(\Di,N)}(\mb \Delta) \;=\; \begin{bmatrix}
 	\mc R_{\mb a_1}^{\bb S^{\Di-1}}(\mb \delta_1) & \mc R_{\mb a_2}^{\bb S^{\Di-1}}(\mb \delta_2) &\cdots & \mc R_{\mb a_N}^{\bb S^{\Di-1}}(\mb \delta_N)
 \end{bmatrix},
\end{align*}
where $\mc R_{\mb a}^{\bb S^{\Di-1}}(\mb \delta)$ is the retractor operator over the sphere as introduced in \Cref{eqn:retract-sphere}. The inverse retraction $\paren{\mc R_{\mb A}^{\mc {OB}(\Di,N)}}^{-1}(\mb B)$ for $\mb B \in \mc {OB}(\Di,N)$ can be constructed similarly by using \Cref{eqn:retract-sphere}.

\begin{comment}
\noindent\yl{Our algorithms currently use the exponential retraction
$$\mc R^{\mc M}_{\mb a}(\bm\delta) = \text{Exp}_{\mb a}(\bm\delta) = \cos(\nrm{\bm\delta})\mb a + \sin(\nrm{\bm\delta}) \frac{\bm\delta}{\nrm{\bm\delta}},$$
with corresponding inverse logarithmic map with $\alpha = \arccos \paren{ \mb a^\top \mb b }$
\begin{align*}
   \paren{\mc R_{\mb a}^{\mc M}	}^{-1} (\mb b) = \text{Log}_{\mb a}(\mb b) = \frac{ \alpha }{ \sin \alpha } \mc P_{\mc T_{\mb a}} \mb b
\end{align*}

$$\left(\mc R^{\mc M}_{\mb a}\right)^{-1}(\mb b) = \text{Log}_{\mb a}(\mb b) = \langle\mb a, \mb b\rangle \frac{\mc P_{\mc T_{\mb a}} (\mb b - \mb a)}{\nrm{\mc P_{\mc T_{\mb a}} (\mb b - \mb a)}}.$$
I'm wondering whether we should use the simpler retraction
$$\mc R^{\mc M}_{\mb a}(\bm\delta) = \frac{\mb a + \bm\delta}{\nrm{\mb a + \bm\delta}},$$
with corresponding inverse
$$\left(\mc R^{\mc M}_{\mb a}\right)^{-1}(\mb b) = \frac{\mc P_{\mc T_{\mb a}} (\mb b - \mb a)}{\innerprod{\mb a}{\mb b}}?$$
This might be easier to follow for the reader, since this retraction mapping is familiar for most users.
}
\end{comment}

\subsection{Brief technical details of solving 2D problems}
The derivative of 2D problems is slightly different from the 1D case, which we briefly introduce below.
\paragraph{Implementations details for SaSD.}  
Let $\mb {\mc Y} = \mb {\mc A}_0 \cconv \mb {\mc X}_0 \in \bb R^{\sample_1 \times \sample_2}$ be a 2D circular convolution of a kernel $\mb {\mc A}_0 \in \bb R^{\Di_1 \times \Di_2}$ and activation map $\mb {\mc X}_0 \in \bb R^{\sample_1 \times \sample_2}$, where $\cconv$ denotes the 2D circular convolution. For the 2D SaSD problem, we consider
\begin{align*}
\min_{\mb {\mc A},\mb {\mc X} } \; \Psi\paren{\mb {\mc A},\mb {\mc X}} \;=\; \underbrace{ \frac{1}{2} \norm{ \mb {\mc Y} - \mb {\mc A} \cconv \mb {\mc X} }{\mathrm{F}}^2 }_{ \psi\paren{\mb {\mc A},\mb {\mc X}} }	\;+\; \underbrace{ \lambda \norm{\mb {\mc X}}{1} }_{g(\mb {\mc X})}, \quad \text{s.t.}\quad \underbrace{\norm{\mb {\mc A}}{\mathrm{F}} = 1}_{\mc M},
\end{align*}
where $\norm{\cdot}{\mathrm{F}}$ denotes the Frobenius norm. Similarly, we have the gradients
\begin{align*}
   \nabla_{\mb {\mc X}}	\psi_{ \mb {\mc A}  }\paren{ \mb {\mc X} } \; &=\; \wt{\mb {\mc A}} \cconv \paren{ \mb {\mc A} \cconv \mb {\mc X} - \mb {\mc Y}  }, \\
   \nabla_{\mb {\mc A}}	\psi_{ \mb {\mc X}  }\paren{ \mb {\mc A} } \;&= \; \res{n_1}{m_1}^* \wt{\mb {\mc X}} \cconv \paren{ \mb {\mc A} \cconv \mb {\mc X} - \mb {\mc Y}  } \res{\Di_2}{\sample_2},
\end{align*}
where $\wt{\mb {\mc Z} }$ denotes a flip operator that flips a matrix $ \mb {\mc Z}$ both vertically and horizontally, i.e.,
\begin{align*}
    	\wt{\mb {\mc Z} }(i,j) = \mb {\mc Z}(\sample_1-i+1, \sample_2 -j+1).
\end{align*}
Note that $\wt{\mb {\mc Z} } \cconv  \mb {\mc V} $ is essentially 2D auto-correlation of $\mb {\mc Z}$ and $\mb {\mc V}$, so that we can rewrite 
\begin{align*}
  	\nabla_{\mb {\mc X}}	\psi_{ \mb {\mc A}  }\paren{ \mb {\mc X} } \;&= \;  \mc F^{-1} \brac{ \mc F^*\paren{\mb {\mc A}} \odot \mc F\paren{\mb {\mc A} \cconv \mb {\mc X} - \mb {\mc Y} } } \\
  	\nabla_{\mb {\mc A}}	\psi_{ \mb {\mc X}  }\paren{ \mb {\mc A} } \;&= \; \res{\Di_1}{\sample_1}^* \mc F^{-1} \brac{ \mc F^*\paren{\mb {\mc X}} \odot \mc F\paren{\mb {\mc A} \cconv \mb {\mc X} - \mb {\mc Y} } } \res{\Di_2}{\sample_2},
\end{align*}
where $\mc F$ denotes the 2D Fourier transform operator, and $\mc F^*$ is its adjoint operator. Finally, we have the Riemannian gradient
\begin{align*}
   \grad \psi_{\mb {\mc X}} \paren{ \mb {\mc A} } \;=\; \mb P_{\mb {\mc A}^\perp } \nabla_{\mb {\mc A}}	\psi_{ \mb {\mc X}  }\paren{ \mb {\mc A} }, \quad \mb P_{\mb {\mc A}^\perp } \mb {\mc Z} \;=\; \mb {\mc Z} -  \frac{\mb {\mc A}}{ \norm{ \mb {\mc A} }{\mathrm{F}}^2  }  \innerprod{ \mb {\mc A} }{ \mb {\mc Z}}.
\end{align*}
The retraction operator and its inversion remain the same as \Cref{eqn:retract-sphere}.

\paragraph{Implementations details for SaS-CDL.} For the multiple kernel deconvolution problem $\mb {\mc Y} =  \sum_{k=1}^N \mb {\mc A}_{0k} \cconv \mb {\mc X}_{0k}$, let the optimization variable $\ol{\mb {\mc A}} \in \bb R^{\Di_1 \times \Di_2 \times N} $ and $\ol{\mb {\mc X}} \in \bb R^{\sample_1 \times \sample_2 \times N} $ be 3-way tensors, with
\begin{align*}
  \ol{\mb{ \mc A} }(:,:,k) = \mb {\mc A}_k, \quad \ol{\mb{ \mc X} }(:,:,k) = \mb {\mc X}_k, \quad 1\leq k \leq N.
\end{align*}
Similar to the 1D case in \Cref{eqn:bilinear-lasso-CDL}, we optimize the following problem
\begin{align*}
   \min_{\ol{\mb {\mc A}}, \; \ol{\mb {\mc X}} }\; 	\Psi\paren{ \ol{\mb {\mc A} }, \ol{ \mb {\mc X} } }  \underbrace{ \frac{1}{2} \norm{ \mb {\mc Y} - \sum_{k=1}^N \mb {\mc A}_k \cconv \mb {\mc X}_k }{\mathrm{F}}^2 }_{ \psi\paren{ \ol{\mb {\mc A}} , \ol{\mb {\mc X}}}} + \lambda \underbrace{ \norm{ \ol{\mb {\mc X}} }{1} }_{ g(\ol{ \mb {\mc X} }) } ,\quad \underbrace{\text{s.t.}\;\norm{ \mb {\mc A}_k }{\mathrm{F}} = 1 \;(1\leq k \leq N)}_{\mc M}.
\end{align*}
The gradient of $\psi_{\ol{ \mb {\mc A} }}\paren{ \ol{ \mb {\mc X} } }$ and $\psi_{\ol{ \mb {\mc X} }}\paren{ \ol{ \mb {\mc A} } }$ can be computed in a similar manner as SaSD. Let $\nabla \psi_{\ol{ \mb {\mc A} }}\paren{ \ol{ \mb {\mc X} } }$ and $\nabla \psi_{\ol{ \mb {\mc X} }}\paren{ \ol{ \mb {\mc A} } }$ denote the gradient of $\psi_{\ol{ \mb {\mc A} }}\paren{ \ol{ \mb {\mc X} } }$ and $\psi_{\ol{ \mb {\mc X} }}\paren{ \ol{ \mb {\mc A} } }$, then we have
\begin{align*}
	\nabla \psi_{\ol{ \mb {\mc A} }}\paren{ \ol{ \mb {\mc X} } }(:,:,i) \;&=\; \wt{\mb {\mc A}}_i \cconv \paren{ \sum_{j=1}^N \mb {\mc A}_j \cconv \mb {\mc X}_j - \mb {\mc Y}},\quad 1\leq i\leq N \\ 
	\nabla \psi_{\ol{ \mb {\mc X} }}\paren{ \ol{ \mb {\mc A} } }(:,:,i) \;&=\; \res{\Di_1}{\sample_1}^* \wt{\mb {\mc X}}_i \cconv \paren{ \sum_{j=1}^N \mb {\mc A}_j \cconv \mb {\mc X}_j - \mb {\mc Y}} \res{\Di_2}{\sample_2},\quad 1\leq i\leq N.
\end{align*}
The Riemannian gradient $\grad \psi_{\ol{\mb {\mc X}}} \paren{ \ol{\mb {\mc A} }}$, and the retraction operator can be generalized from 1D case in a very similar fashion. We omit the details here.

%\section{Supplementary of Experiments}

\end{document}